\documentclass{agujournal}

\journalname{JGR-Planets}

\begin{document}

%
%

\title{Impact of gravity waves on the middle atmosphere of Mars: a non-orographic gravity wave parameterization based on Global Climate modeling and MCS observations}

%
%




\authors{G. Gilli\affil{1,2}, F. Forget\affil{2}, A. Spiga\affil{2}, T. Navarro\affil{3}, E. Millour\affil{2}, L. Montabone\affil{2,4}, 
	  A. Kleinb\"ohl\affil{5}, D. M. Kass\affil{5}, D. J. McCleese\affil{6}, J. T. Schofield\affil{5}}

\affiliation{1}{Instituto de Astrof\'isica e Ci\^{e}ncias do Espa\c{c}o (IA), Universidade de Lisboa, OAL, Tapada da Ajuda, PT1349-018 Lisboa, Portugal}
\affiliation{2}{Laboratoire de M\'et\'eorologie Dynamique (LMD/IPSL), Centre National de la Recherche Scientifique, Sorbonne Universit\'e, \'{E}cole Normale Sup\'erieure, \'{E}cole Polytechnique, Paris, France}
\affiliation{3}{Department of Earth, Planetary, and Space Sciences, University of California, Los Angeles, CA 90095-1567, USA}
\affiliation{4}{Space Science Institute, Boulder, Colorado, USA}
\affiliation{5}{Jet Propulsion Laboratory, California Institute of Technology,
	Pasadena, California, USA}
\affiliation{6}{California Institute of Technology, Pasadena, California, USA}



\correspondingauthor{Gabriella Gilli}{ggilli@oal.ul.pt}




\begin{keypoints}

\item A Stochastic non-orographic gravity wave (GW) scheme is implemented into the LMD-MGCM
\item Non-orographic GW generated above typical convective layers control diurnal tides
\item The implemented GW scheme improves the accuracy of the LMD-MGCM between 1 and 0.01 Pa 
in comparison with MCS

\end{keypoints}

%
%


\begin{abstract}
The impact of gravity waves (GW) on diurnal tides and the global circulation in the middle/upper atmosphere of Mars is investigated using a General Circulation Model (GCM). 
We have implemented a stochastic parameterization of non-orographic GW into the Laboratoire de M\'et\'eorologie Dynamique (LMD) Mars GCM (LMD-MGCM) following an innovative approach. The source is assumed to be located above typical convective cells ($\sim$ 250 Pa) and the effect of GW on the circulation and predicted thermal structure above 1 Pa ($\sim$ 50 km) is analyzed. We focus on the comparison between model simulations and observations by the Mars Climate Sounder (MCS) on board Mars Reconnaissance Orbiter during Martian Year 29. MCS data provide the only systematic measurements of the Martian mesosphere up to 80 km to date.
The primary effect of GW is to damp the thermal tides by reducing the diurnal oscillation of the meridional and zonal winds. The GW drag reaches magnitudes of the order of 1 m/s/sol above 10$^{-2}$ Pa in the northern hemisphere winter solstice and produces major changes in the zonal wind field (from tens to hundreds of m/s), while the impact on the temperature field is relatively moderate (10-20K). It suggests that  GW induced alteration of the meridional flow is the main responsible for the simulated temperature variation.
The results also show that with the GW scheme included, the maximum day-night temperature difference due to the diurnal tide is around 10K, and the peak of the tide is shifted toward lower altitudes, in better agreement with MCS observations.

\end{abstract}

%
%

\section{Introduction }
Gravity waves (GW) are small-scale atmospheric waves frequently detected in terrestrial planet atmospheres. They are an intrinsic feature of all stably stratified planetary atmosphere and play an important role in the large-scale circulation and variability of the middle/upper atmospheres of the Earth and Mars \citep{Barnes1990, Joshi1995, Forget1999, Angelatsicoll2005, Fritts2006, FrittsAlexander2003, Alexander2010, Medvedev2011}.
Depending on the source, GW can be excited in the troposphere by a variety of mechanisms, 
 which perturb the stratified atmospheric fluid and produce GW oscillations. The sources
include flow over topography (orographic GW) or atmospheric convection, front systems and jet-streams (non-orographic GW). The restoring force is the buoyancy that results from the adiabatic displacements of air parcels characteristic of these disturbances \citep{FrittsAlexander2003}.
GW propagation into the atmosphere  provides a significant source of momentum and energy to the mean flow when they break or encounter critical values. Thus, the gravity wave drag (i.e. gravity wave momentum deposition to the mean flow) may cause wind acceleration or deceleration. 
\\On the Earth, the observed reversal of the temperature gradient in the mesopause region, resulting in a winter polar warming around the mesopause level ($\sim$ 80-90 km), is related to the effects of GW drag on the global circulation \citep{Lindzen1981, Holton1982, Hauchecorne1987}. 
It is also well established that non-stationary (i.e. non-orographic)  GW are a substantial driver of the quasi-biennial oscillation (QBO) in the equatorial regions of the Earth's atmosphere \citep{LindzenHolton1968, Lott2012}, complementing the forcing from the synoptic and planetary scale equatorial waves \citep{Lott2013}.
\\For Mars, several studies reported large density and temperature fluctuations ($\sim$ 5 to 50 $\%$) on small vertical scales height (<10 km vertically) from atmospheric entry profiles measured by Opportunity, Spirit, and Mars Pathfinder  \citep{Magalhaes1999, WithersSmith2006, Holstein-Rathlou2016}, as well as small horizontal scales (typically 20-300 km) in various Mars accelerometer data sets at aerobraking altitudes \citep{Keating1998, Withers2006, Creasey2006b, Fritts2006} and from mass spectrometer measurements [\citealp{Yigit2015} and, \citealp{Terada2017}].
Radio occultation temperature profiles by Mars Global Surveyor (MGS) obtained from the surface up to 35 km altitude also showed significant wave activity over the tropics, also partly attributed to zonally modulated thermal tides \citep{Hinson1999, Creasey2006a}. 
Furthermore, "rocket dust storms" during which rapid and efficient vertical transport takes place by injecting dust particles at high altitudes in the Martian troposphere (30 to 50 km), may generate GW \citep{Spiga2013}.
The role of mesoscale GW is supposed to be crucial for local CO$_2$ condensation, responsible  for the formation of mesospheric CO$_2$ clouds observed by Mars Express between 60 and 80 km altitude  [\citealp{Spiga2012} and \citealp{Montmessin2007}].
\\MGS, Mars Odyssey (ODY) and Mars Reconnaissance Orbiter (MRO) aerobraking measurements were used to analyze density and temperature perturbations in the Martian thermosphere, where small-scale variability has been systematically observed whenever in situ data have been obtained (e.g. \citet{Keating2007AGU, Fritts2006, Creasey2006b, ZurekSmekar2007}). MGS results (both radio profile and accelerometer data) leave open the question of the primary source mechanisms and source spectra, such as the distribution of phase speeds.
Regarding the source, convection is often invoked to generate non-orographic GW (e.g. non stationary waves with non-zero phase speed) which propagate upwards \citep{Yigit2015, Medvedev2011}. 
\citet{Fritts2006} and \citet{Creasey2006b} found that GW amplitudes vary significantly with time and season, being generally larger in Northern autumn and winter (i.e. second half of Martian year), and at middle to high latitudes, apparently reflecting mean source and filtering conditions. Their amplitudes also appear to vary with longitude and time and may provide clues to interactions with larger-scale motions \citep{Fritts2006}. According to \citet{Creasey2006a}, westward-propagating waves may be encountering a critical level in spring and summer, but as the jet decreases towards autumn equinox, these waves may propagate to the thermosphere.
More recently, NASA's Neutral Gas Ion Mass Spectrometer (NGIMS) instrument aboard the Mars Atmosphere and Volatile EvolutioN (MAVEN) satellite retrieved large (20-40 $\%$) GW-induced CO$_2$ density perturbations in the Mars upper atmosphere between 180 and 220 km \citep{Yigit2015}. Wave features were found with apparent wavelengths of $\sim$ 100 and 500 km in the Ar density around the exobase \citep{Terada2017}.

Many different parameterizations including those of \cite{Lindzen1981, Hines1997} have been used to represent the impact of subgrid scale gravity wave processes in the terrestrial atmosphere.  For Mars, the inclusion of orographic GW effects improves the performance of models at lower and higher altitudes \citep{Barnes1990, Collins1997, Forget1999, Rafkin2001, Angelatsicoll2005}. Also, the thermal effect of non-orographic gravity waves has been proposed to explain some of the puzzling model-observation discrepancy identified in the Martian atmosphere temperatures between 100 and 140 km \citep{MedvedevYigit2012}, although a major part of the discrepancy can be attributed to issues in the calculation of the non-LTE radiative cooling by CO$_2$ \citep{ Forget2009, Medvedev2015}. 
\citet{Medvedev2015} implemented a nonlinear spectral gravity wave parameterization \citep{Yigit2010} in their GCM, extended up to 130 km, and their simulations showed that: 1. GW decelerate zonal winds at all seasons, 2. they produce jet reversals  similar to those observed in the terrestrial mesosphere and lower thermosphere, 3. GW weaken the meridional wind and modify the zonal mean temperature by up to $\pm$ 15 K. However, modeling efforts for Mars suffer from lack of measurements to validate predicted wind fields and from no observational constraints on the GW forcing \citep{Creasey2006b}
Other studies performed  high-resolution simulations ($\sim$ 60 km grid size) with a general circulation model in order to resolve a significant portion of small-scale GWs and to capture the impact of GWs on the dynamics and energetics of Mars atmosphere without any parameterizations \citep{Kuroda2015, Kuroda2016, Kuroda2019}. Those authors showed that the GW activity varies greatly with season and geographical location, with stronger wave generation in the northern hemisphere winter in the mesosphere and  smaller activity in polar regions of the troposphere throughout all seasons. Given the lack of global observations of GW the climatology of the small-scale disturbances obtained in \citet{Kuroda2019} can serve as a proxy for gravity waves that are largely not resolved by GCMs with conventional grid resolution.

In this paper we describe the results of the implementation of a non-orographic GW parameterization into the Mars Global Climate Model (MGCM) developed at the Laboratoire de M\'et\'eorologie Dynamique (LMD), which already included an orographic GW scheme \citep{Forget1999}.
Although the LMD-MGCM was already able to reproduce temperature observations with an error of less than 10 K at most locations and times, we found that it was necessary to update the representation of several physical processes at work in the Martian atmosphere  to further improve its accuracy. Among them: 1) the vertical distribution of the dust, characterized by detached layers (between 20 and 30 km) and by large day-night variation \citep{McCleese2010, Heavens2011a, Heavens2011b, Heavens2011c, Madeleine2011, Navarro2014},  2) the radiative impact of clouds, which remains challenging to be well represented in the GCM \citep{Navarro2014}, 3) the effect of GW (orographic and/or non-orographic), 4) the phasing of the thermal tide wave in the vertical compared to data from the Mars Climate Sounder (MCS) on board MRO \citep{Navarro2017}.  
MCS-MGCM biases probably resulted from the incorrect representation of the three previous processes mentioned above \citep{Forget2017mamo} and several improvements are currently being implemented.
\\Here we took advantage of the previous development work carried out by our team for the LMD Earth \citep{Lott2012} and Venus GCM \citep{Gilli2017} to also implement the same non-orographic GW scheme in the MGCM.
According to \citet{Lott2012}, the stochastic parameterization represents the unpredictable aspects of the sub-grid dynamics, considering that convection and fronts can generate waves throughout the full range of phase speeds, wave frequency, vertical and horizontal scales.In a context where the sources are not known precisely, we consider this is a reasonable choice.
One of the main goals of this study is to understand the impact of non-orographic GW on the global circulation and  the thermal structure of the Martian middle atmosphere (50-100 km altitude). What is the magnitude of GW-induced drag? Where do GW break/saturate and deposit the maximum momentum? Can GW explain the remaining discrepancy between the MCS observations and the GCM simulations? If so, what is their impact on the predicted winds?

Section \ref{gw_param} describes the non-orographic GW scheme and the main tunable parameters adopted in this work. The LMD-MGCM model and MCS dataset used here are described in Sections 3 and 4, respectively. The impact of our non-orographic GW scheme on the simulated wind and temperature is described in Section \ref{impact_jets}. The method implemented here to identify the set of "best-fit" GW parameters is explained in Section \ref{results}, together with the comparison with MCS results. A number of sensitivity tests is discussed in Section 7 and concluding remarks are given in the Section \ref{conclusions}.

\section{Non-orographic GW parameterization: formalism}
\label{gw_param}

The scheme implemented here is based on a stochastic approach, as fully described  in \citet{Lott2012} and \citet{Lott2013}. In such a scheme a finite number of waves (say M = 8), with characteristics chosen randomly, are launched upwards at each physical time-step ($\delta t$ = 15 min) and at each horizontal grid point.The question of the location of the non-orographic GW sources is a difficult one and given the lack of information on such location, launching the gravity waves at each grid point is probably the simplest assumption. The number of waves is given as M = NK $\times$ NO $\times$ NP, with NK=2 values of GW horizontal wave-numbers, NO=2 absolute values of phase speed, and NP= 2 directions (westward and eastward) of phase speed.
This approach allows the model to treat a large number of waves at a given time $t$ by adding the effect of those M waves to that of the waves launched at previous steps, to compute the tendencies. 
We need to parameterize GWs whose life cycle (i.e. from its generation to wave break) is contained in a characteristic time interval  $\Delta t$ that has to significantly exceed the GCM time step $\delta t$. On the Earth,  GW theory indicates that atmospheric disturbances induced by convection have life cycles with duration $\Delta t$ around 1 day ($\Delta t$ = 24h) \citep{Lott2013}. This typical time scale is likely to be relevant for Mars too: the choice for this timescale on Earth is made considering a mix of inertio-gravity waves (influenced by the planetary rotation, which is similar on both planets) and convectively-generated gravity waves, which have shorter frequency than the inertio-gravity waves, but which can propagate for several hours before dissipation or breaking occurs.
The spectrum is discretized in 770 stochastic harmonics ($\approx$ M x $\Delta t/ \delta t$) which contribute to the wave field each day and at a given horizontal grid point. At each time $t$ the vertical velocity field $w'$ of an upward propagating GW can be represented by this sum:
\begin{equation}
	w' =\sum_{n=1}^{\infty} C_{n}  w'_{n}
\end{equation}  where $C_{n}$  are normalization coefficients such that $ \sum_{n=1}^{\infty} C_{n}^{2}  = 1$. It is assumed that each of the $w'_{n}$ can be treated independently one from the others, and each  $C_{n}^{2}$ can be viewed as the probability that wave field is given by the GW $w'_{n}$ (see also the expression for $C_{n}$ in Equation \ref{Cn_equation}). This formalism is then applied to a very simple multi-wave parameterization, in which $w'_{n}$ represents a monochromatic wave as follows
\begin{equation}
\label{wn}
	w'_{n} =  \Re  \left\lbrace \hat{w}_{n}(z)e^{z/2H}e^{i(k_{n}x+l_{n}y- \omega_{n}t)}  \right\rbrace 
\end{equation}

where the wavenumbers $k_{n}$, $l_{n}$ and frequency $\omega_{n}$ are chosen randomly. In Equation \ref{wn}, H$\sim$ 11 km is a middle atmosphere characteristic vertical scale for Mars and z is the log-pressure altitude z = H $\ln$(P$_r$/P), with $P_r$ a reference pressure (P$_r$ = 250 Pa), taken here at the height of the source (i.e. above typical convective cells $ z \sim$ 8 km). To evaluate the amplitude of $\hat{w}_{n}$ we randomly impose it at a given launching altitude $z_0$, and then iterate from one model level $z_1$ to the next $z_2$ by a Wentzel-Kramers-Brillouin (WKB) approximation (see Equation 4 of \citet{Lott2012} for details). Using that expression plus the polarization relation between the amplitudes of large scale horizontal wind $\hat{u}$ and vertical wind $\hat{w}$ (not shown here), we can deduce the Eliassen-Palm (EP) flux (vertical momentum flux of waves),
\begin{equation}
\label{EPflux}
\vec{F}^z (k,l, \omega) = \Re \{ \rho_{r} \vec{\hat{u}} \hat{w^*} \}   = \rho_{r} \frac{\vec{k}}{| \vec{k}|^2} m(z) ||\hat{w}(z)||^2
\end{equation}
with $k,l$ the horizontal wavenumber and $\omega$ the frequency of the vertical velocity field. The latter is included in the vertical wavenumber $m = \frac{N| \vec{k}|}{\Omega}$ taken as the WKB non-rotating approximation in the limit $H \rightarrow \infty$, with $\Omega = \omega - \vec{k} \vec{u} $ and N the Brunt-Vaisala frequency (see \citet{Lott2012} for details). 
In Equation (\ref{EPflux}) $\rho_{r}$ is the density at the reference pressure level $P_{r}$ ($\rho_{r} \sim $ 0.007 kg m$^{-3}$).
\\In our scheme we randomly impose both $\hat{w}_{n}$ and the EP-flux at a given launching altitude $z_0$ (see Section \ref{Sec_inputs}). We also have explicitly introduced a constant vertical viscosity $\mu$ (see Eq. 4 in \citet{Lott2012}) which controls the GW drag vertical distribution near the model top. To move from one model level to the next model level above, we essentially conserve the EP-flux, but allow a small diffusivity, $\nu = \mu / \rho_0$, which can be included by replacing  $\Omega$ by   $\Omega + i \nu m^2$. This small diffusivity is here to guarantee that the waves are ultimately dissipated over the few last model levels, if they have not been before (hence the division by the density $\rho_0$). In addition, this new EP-flux amplitude is limited to that produced by a saturated monochromatic wave $\hat{w}_s$ following \citep{Lindzen1981}:
\begin{equation}
\label{w_sat}
\hat{w}_s = S_c \frac{\Omega^2}{|\vec{k}| N} e^{-z/2H} k^*/	|\vec{k}| 
\end{equation}
or either $\hat{w}$ = 0 when $\Omega$ changes sign, to treat critical levels.
In Equation \ref{w_sat} $S_c$ is a tunable parameter and $k^*$ a characteristic horizontal wavelength corresponding to the longest wave being parameterized (see more details in Sec. \ref{Sec_inputs})  
\\Finally, we calculate the tendencies $\rho^{-1} \delta_z \vec{F}^{z}_{n'}$   ($n'$= 1, M) produced by the GW drag on the winds, computing the tendencies due to the M generated waves. Since we assumed that $w'_n$ are independent realizations, the mean tendency they produce is the average of these M tendencies. Thus,  we first redistribute the averaged tendency over the longer time scale $\Delta t$ by re-scaling it by $\delta t$/$\Delta t$  and second, we use the auto-regressive (AR-1) relation described in \citep{Lott2012} as follows:

\begin{equation}
\label{ARrelation}
  \left( \frac {\delta \vec{u}}{\delta t} \right)_{GWs}^{t} = \frac{\delta t}{\Delta t} \frac{1}{M} \sum_{n'=1}^{M} \frac{1}{\rho_0} \frac{\delta \vec{F}^{z}_{n'} }{\delta z} +  \frac{\Delta t - \delta t}{\Delta t}  \left( \frac {\delta \vec{u}}{\delta t} \right)_{GWs}^{t- \delta t} 
\end{equation}
This indicates that, at each time step, we promote M new waves by giving them the largest probability to represent the GW field, and degrade the probabilities of all the others by the multiplicative factor $(\Delta t - \delta t )/\Delta t$. As explained in \citet{Lott2012}, by expressing the cumulative sum underneath the AR-1 relation in Equation \ref{ARrelation}, we recover the formalism for infinite superposition of stochastic waves by taking: 
\begin{equation}
\label{Cn_equation}
 C^{2}_{n} = \left( \frac{\Delta t - \delta t}{\Delta t}\right)^p \frac {\delta t}{M \Delta t}
\end{equation}
where $p$ is the nearest integer that rounds $(n-1)/M$ (i.g. toward lower values).

\subsection{GW parameters setup}
\label{Sec_inputs}
In this section we describe the main tunable parameters used in the non-orographic GW scheme implemented here. The characteristics of every wave launched in the GCM are selected randomly with a prescribed box-shaped probability distribution, whose boundaries are key model parameters. These are chosen on the basis of observational constraints (whenever available) and theoretical considerations, and tuned using MCS data comparison.
A total of 26 runs for the full Martian year (MY) 29 have been performed to cover a large range of possible combinations of the GW characteristics, based on those constraints. One of the advantages of the GW scheme used here is that, contrarily to other GW parameterizations, each parameter has a physical meaning, as described below. 
\begin{description}
\item [Source height and duration]  
First, we assumed that the non-orographic GW source is placed above typical convective cells (i.e. around 8 km, depending on the topography). Mars Express Radio Occultation (RO) profiles provided good coverage at latitude  and local times where planetary boundary layer (PBL) convection is occurring, giving an accurate determination of the depth and the spatial variation of the convective boundary layer (CBL) at fixed local time ($\sim$ 17 h) \citep{Hinson2008}. Those authors found that the CBL extends to an height of 3-10 km above the surface at the season of the measurements (mid-spring in the northern hemisphere). \cite{Spiga2010} compared the RO temperature profiles to large-eddy simulations performed with the LMD Martian mesoscale model and found intense CBL dynamics within the measured depths (up to 9 km).
In our scheme the parameter controlling the wave launching level $z_0$ is $\sigma =P/P_{s}=0.4$, that corresponds to a region centered at about 250 Pa ($\sim$ 8 km), covering pressures from 160 Pa to 348 Pa, depending on the surface pressure $P_{s}$ and varying with season. The source is chosen to be uniform, without latitudinal variation, and aside from the atmospheric profiles through which the GWs' propagation is modeled, nothing else in the scheme varies with location or season. Considering that non-orographic GW  are expected to be generated by multiple sources (e.g. PBL convection, jet acceleration, dusty convection etc) and that represents a complex interplay of timescale for GCMs, we assume here that the source is turned on all day. In future developments, we foresee to implement in our scheme a characteristic time interval more representative of the life cycle of GW produced by PBL convection on Mars, but this is beyond the scope of this paper.

 \item [EP-flux amplitude]  
$F^z$ from  Equation \ref{EPflux} gives the vertical rate of transfer of wave horizontal momentum  per unit of area. This value has never been measured in Mars' atmosphere and it represents an important degree of freedom in the parameterization of gravity waves. Thus, in our scheme we impose the maximum value of the probability distribution  $F^{z}_{max}$ at the launching altitude z$_0$ (see previous point), for every set of MY29 runs. 
In order to define $F^{0}_{max}$ we have explored typical values used in the literature to evaluate the order of magnitude of the EP-flux within the realm of what is realistic.
For instance, using aerobraking data \citet{Fritts2006} found an estimation for momentum flux about 2000 m$^2$ s$^{-2}$ at altitudes 100-120 km where the density is typically from 10$^{-7}$ to 10$^{-9}$ kg m$^{-3}$, and this yields values for EP-flux of the order 2$ \cdot$10$^{-6}$ to 2$ \cdot$10$^{-4}$ kg m$^{-1}$ s$^{-2}$.
Calculations by a one-dimensional full-wave model performed by \cite{Parish2009} gave mean momentum flux (e.g. $\bar\rho \overline{u'w'}$)  $\approx$ 7$ \cdot$10$^{-7}$ kg m$^{-1}$ s$^{-2}$ below  100 km, at 82$^\circ$ N latitude during winter solstice, for a GW packet propagating with horizontal wavelength between 38 km and 150 km  (see their Figure 6).
Other authors \citep{Medvedev2011, Medvedev2015} also implemented a non-orographic gravity wave scheme \citep{Yigit2010,Yigit2015} in their GCM using an approach different from ours. They adopted an analytical form of the GW spectrum, in which the vertical propagation  of horizontal momentum fluxes is given by
\begin{math}
\overline{u'w'}_j = sgn(c_j -\bar{u}_0) \overline{u'w'}_{max} exp [-(c_j - \bar{u}_0)^2/c^2_w] 
\end{math}
, with $\bar{u}_0$ the mean wind at the source level, $c_j$ the phase speed of the harmonic $j$, and $c_w$ the half-width at half maximum of Gaussian distribution, fixed at $c_w$ = 35 m s$^{-1}$. In their experiments the spectrum is discretized with 28 harmonics for two values of $\bar{u}_0$= 0 and 20 m s$^{-1}$. Although they employed a geographically uniform flux per unit mass $\overline{u'w'}_{max}$= 2.5$ \cdot 10^{-3}$ m$^{2}$ s$^{-2}$, the wave source varies in time and space because of the modulation by the simulated local wind $\bar u_0$ in the lower atmosphere. In their scheme the direction of propagation of GW harmonics coincides with that of local wind at $z_0$ for $c_j > 0$ (positive momentum fluxes) and it is against it for $c_j < 0$ (see \citet{Medvedev2011} for more details.)
Assuming typical mean densities from 10$^{-3}$ to 10$^{-2}$ kg m$^{-3}$ around 250 Pa (i.e. the GW source level of our simulations) this gives a momentum flux $F^{0}_{max}$ above typical convective cells of the order of $10^{-6}$ to $10^{-5}$ kg m$^{-1}$ s$^{-2}$, which is in the range of values explored in our study.
\\It should be stressed here that the stochastic approach implemented in our scheme has the advantage of allowing to treat a wide diversity of emitted gravity waves, thereby a wide diversity of momentum fluxes. Our only setting is the maximum EP-flux amplitude at the launching altitude, therefore we tested several values of  $F^{0}_{max}$ by performing sensitivity tests for values of $F^{0}_{max}$ ranging from 10$^{-10}$ kg m$^{-1}$ s$^{-2}$ to 10$^{-5}$ kg m$^{-1}$ s$^{-2}$ (see also Section \ref{sensTest}). Then the amplitude of the EP-flux for each GW is chosen randomly between 0 and $F^{0}_{max}$. We set 10$^{-10}$ kg m$^{-1}$ as a lower limit for our tests because GW impact is negligible in the LMD-MGCM for momentum flux smaller than that value.

\item [Horizontal wavenumber]	 
In our scheme the horizontal wave number amplitude is defined as in \cite{Lott2012} $ k* < |k| < k_s $. The minimum value is $k* = 1/ \sqrt{ \Delta{x} \Delta {y}}$, where  $\Delta x$ and $\Delta y$ are comparable with the GCM horizontal grid ($\delta$x and $\delta$y $\approx$ 600 km in this study). This value corresponds to the longest waves that one parameterizes with the current  GCM horizontal resolution. The maximum (saturated) value is $k_s < N/u$, $N$ being the Brunt-Vaisala frequency associated to the mean flow and $u_0$ the mean zonal wind at the launching altitude. The  corresponding min/max values for the GW horizontal wavelength ($\lambda_h$ = $2\pi/ k$ ) are between 10 km and 300 km. Those values are within the observed range of GW wavelengths  \citep{Magalhaes1999,Hinson1999, Fritts2006}. 

\item [Phase speed] Another key  parameter is the amplitude of absolute phase speed $|c| = |\omega / k|$. As for the other tunable parameters, we impose the minimum $c_{min}$ and maximum  $c_{max}$ values of the probability distribution at the beginning of the runs, and the model chooses randomly $|c|$ between $c_{min}$ and  $c_{max}$.  Here $c_{min}$ is set to 1 m/s (i.e. for non-stationary GW)  and $c_{max}$ is of the order of the zonal wind speed at the launching altitude. The range of  $c_{max}$ is consistent with previous values used in the literature \citep{Medvedev2015}. Both eastward (c > 0) and westward (c < 0) moving GWs are considered. Three values of maximum probable $c_{max}$ have been tested to evaluate the sensitivity of winds and temperature to this parameter (i.e. 10 m/s, 30 m/s and 60 m/s). See Section \ref{sensTest} for details. 

\item [Saturation parameter] $S_c$ is a tunable parameter in our scheme, on the right hand side of Equation \ref{w_sat}, which controls the breaking of the GW  by limiting the amplitude $w_s$. 
In the baseline simulation we set $S_c = 1$.

\end{description}

\section{LMD-MGCM description}
The LMD-MGCM is a finite-difference model based on the discretization of the horizontal domain fields on a latitude-longitude grid \citep{Forget1999}. The horizontal resolution used in this work is 64 longitude x 48 latitudes (3.75$^\circ$ x 5.62$^\circ$). Vertical levels are hybrid coordinates, with 32 levels up to 0.05 Pa (about 100 km altitude), above the top of MCS profiles. This is a standard vertical resolution, as used in all previous recent papers describing LMD-GCM and MCD results \citep{Navarro2017} and \citep{Madeleine2014}. The GW parameterization is precisely designed to cope with the coarse vertical resolution of the GCM: the propagation of the GW is not resolved by the GCM, it is a subgrid-scale phenomenon whose impact on the large-scale flow (heat and momentum transfer when they break) is computed by the parameterization, then passed to the dynamical core of the GCM as a tendency for temperature and wind. For each physical timestep of 15 Martian minutes (1 minute is about 1/1440th of a Martian sol), 10 dynamical time steps occur. The model includes the CO$_2$ cycle, with condensation and sublimation of CO$_2$ and the global change of the total atmospheric mass \citep{Forget1998}. The cycles of dust and water and their interactions are also represented. Dust is modeled with a two moments scheme, that transports both the mass mixing ratio and the number of dust particles, with a distribution for particle size assumed to be log-normal with a fixed effective variance \citep{Madeleine2011}. Lifting is global and constant with a rescaling applied to the total column quantity at each physical timestep to match the observed column of dust during Martian Year 29 \citep{Montabone2015}. The water cycle includes a microphysical scheme for the sublimation and condensation of water ice clouds on the dust particles and interactions between the ice at the surface and the atmosphere \citep{Navarro2014}. Dust and water ice in the atmosphere are 4D variables, transported and radiatively active \citep{Madeleine2011, Madeleine2012}. Another key improvement was the parameterization of convection and near surface turbulence using a thermal plume model, which is coupled to surface layer parameterizations taking into account stability and turbulent gustiness to calculate surface-atmosphere fluxes \citep{Colaitis2013}.
\\The LMD-MGCM also includes a parametrization of the orographic gravity waves \citep{Forget1999, Angelatsicoll2005} based on the gravity wave drag scheme of \cite{LottMiller1997} and \cite{Miller1989}. At the time the scheme parameters were chosen conservatively and the effect of the orographic wave drag was probably underestimated. In this work, we add a more general scheme which includes non-orographic gravity waves, but it is possible that this also accounts for the previously underestimated effect of orographic gravity waves. At this stage it is very difficult to separate the contribution of orographic sources to the wind drag from non-orographic ones, but this will be certainly the next step in a future theoretical work.

\section{MCS Data set description}
\label{MCS_description}
MCS \citep{McCleese2007} is a limb-viewing infrared radiometer aboard the sun-synchronous MRO spacecraft. Its nominal local times of observations at most latitudes (except when the orbit crosses high latitudes) are fixed around 03:00 and 15:00 $h$ Local Time (LT). \citet{Kleinboehl2009} obtained profiles of temperature (using the CO$_2$ 15 $\mu m$ absorption band) as well as profiles of dust and water ice extinction opacity (at wavelengths centered respectively around 21.6 $\mu m$ and 11.9 $\mu m$), nominally from the surface to about 0.1 Pa with 5 km vertical resolution. MCS data \citep{McCleese2010} currently provide the only systematic measurements of temperature in Mars' mesosphere up to about 80 km. For several technical reasons explained in the above-cited papers, though, a large number of profiles stop well before reaching the surface. Although the nominal vertical resolution of the instrument is 5 km (corresponding to the separation of the peaks of the weighting functions), the profiles are over-sampled using information from more than one weighting function. Both temperature and aerosol extinction opacity profiles are standard MCS products. 
Most recently, \citet{Kleinboehl2017} developed a new scheme for the retrievals of these quantities, based on a two-dimensional rather than one-dimensional radiative transfer. Instead of simply assuming spherical symmetry, the 1-D retrieved fields are interpolated along the orbit to provide correction factors. Further retrievals are then performed using the interpolated fields and a 2-D radiative transfer scheme, in order to correct for gradients along the line-of-sight. This new retrieval methodology (producing retrieval version $5.2$) is particularly useful to mitigate dayside-nightside differences in retrievals at high latitudes, where horizontal gradients may be strong. 

In this paper, we use these latest retrievals (version $5.2$). Temperature uncertainties are typically around 0.5 K and only increase at low altitudes (below about 5 km altitude), where the atmosphere starts to become opaque, and at altitudes above about 60 km, where the instrument signal-to-noise ratio starts to decrease. Since we are interested in comparing gridded GCM temperature fields with data, we have carried out binning of MCS observed temperatures in 4-D bins (latitude, longitude, solar longitude, and local time). We separate observations into dayside (06:00 h < LT $\le$ 18:00 h) and nightside (00:00 h < LT $\le$ 06:00 h and 18:00 h < LT $\le$ 24:00 h) bins. As mentioned above, most MCS observations at latitudes equatorward of |80$^\circ$| are provided around 03:00 h in the nightside and 15:00 h in the dayside. Exceptions are cross-track observations that span additional local times, but this kind of observation only started after September 13, 2010 (L$_S$ = 146$^\circ$, MY 30) \citep{Kleinbohl2013}, therefore it does not affect the results of this paper, in which we only use MY29 observations. This particular year was chosen because at the time of the preparation of the manuscript it was the most complete year after the global dust storm in MY28.  A comparison with less dusty years (but with worst coverage) such as MY30 or MY31 is left for a future study. Furthermore, we limit our comparison between model results and data to latitudes equatorward of |80$^\circ$|. By doing this, we can safely use GCM results at local times 03:00 h in the nightside and 15:00 h in the dayside, and avoid comparison at non-homogeneous local times. The width of the latitude, longitude, and solar longitude bins of MCS observations is, respectively, 3.75$^\circ$, 5.625$^\circ$, and 5$^\circ$, in order to match MGCM horizontal resolution.
The binned values are provided on the same pressure grid as the one used by the MCS team for their standard products.
It is worth noting that MRO entered a long period of safe mode in late 2009, therefore there are no MCS observations in MY 29 after L$_s$ = 328$^\circ$.

\section{Impact of non-orographic GW drag on LMD-MGCM zonal winds}
\label{impact_jets}

\begin{figure}[!htb] 
	\centering
	\includegraphics[width=1.1\linewidth]{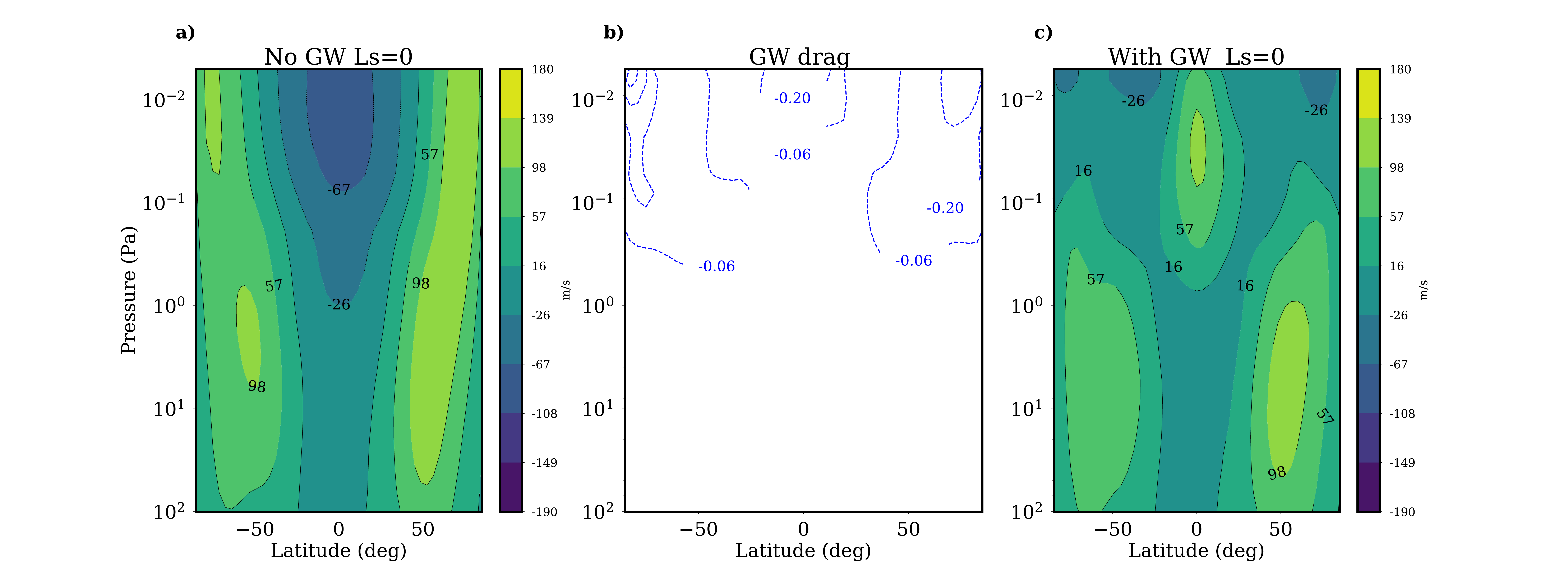}
	\caption{Zonal mean wind (m/s) simulated by the LMD-MGCM with the non-orographic GW scheme turned off (panel a) and on (panel c), for early northern hemisphere spring (L$_s$=0$^\circ$). Panel b: daily averaged gravity wave drag on the zonal wind in m/s/sol (negative values indicate deceleration of zonal wind). Only values above $|0.001|$ m/s/sol have been plotted. The pressure range indicated in these figures corresponds to altitude levels between 0 and 100 km, approximately.}
	\label{Wind_GWdrag1}
\end{figure}

\begin{figure}[!htb] 
	\centering
	\includegraphics[width=1.1\linewidth]{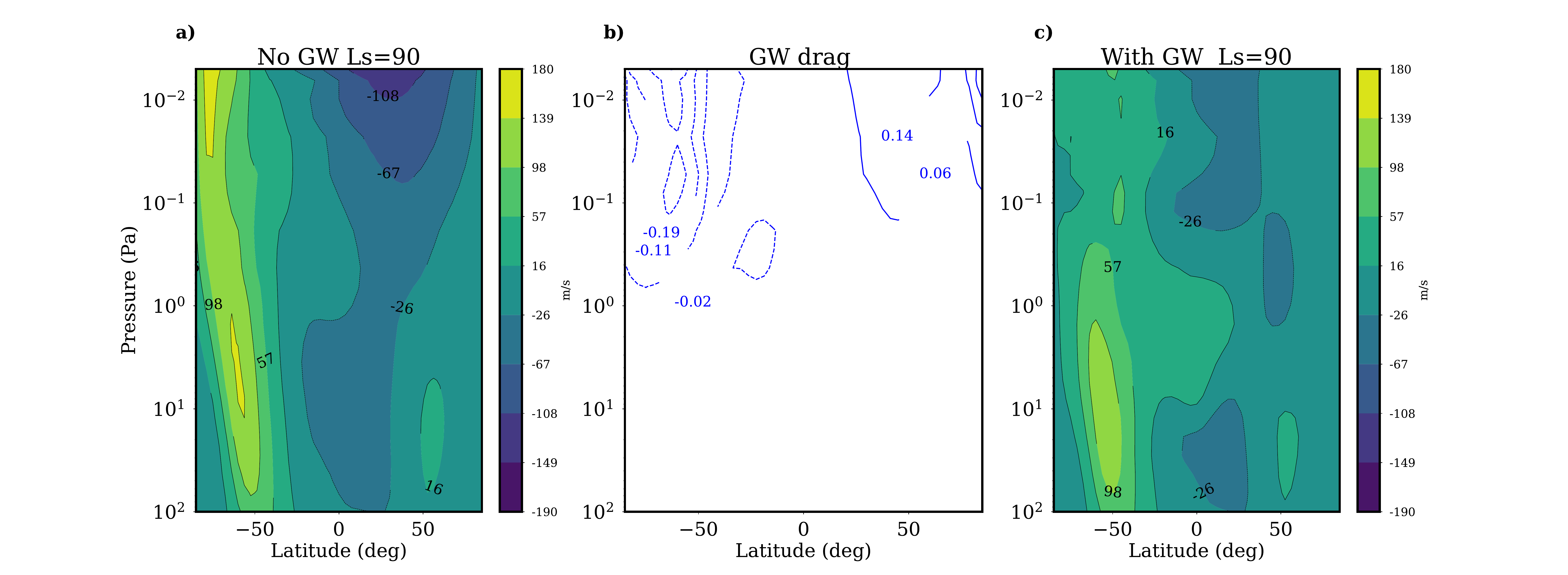}
	\caption{Same as Figure \ref{Wind_GWdrag1} but for early northern hemisphere summer L$_s$=90$^\circ$}
	\label{Wind_GWdrag4}
\end{figure}

\begin{figure}[!htb] 
	\centering
	\includegraphics[width=1.1\linewidth]{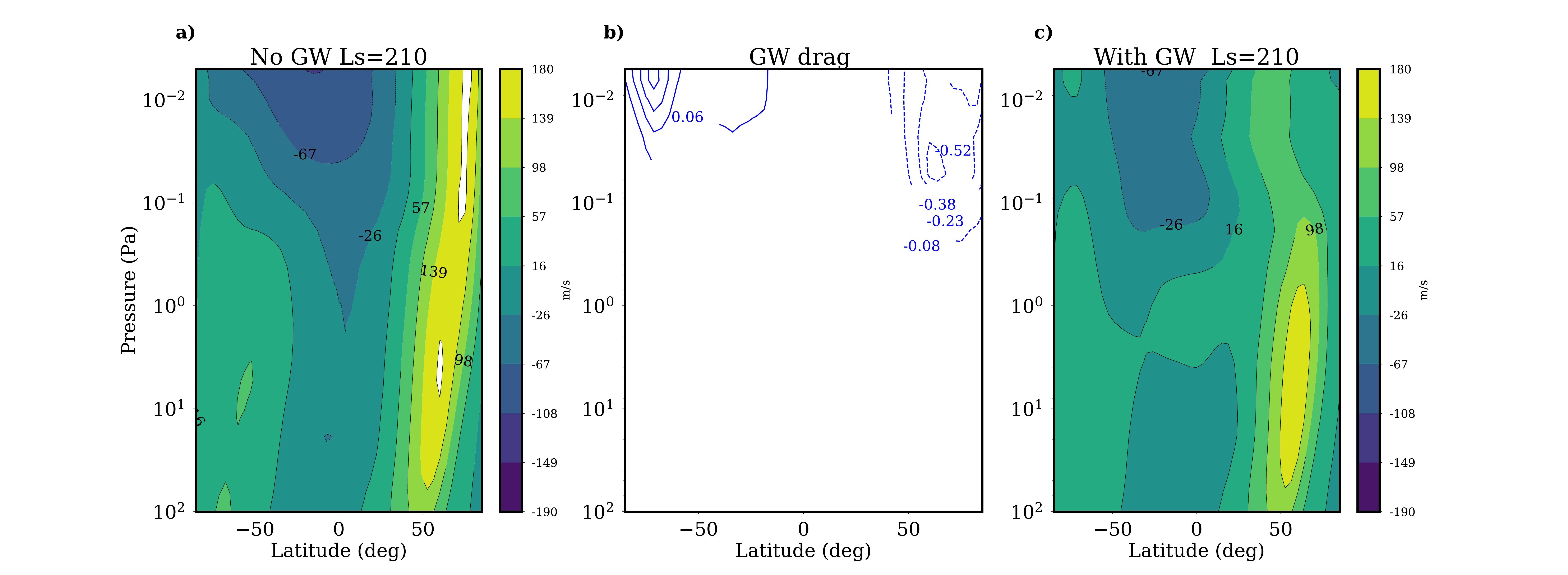}
	\caption{Same as Figure \ref{Wind_GWdrag1} but for early northern hemisphere fall L$_s$=210$^\circ$}
	\label{Wind_GWdrag3}
\end{figure}

\begin{figure}[!htb] 
	\centering
	\includegraphics[width=1.1\linewidth]{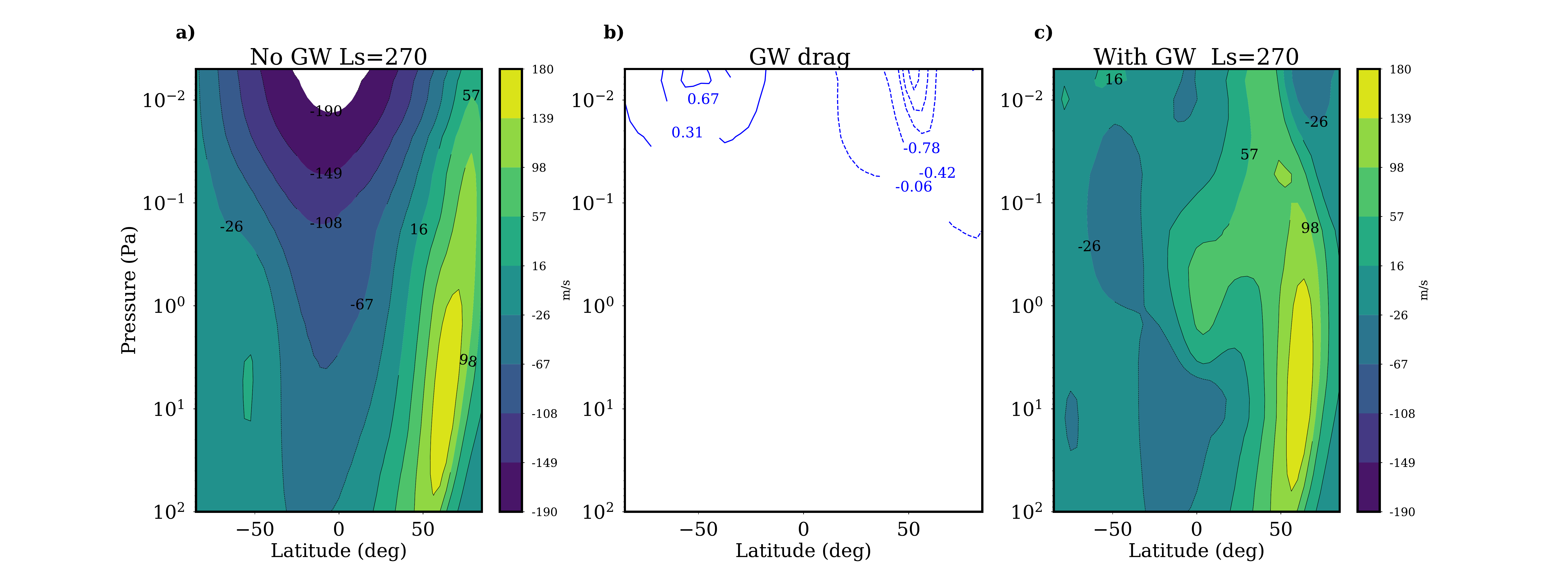}
	\caption{Same as Figure \ref{Wind_GWdrag1} but for early northern hemisphere winter L$_s$=270$^\circ$. The white spot in panel a) represents values lower than -190 m/s.}
	\label{Wind_GWdrag2}
\end{figure}

Examples of simulated zonal mean winds before and after the implementation of the non-orographic GW parameterization in the LMD-MGCM  are plotted in Figures \ref{Wind_GWdrag1}-\ref{Wind_GWdrag2} at four solar longitudes: L$_s$=0$^\circ$, L$_s$=90$^\circ$, L$_s$=210$^\circ$ and L$_s$=270$^\circ$. The runs described here correspond to the MGCM with the GW scheme on, using the reference parameters listed in Table \ref{tab1}. The choice of this set of parameters will be discussed in section \ref{results}.
\\ Without the non-orographic GW  scheme activated the simulated wind structure is comparable to the results described in previous theoretical works \citep[e.g.][]{Forget1999, Haberle1999, Hartogh2005}. 
It shows a strong (retrograde) easterly jet at northern hemisphere spring equinox (L$_s$=0$^\circ$), in the middle atmosphere above $\sim$ 0.5 Pa (about 60 km) at equatorial regions, and strong (prograde) westerly wind in both hemispheres at high latitudes with velocities increasing with height (see panel \textit{a} in Figure \ref{Wind_GWdrag1}). At the northern winter solstice the circulation of Mars atmosphere is dominated by a quasi-global Hadley cell extending from 60$^\circ$ S almost to the north pole and model simulations predict a strong prograde wind jet corresponding to the pole-to-pole heating gradient (panel \textit{a} in Figure \ref{Wind_GWdrag2}). When the atmospheric dust loading has one of its peaks (L$_s$=210$^\circ$), the ascending branch of the Hadley cell reaches higher latitudes if the thermal forcing is very strong, for instance when the dustiness increases, and the prograde winds are also stronger than at other seasons (panel \textit{a} in Figure \ref{Wind_GWdrag3}).  
At northern summer solstice instead (panel a in Figure \ref{Wind_GWdrag4}), MGCM simulations show a weaker thermal forcing than at northern winter solstice, because of the smaller amount of dust in the atmosphere and the reduced solar flux near aphelion \citep{Forget1999}. Consequently, the Hadley circulation is much less intense and the simulated southern winter polar warming also weaker than at northern winter solstice.
\\When the non-orographic GW scheme is on, the retrograde wind jet would prevent most of the westward (c < 0) propagating GW from reaching the mesosphere, leaving primarily eastward waves with positive momentum flux to propagate, with the reverse being true for westerly jets at mid-high latitudes. The complete wave momentum deposition occurs in those layers where upward propagating non-orographic GW energy with phase speed $c$ meets a zonal wind vector opposite to its direction. In other words, when $c$  is close to the mean zonal wind  ($c \rightarrow \bar{u}$) or if its amplitude is large enough to create shear instability ($c -  \bar{u} \rightarrow \bar{u} \neq 0$), the saturation of a gravity wave occurs and the wave momentum flux is transferred to the mean flow \citep{Lindzen1981, Terada2017}.
For non-orographic GW with a given phase speed $c \neq \bar{u} $, the intrinsic speed $|\bar{u}-c|$ becomes larger in amplitude in presence of wind jets, inducing a stronger GW drag (i.e. deceleration if c < $\bar{u}$ or acceleration c > $\bar{u}$).  Therefore  GW drag is especially strong in the upper part of the zonal winter jets, where the contrast between the mean flow velocity and the given phase speed (toward which the speed of the flow is decelerated/accelerated) is high. 
The induced drag due to the non-orographic GW momentum deposition to the mean state changes  the large-scale horizontal winds, hence the large-scale circulation and vertical winds. Consequently, those waves drive large-scale heating/cooling by adiabatic downward/upward large-scale vertical motion associated with the altered circulation.
\\The GW parameterization scheme implemented here allows for quantifying the "GW drag" (e.g. the zonal and time averaged gravity wave momentum deposition to the zonal flow) in order to gain further insight into the nature of changes in the modeled mean fields (see panels \textit{b} in Figures \ref{Wind_GWdrag1}-\ref{Wind_GWdrag4}). In all simulations the GW drag is significantly large (> $|0.001|$ m/sol/s) above $\sim$ 1 Pa (about 50 km) and it predominantly reduces the westerly jets at high latitudes (negative drag) and weakens the equatorial easterly jet by accelerating  the zonal wind (positive drag). In most seasons the negative GW drag on the retrograde zonal wind is about two orders of magnitude larger than the positive one for the same period. This is due to the damping of diurnal tides by the GW (see Section \ref{diurnal_tides}), and thus of the mean zonal wind: the diurnal tides tend to drag the zonal wind toward the phase velocity of the tides (i.e. the motion of the sub-solar point in a westward direction) which is about -240 m/s, and the retrograde jet is mainly forced by the interaction between thermal tides and wave mean flow \citep{Forbes2002}.
During the dust season (L$_s$=210$^\circ$) westward zonal wind jets are stronger than at other seasons above 1 Pa, therefore the impact of GW on the general circulation is expected to be larger. Our results in Figure \ref{Wind_GWdrag3} show instead that the GW-induced deceleration during this season is similar to that in Figure \ref{Wind_GWdrag2} (L$_s$=270$^\circ$) at same pressure, reaching approximately 0.55 m/s/sol between 10$^{-1}$ Pa and 10$^{-2}$ Pa.
However, those levels are close to the upper limit of our runs, and further studies with a vertically extended version of the LMD-MGCM should be performed in order to better quantify the GW drag at those altitudes. 

A tentative comparison with previous works is give here. Unfortunately, there are very few published studies on non-orographic GW parameterization implemented into Mars GCMs, and they are mostly focused on the theoretical impact of GW in the thermosphere \citep{Medvedev2011, MedvedevYigit2012}. 
Those authors also claimed that the main influence of GW in the winter hemisphere is near the edge of westerly jet, and that the net effect on the mean zonal wind is to decelerate and even reverse it increasingly with height (between 100 and 130 km). The magnitudes of the drag they found vary from tens at 0.1 Pa ($\sim$ 80 km) to hundreds of m/s/sol above, depending on the shape of simulated jets, and the assumed wave sources. In our simulations GW drag reaches maximum magnitudes of the order of 1 m/s/sol around 10$^{-2}$ Pa in the northern hemisphere (NH) winter solstice, about a factor 100 smaller than values estimated by \citet{Medvedev2011} at the same season (see their Figure 1). This discrepancy may be related to the difference between the maximum amplitude of the EP-Flux at the launching altitude used in our best-fit simulations ($F^{0}_{max} = 7\cdot 10^{-7} $ kg m$^{-1}$ s$^{-2}$) and the EP-flux values used in \citet{Medvedev2011}, which are about 2 order of magnitude larger ($10^{-4}$ - $10^{-5}$ kg m$^{-1}$ s$^{-2}$). Our maximum value is however consistent with the one estimated by \citet{Parish2009} and with the lower limit estimated by \citet{Fritts2006}).
Nevertheless, a quantitative comparison of GW-induced wind drag should be taken with caution, because it would require first a detailed inter-comparative study between the two different GW parameterizations, which is not straightforward and should be addressed in a future dedicated paper. Here we have investigated the possible impact of the vertical resolution on the GW drag by increasing the vertical resolution from 6-7 km (32 levels) to 2-3 km (54 levels). The results (see Figure in Supplement Material) show that GW drag values are of the same order of magnitude with higher resolution than with our standard GCM model resolution, confirming that our GW parameterization does not depend on the vertical resolution."

\begin{figure}[!htb] 
	\centering
	\includegraphics[width=1.1\linewidth]{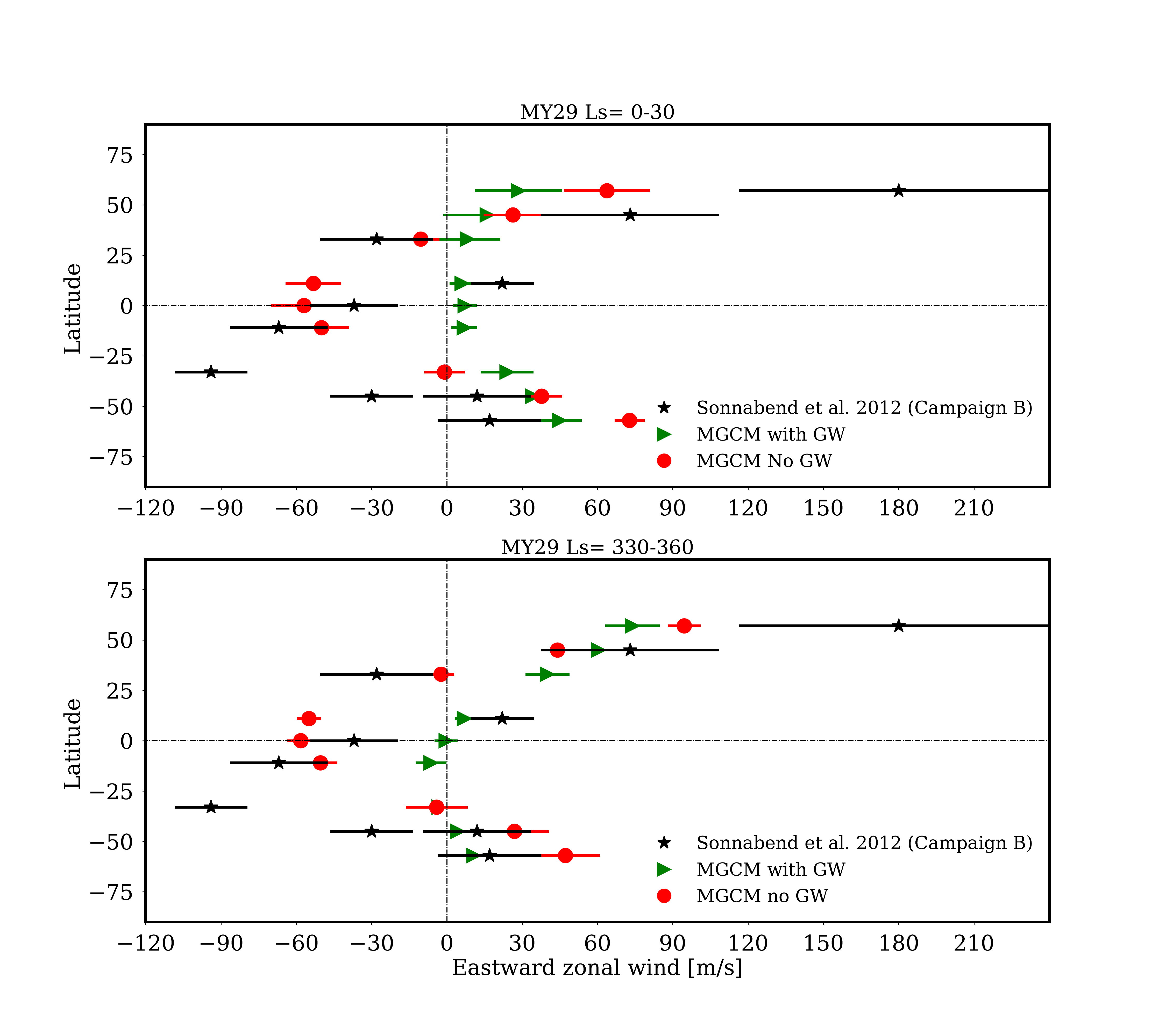}
	\caption{Comparison of retrieved wind values as in \citet{Sonnabend2012} (see their Figure 8) indicated with stars, with predicted daytime values from the LMD-MGCM with the non-orographic GW scheme on (green triangles) and off (red dots). The comparison is given for campaign B occurred during two martian months of MY29: L$_s$=0-30$^\circ$ (upper panel) and L$_s$=330-360$^\circ$ (bottom panel).  The wind values from the MGCM were averaged between 50 and 100 km (top of the model), according to the contribution function described in \citet{Sonnabend2012}. The error plotted for the MGCM dataset is the full month standard deviation at the exact latitudes of the measurements. }
	\label{Wind_comparison}
\end{figure}

\subsection{Comparison with wind observations}
As shown in the previous section, the non-orographic GW parameterization implemented in this work produces major changes on the simulated zonal winds. Overall, the GW slow down zonal winds everywhere, and in the tropics they become nearly zero (panels c in Figures \ref{Wind_GWdrag1}-\ref{Wind_GWdrag4}).
Unfortunately, there are no direct wind observations to compare with, except for rare and sparse observations from ground based telescopes \citep{Sonnabend2012, Valverde2016}. The wind fields derived from MCS temperature fields using geostrophic balance are based on several assumptions (see \citet{McCleese2010}) and they are certainly not the gold standard to validate model simulations. They are estimates of zonal gradient wind calculated via thermal wind, but we can expect the atmosphere of Mars to be more complicated than that.
Here instead we compare our results with the retrieved wind velocity in the mesosphere of Mars during one observing campaign occurred in MY29 (campaign B, in November and December 2007 described in \citet{Sonnabend2012}.) Figure \ref{Wind_comparison} shows extracted simulated daytime zonal mean wind values for the corresponding seasons (L$_s$=0-30$^\circ$ and L$_s$=330-360$^\circ$), latitudes and altitude range. MGCM wind were averaged between 50 and 116 km, according to the contribution function of the measurements. The results of \citet{Sonnabend2012} (their Figure 8) are also plotted in Figure \ref{Wind_comparison}.
When comparing the results, there are several discrepancies between model and data, particularly in the equatorial region (30$^\circ$N-30$^\circ$S), where the inclusion of the GW scheme does not seem to improve the wind values. MGCM averaged field without the non-orographic GW are retrograde near the equator, with values within the error bars of most of the measured values, while the results extracted from the MGCM after the implementation of the GW scheme are close to zero. 
The only exception is the measurement around the equator (latitude = 11$^\circ$ N) showing zonal wind $\pm$ 30 m/s, where a MGCM with zero wind in that region is consistent. The same is true for the campaign A occurred in MY30 (not shown here), where low westerlies winds around 20-30 m/s at southern latitudes are also consistent with values simulated in this work. In northern latitudes the fit is not very good (both with or without GW) but there the data is also all over the place, with 30 m/s westward in campaign A and 180 m/s eastward in campaign B, at a similar season, and measured uncertainties are also larger.

\subsection{Impact of GW mean flow forcing parameters}
\label{GWmean_flow_forcing}

\begin{figure}[!htbp] 
	\centering
	\includegraphics[width=1.2\linewidth]{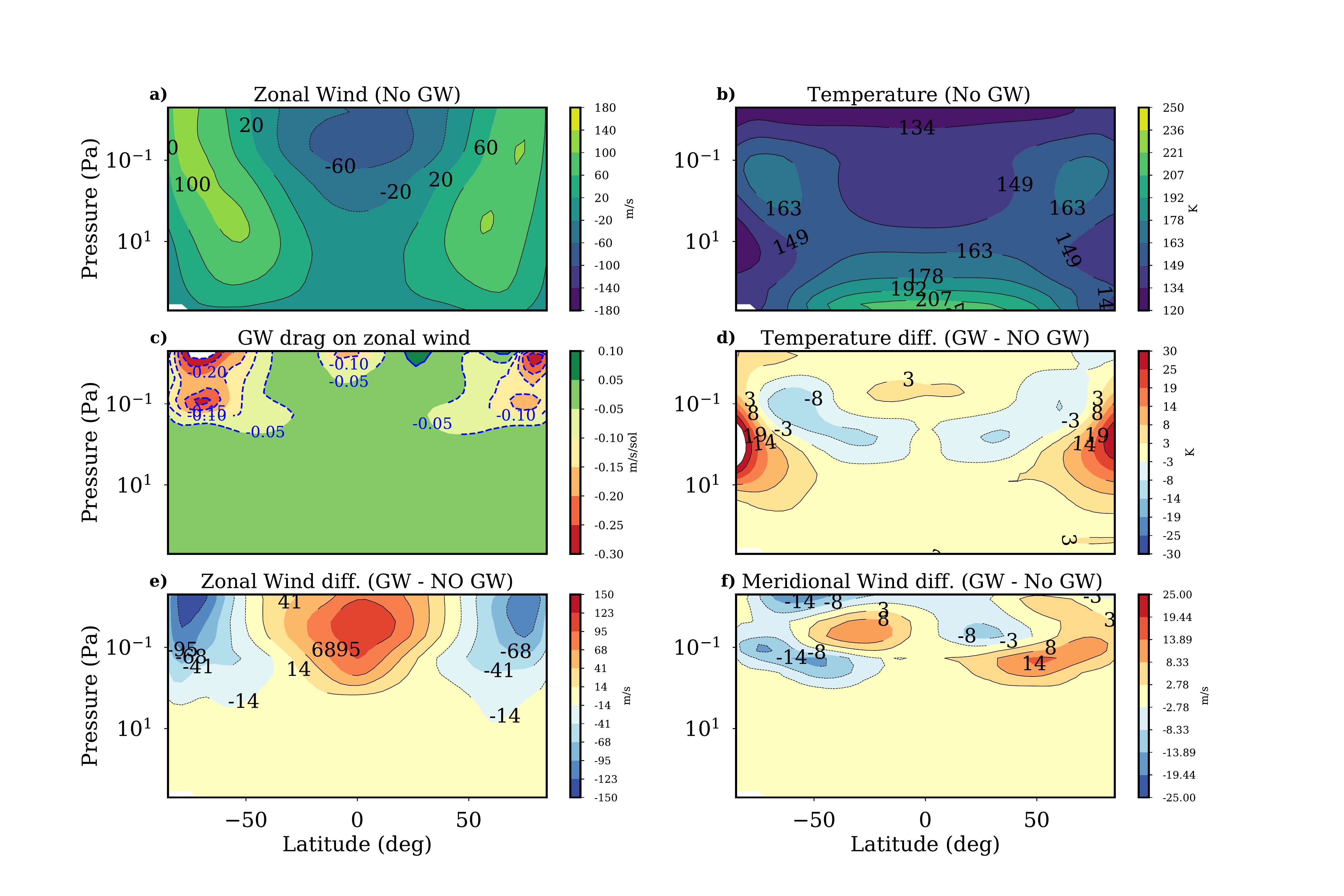}
	\caption{Mean zonal wind in m/s (panel $a$) and temperature (panel $b$) simulated by the MGCM before the implementation of the non-orographic GW parameterization in the period L$_s$ = 0-30$^\circ$. Panel c shows the mean GW drag on the zonal wind in m/s/sol when the non-orographic GW scheme is on: negative values indicate deceleration of zonal wind and positive field is acceleration. Only values above |0.001| m/s/sol have been plotted. Panels $d$, $e$ and $f$ represent the simulated difference "With GW - Without GW" of temperature field in K, zonal winds in m/s, and meridional winds in m/s, respectively.}
	\label{U_V_T_diffGW_L030}
\end{figure}

\begin{figure}[!htbp] 
	\centering
	\includegraphics[width=1.2\linewidth]{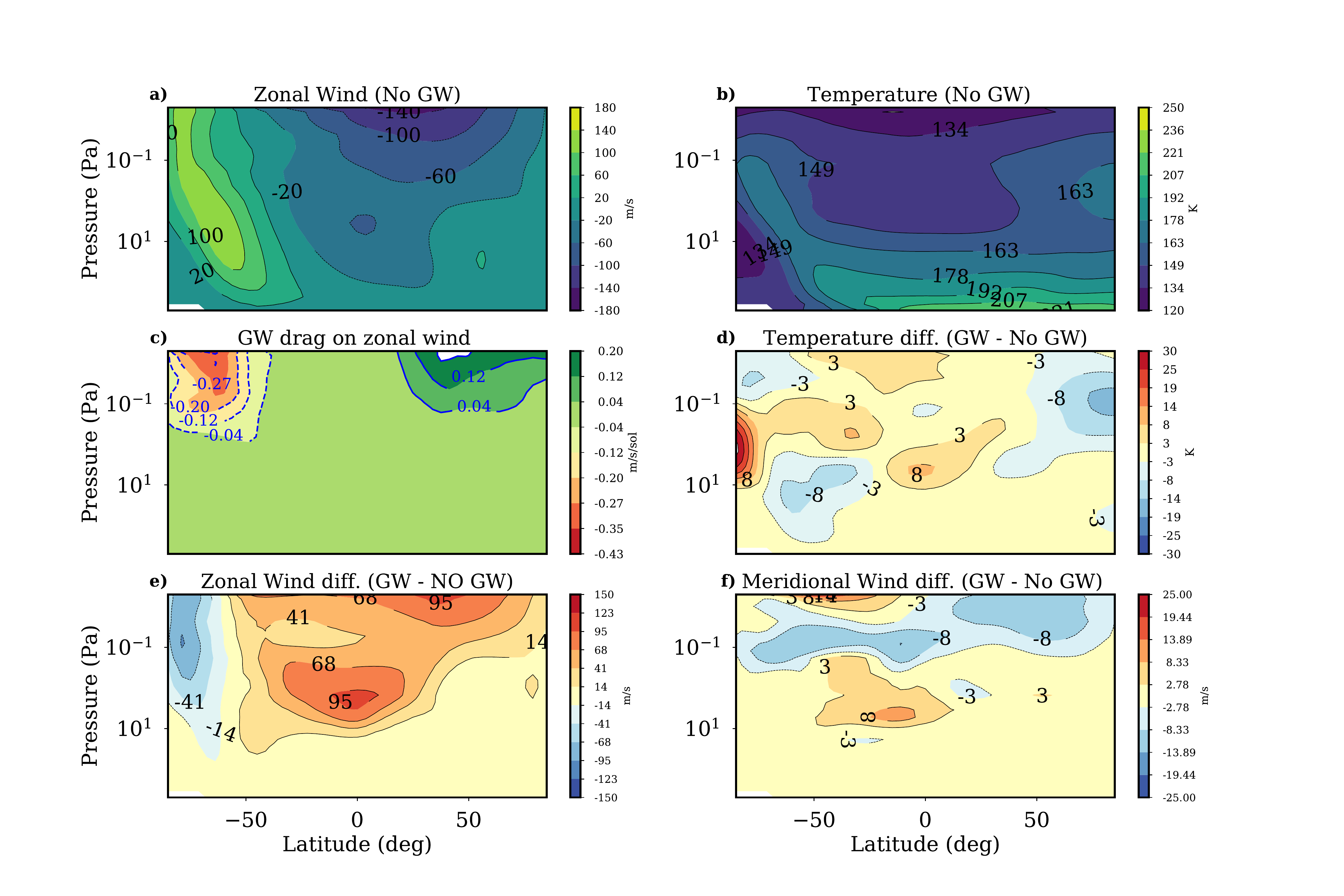}
	\caption{Same as Figure \ref{U_V_T_diffGW_L030} but for L$_s$ = 90-120$^\circ$.}
	\label{U_V_T_diffGW_L90120}
\end{figure}

\begin{figure}[!htbp] 
	\centering
	\includegraphics[width=1.2\linewidth]{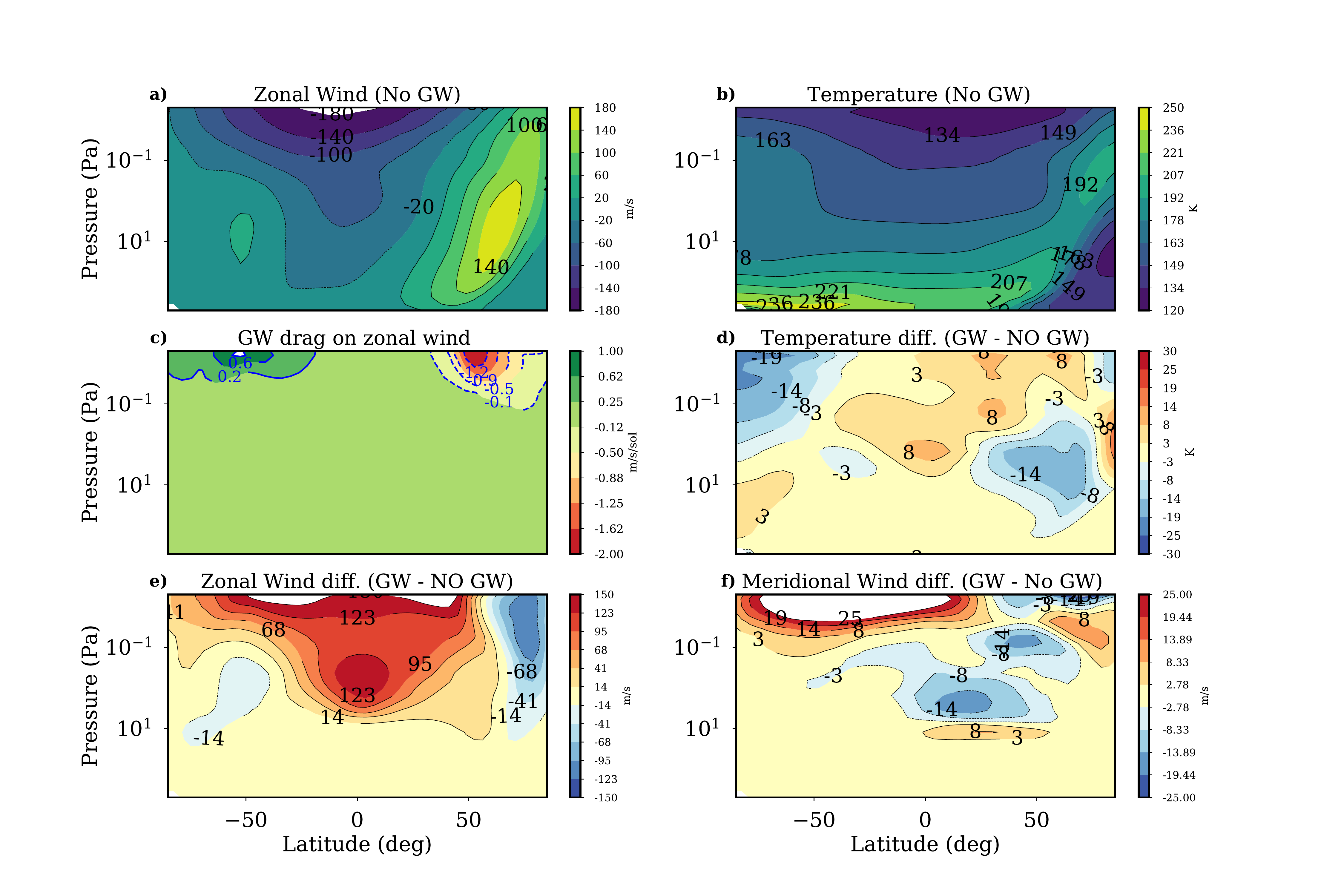}
	\caption{Same as Figure \ref{U_V_T_diffGW_L030} but for L$_s$ = 270-300$^\circ$}
	\label{U_V_T_diffGW_L270300}
\end{figure}

The momentum deposited by GW on the mean state not only may reverse easterly solsticial equatorial zonal jets in the middle atmosphere ($\sim$ 5-0.05 Pa) and weaken the westerly zonal jets at mid-high altitudes, as described in section \ref{impact_jets}, but it also drives changes in the meridional flow. Figures \ref{U_V_T_diffGW_L030}, \ref{U_V_T_diffGW_L90120}, \ref{U_V_T_diffGW_L270300} show the alteration of zonal wind, meridional circulation and temperature due to parameterized GWs, and the GW drag on zonal wind for the northern hemisphere (NH) Spring (L$_s$=0,30$^\circ$) , NH summer (L$_s$=90$^\circ$,120$^\circ$) and NH Winter (L$_s$=270$^\circ$,300$^\circ$), respectively. Note that the contour maps in panels $c$ of Figures \ref{U_V_T_diffGW_L030}-\ref{U_V_T_diffGW_L270300} represent monthly averaged GW drag, while Figures \ref{Wind_GWdrag1}, \ref{Wind_GWdrag4} and \ref{Wind_GWdrag2}, show daily averaged at the corresponding solar longitude L$_s = 0^\circ$, L$_s = 90^\circ$ and L$_s = 270^\circ$, respectively. Also note that the contour scales are different from Figures \ref{Wind_GWdrag1}-\ref{Wind_GWdrag2}.
\\Focusing on Figure \ref{U_V_T_diffGW_L030}, the equator-to-pole flow increases by ten m/s at about 0.5 Pa (60 km approximately) with respect to the runs without GW (see panel f), thus increasing convergence at the pole, hence downward motions. As a consequence, the polar warming increases up to 30 K in both hemispheres, and the adiabatic cooling increase up to 10 K in the layers above, at mid-latitudes. At the equator, local heating above 10$^{-1}$ Pa (around 80 km) results from the reduction of ascending branch of the Hadley cell at these altitudes. 
For the period L$_s$=90$^\circ$,120$^\circ$ (Figure \ref{U_V_T_diffGW_L90120}), the dynamical mechanism is analogous, but the impact is smaller: the southern polar warming is increased by more than 15 K, caused by the GW induced acceleration of the north-to-south pole meridional flow above 1 Pa (see panel f). Similarly, in the NH winter (Figure \ref{U_V_T_diffGW_L270300}) the effect of GW is to accelerate the upper branch of the solsticial Hadley cell up to 30 m/s, and adiabatic heating takes place due to the intensification of poleward circulation cells, thus increasing the northern polar warming in the middle atmosphere of Mars by about 15 K on average. GW drag reaches magnitudes of the order of 1 m/s/sol above 10$^{-2}$ Pa in the NH winter solstice (panel $b$ in Figure \ref{Wind_GWdrag4}), and produces a major change in the zonal wind field ($\sim$ 100 m/s), while the impact on the temperature field is relatively moderate ($\sim$ 10-20 K).
Those considerations indicate that GW induced alteration of the meridional flow is responsible for the simulated temperature variation. The impact on the temperature field will be also discussed in the next session.

\section{Results: comparison with MCS data}
\label{results}

Since the characteristics of GW spectrum are not well known on Mars, the strategy adopted in this work was to identify a set of reference tunable GW parameters that reduced model data biases by the greatest amount by comparing the LMD-MGCM and MCS thermal structure, specifically using diurnal tides as diagnostics. We run 26 simulations of a full Martian Year (MY 29) with non-orographic GW parameterization included. The subset of "best-fit" GW characteristics listed in Table \ref{tab1} was selected with the help of sensitivity tests (see Section \ref{sensTest}).
Both MCS data and LMD-MGCM simulations were binned in boxes of 3.75$^\circ$ latitude x 5.625$^\circ$ longitude.
\\In this section we will focus only on the LMD-MGCM simulations performed with the non-orographic GW scheme activated using the wave characteristics as in Table \ref{tab1}. Note that they correspond to Case 2 in Table \ref{tab2}.

\begin{table}[!htbp]
	\centering
	\begin{tabular}{| c | c | c | c |}
		\hline 
		c          &  k$_h$                          & F$^{0}$     & S$_c$                               \\
		\small [m/s]
		&       \small   [km]        & \small kg m$^{-1}$ s$^{-2}$   &    \\
		\hline
		[1 - 30]                 &    [10 - 300]                              &  [0 - 7 $\cdot$ 10$^{-7}$]   &  1                                   \\
		\hline
	\end{tabular}
	\caption{Best-fit wave characteristics in the GW scheme implemented in this work: $c$ the absolute phase speed, k$_h$ the horizontal wavelength amplitude,  $F^{0}$ the vertical momentum (EP-flux) at the source and S$_c$ the saturation parameter. Values in the bracket indicate the extremes of the probability distribution used here for the reference simulations.}
	\label{tab1}
\end{table} 

\begin{figure}[!htbp] 
	\centering
	\includegraphics[width=0.9\linewidth]{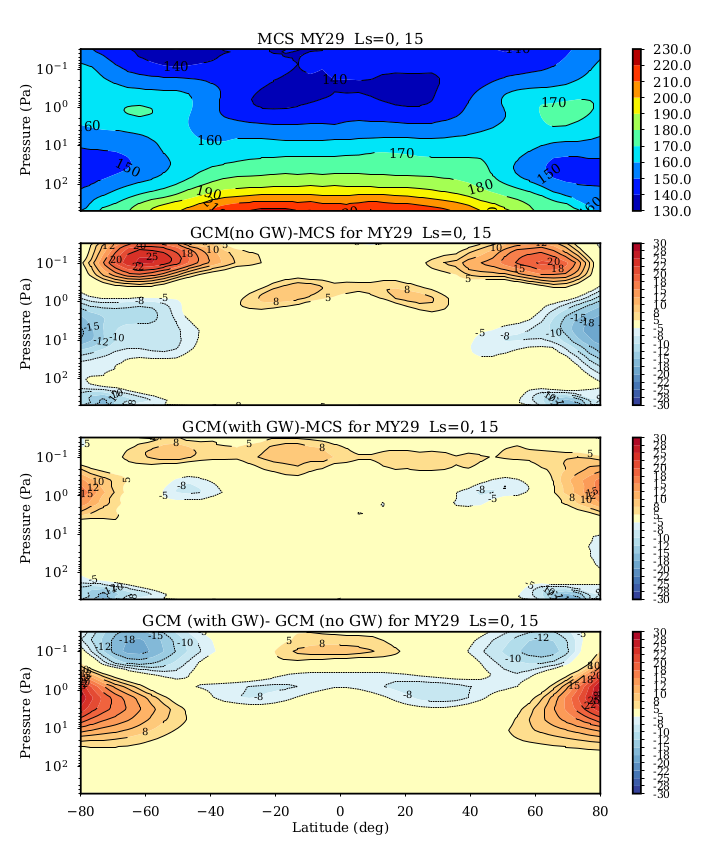}
	\caption{Data-model comparison of zonal mean day-night temperature  (T$_{am}$ + T$_{pm})/2$  averaged over the NH Spring Equinox (L$_{s}$=0$^\circ$-15$^\circ$) for MY29. Measurements by MCS/MRO are in top panel. Differences MGCM-MCS without and with non-orographic GW parameterization are shown in the second  and third panels from the top, respectively. The bottom panel shows temperature differences between model simulations (with GW - without GW).}
	\label{Tmean_Ls0}
\end{figure}

\begin{figure}[!htbp] 
	\centering
	\includegraphics[width=0.9\linewidth]{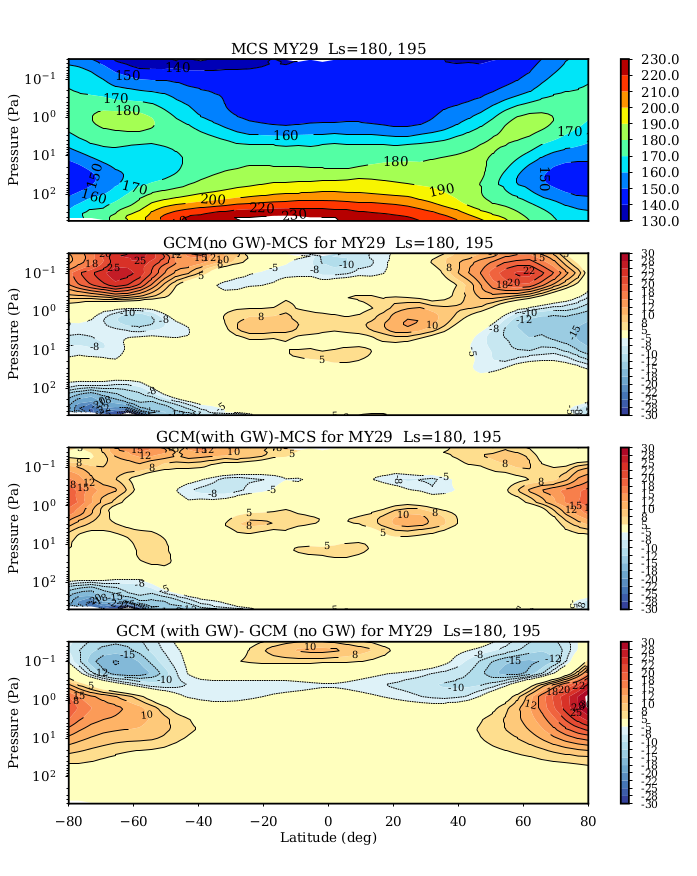}
	\caption{As in Figure \ref{Tmean_Ls0} but for NH  autumn equinox (L$_{s}$=180$^\circ$-195$^\circ$)}
	\label{Tmean_Ls180}
\end{figure}

\begin{figure}[!htbp] 
	\centering
	\includegraphics[width=0.9\linewidth]{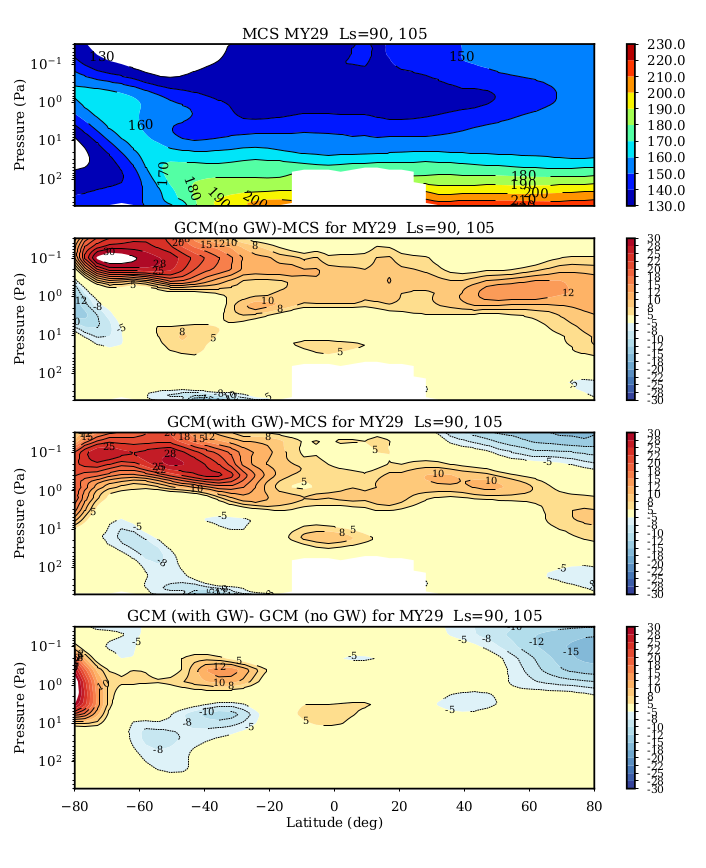}
	\caption{As in Figure \ref{Tmean_Ls0} but for NH summer solstice (L$_{s}$=90$^\circ$-105$^\circ$). Note that the white spot at equatorial latitudes near the surface is due to the topography.}
	\label{Tmean_Ls90}
\end{figure}

\begin{figure}[!htbp] 
	\centering
	\includegraphics[width=0.9\linewidth]{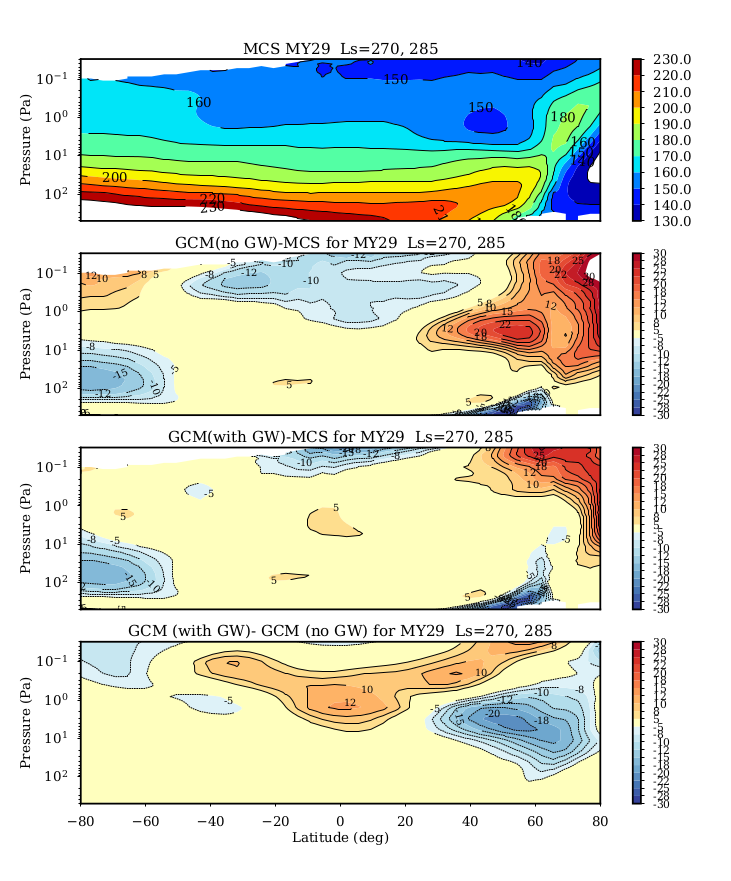}
	\caption{As in Figure \ref{Tmean_Ls0} but for NH winter solstice (L$_{s}$=270$^\circ$-285$^\circ$)}
	\label{Tmean_Ls270}
\end{figure}

\subsection{Impact of non-orographic GW drag on LMD-MGCM thermal structure}

Examples of the impact of our GW scheme on the thermal structure of the Martian mesosphere predicted by the LMD-MGCM (up to about 100 km) are given in Figures \ref{Tmean_Ls0}-\ref{Temp_Lsvar_60N-80N}.
Figures \ref{Tmean_Ls0}-\ref{Tmean_60S-80S_season} compare zonal mean day-night temperature averages T$_{mean}$  from MCS data (top panels) with model results from the LMD-MGCM with and without the non-orographic GW parameterization (middle panels). Differences between the model runs with and without the parameterization are also given. T$_{mean}$  stands for (T$_{am}$ + T$_{pm})/2$, where T$_{am}$ and T$_{pm}$ are the temperature at 03:00 h and 15:00 h local time, respectively.  As discussed in Section \ref{MCS_description}, latitudes larger than 80$^\circ$ (North and South) are excluded from this study to avoid comparison at non-homogeneous local time.
In Figures \ref{Temp_Lsvar_20S-20N} and \ref{Temp_Lsvar_60N-80N} we plot examples of nightime (T$_{am}$) and daytime (T$_{pm}$) zonal mean temperatures at pressure z= 1 Pa, for latitude ranges 20$^\circ$S-20$^\circ$N and 60$^\circ$N-80$^\circ$N, respectively,  during MY29 and for all seasons in which MCS data are available (L$_s$ = 0-330$^\circ$).
\\It is assumed that the state of the model before including the new GW parameterization was consistent but incomplete to represent the reality, and we add the new GW parameterization hoping to get a more realistic emulation of the Martian atmosphere by our GCM. However, there is a chance that the new parameterization is instead compensating for errors in other routines. This weakness applies to any GCM study and we have to make the underlying assumption that there is no unphysical error in the existing physics routines implemented in our model. Notably, the parameterization by \citet{Madeleine2014} and \citet{Navarro2014} indeed improved the comparison between the LMD-MGCM and the observations for good physical reasons (e.g. coupled dynamical-radiative processes for dust, and radiative effect of clouds), but we have no choice other than to conduct our GCM work under the above-mentioned underlying assumption. 

\subsubsection{Equinoxes}
The general circulation at the Mars equinoxes is dominated by two prograde mid-latitude jets corresponding to the equator-to-pole heating gradients at low altitude with a single Hadley cell in each hemisphere. Above 10 Pa ($\sim$ 40 km), the LMD-MGCM predicts a dynamically driven temperature inversion around 1 Pa ($\sim$ 60-70 km) above both poles \citep{Forget1999}.
When the GW scheme is off, we found that during the Northern Hemisphere (NH) Spring Equinox (see Figure \ref{Tmean_Ls0}) the MGCM-MCS difference exceeds 20 K; in particular the model is colder at middle to high latitudes below 1 Pa, and warmer in the region between 1 and 0.1 Pa (about 50-75 km altitudes), mainly around 60$^\circ$ latitude (North/South). In the MGCM version with the non-orographic GW scheme on, the region between 50$^\circ$ and 80$^\circ$ latitude (North/South) is up to 22 K warmer below 1 Pa and cooler above between 40$^\circ$ and 70$^\circ$ latitude (North/South), thus reducing the differences with MCS data. The bottom panel in Figure \ref{Tmean_Ls0} illustrates the net effect of our non-orographic GW parameterization  on zonal mean average temperature. Temperature in the tropical region (40$^\circ$S-40$^\circ$N) is also reduced by about 8 K around 1 Pa, in better agreement with the data.  
Similar conclusions hold when comparing results for NH   Autumn (Figure \ref{Tmean_Ls180}). However, in this case the warmer region around 1 Pa around the equator is only partially reduced when the non-orographic GW scheme is activated. 
In Figures \ref{Tmean_Ls0} and \ref{Tmean_Ls180}, the impact of GW on the temperature field is the one described for orographic (i.e. low phase speed) GW on Mars by early modeling studies like \citet{Barnes1990, Collins1997, Forget1999}: the adiabatic cooling at low latitudes and the polar warming is enhanced and shifted to lower altitudes (10 to 100 Pa level) as a results of the GW friction acting on the zonal wind. Lowering the zonal wind reduces the Coriolis forces that limit the poleward meridional winds. In particular, this enhances the mass convergence at high latitudes and strengthens the polar warming.

\subsubsection{Solstices}
\begin{figure}[!htbp] 
	\centering
	\includegraphics[width=1\linewidth]{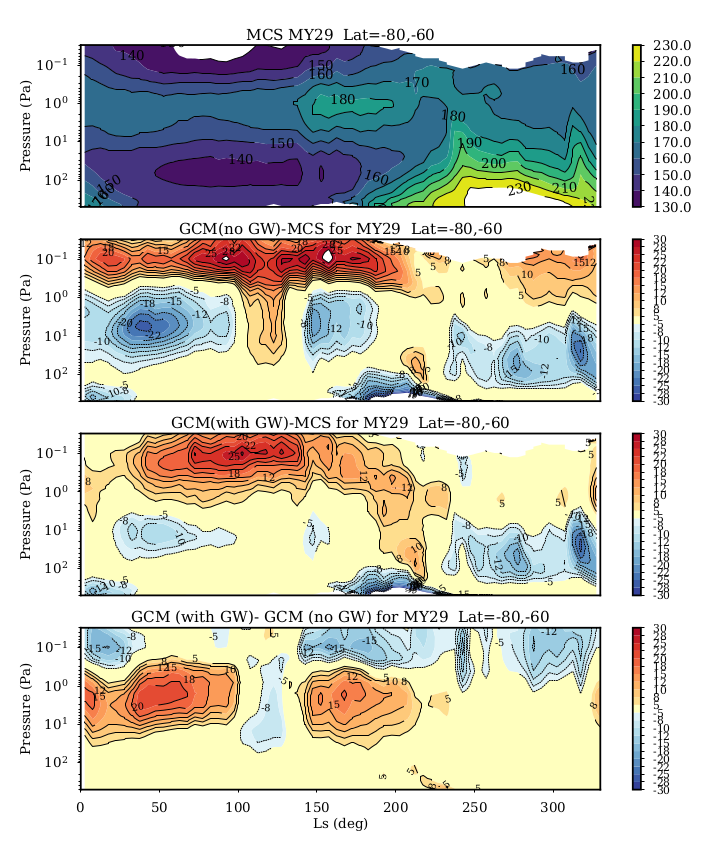}
	\caption{Zonal mean temperature  as in Figures \ref{Tmean_Ls0}-\ref{Tmean_Ls270} but as function of solar longitudes,  in the latitudinal band 60$^\circ$S-80$^\circ$S.}
	\label{Tmean_60S-80S_season}
\end{figure}

\begin{figure}[!htbp] 
	\centering
	\includegraphics[width=1\linewidth]{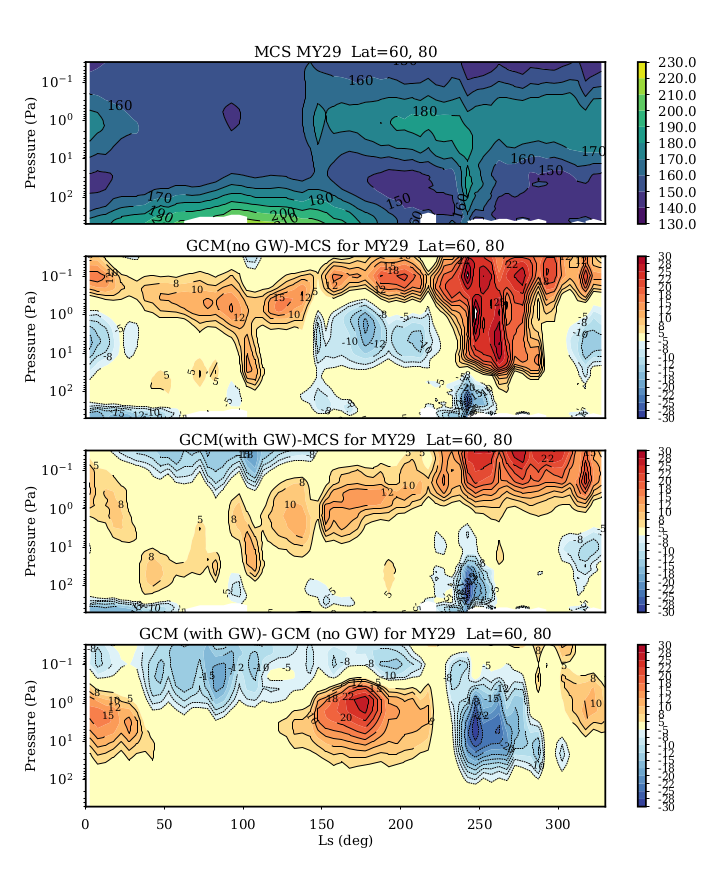}
	\caption{Zonal mean temperature  as in Figures \ref{Tmean_Ls0}-\ref{Tmean_Ls270} but as function of solar longitudes,  in the latitudinal band 60$^\circ$N-80$^\circ$N.}
	\label{Tmean_60N-80N_season}
\end{figure}

Concerning the solstice, the impact of non-orographic GW in MGCM simulations is also significant, during both the NH summer solstice (L$_{s}$=90$^\circ$-105$^\circ$) and NH winter (L$_{s}$=270$^\circ$-285$^\circ$), as shown in Figure \ref{Tmean_Ls90} and  \ref{Tmean_Ls270}, respectively.
As discussed in \citet{Forget1999}, around Northern winter solstice the strong pole-to-pole diabatic forcing creates a quasi-global Hadley cell which extends to 0.05 Pa ($\sim$ 80 km). In such a cell the Coriolis force contributes to accelerate the poleward meridional motion on Mars, thus inducing a mass convergence and strong warming of the middle polar atmosphere down to about 5 Pa ($\sim$ 25 km), as also observed \citep{JakoskyMartin1987, Theodore1993}. \citet{Forget1999} suggested that thermal inversions can generally be expected around 1 Pa (60-70 km) above the winter polar regions near solstice, and above both poles near equinox.
However, we note here that the effect of GW is not an enhancement of high latitude (60$^\circ$-80$^\circ$) warming (see also Figures \ref{Tmean_60N-80N_season} and \ref{Tmean_60S-80S_season}), as during the equinoxes. On the contrary, around Northern winter solstice the GW drag tends to reduce the high latitude warming by more than 15 K (e.g. Figure \ref{Tmean_Ls270}), while heating the atmosphere above 1 Pa ($\sim$ 60 km) at low/middle latitudes. This appears different than in most published results on orographic and non-orographic GW, and the opposite of what was simulated at the equinoxes (compare bottom panels in Figures \ref{Tmean_Ls0} and \ref{Tmean_Ls270}). In reality, as discussed in Section \ref{GWmean_flow_forcing}, the polar warming (poleward of 80$^\circ$N, not shown in Figures \ref{Tmean_Ls270}) is increased as expected (see Figure \ref{U_V_T_diffGW_L270300}), and it is only between 30$^\circ$N and 80$^\circ$N that the warming is reduced. This results from the poleward shift of the warming subsidence, but also from the direct effect of the GW drag on the meridional wind and on the damping of thermal tides (see section \ref{diurnal_tides}), which partly controls the high latitude warming around northern winter solstice \citep{WilsonHamilton1996}.
Data-model comparison improves mostly in NH winter, with biases at pressure range 10-1 Pa reduced up to 20 K, especially between 40$^\circ$N and  80$^\circ$N  latitude (see Figure \ref{Tmean_Ls270}). There are still differences between MGCM and MCS results: at similar pressure levels and around the equator simulations are warmer than MCS data by 5 K, thus increasing model biases.  
Possible causes for those remaining discrepancies are related to: i) the uncertainties on the dust distribution in the model during the dust season, ii) the radiative effects of water-ice clouds on the thermal structure during the NH summer, and/or iii) the vertical resolution in the water ice cloud structure, during the northern summer season. At the aphelion the uncertainties related to the cloud modeling are larger than the effect of non-orographic GW on the large-scale circulation, and it is still challenging to reach a good accuracy to represent it well with GCMs.


\subsubsection{Thermal structure seasonal variations}

\begin{figure}[!htbp] 
	\centering
	\includegraphics[width=1\linewidth]{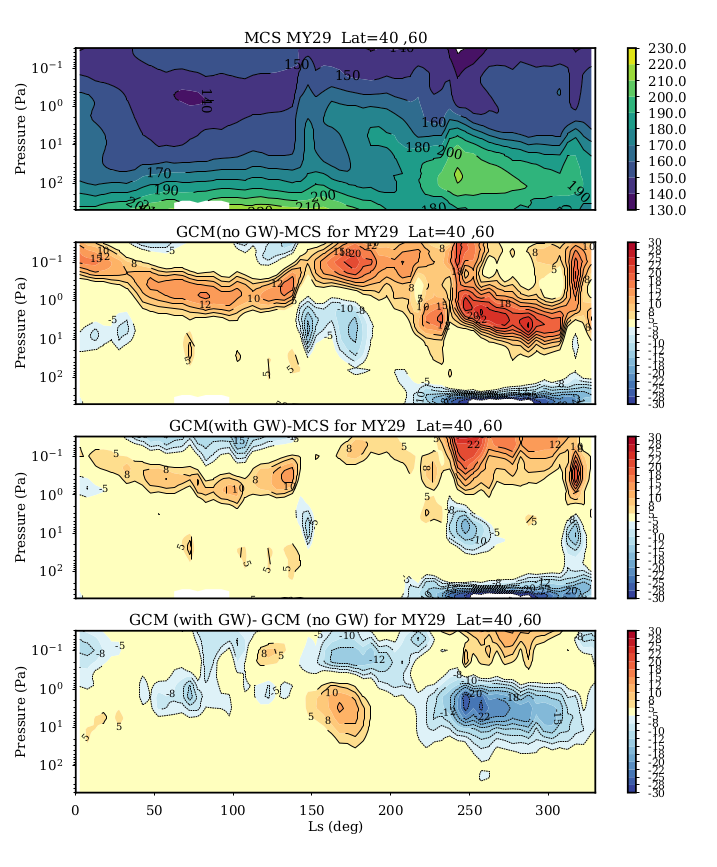}
	\caption{Zonal mean temperature  as in Figures \ref{Tmean_Ls0}-\ref{Tmean_Ls270} but as function of solar longitudes,  in the latitudinal band 40$^\circ$N-60$^\circ$N.}
	\label{Tmean_40-60_season}
\end{figure}

\begin{figure}[!htbp] 
	\centering
	\includegraphics[width=1\linewidth]{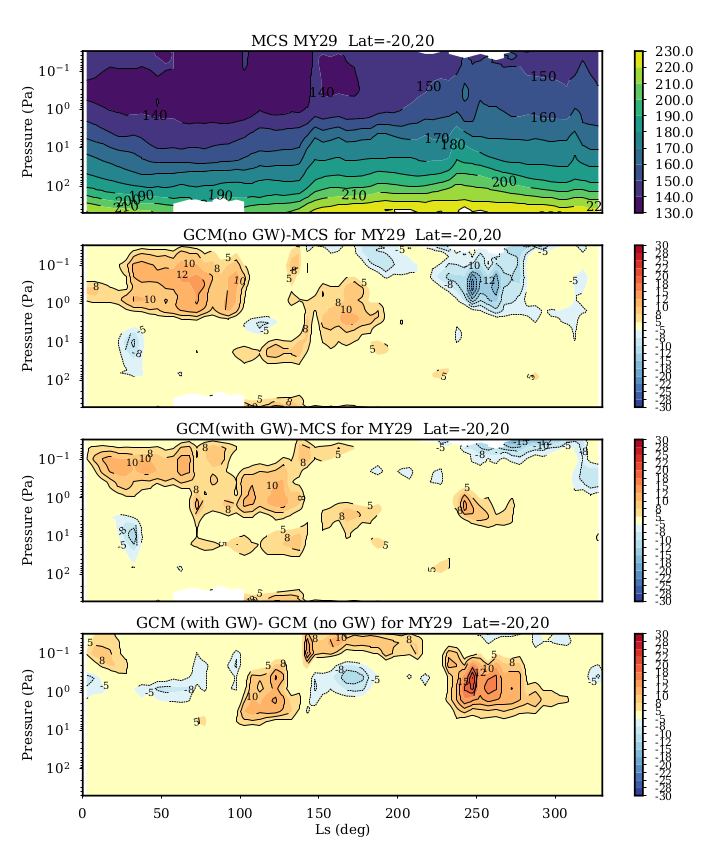}
	\caption{Zonal mean temperature  as in Figures \ref{Tmean_Ls0}-\ref{Tmean_Ls270} but as function of solar longitudes,  in the equatorial latitudinal band 20$^\circ$S-20$^\circ$N. The white spot near the surface is due to the topography.}
	\label{Tmean_20-20_season}
\end{figure}

To illustrate the seasonal evolution of simulated temperature after the implementation of the GW scheme into the LMD-MGCM using the baseline GW parameters as in Table \ref{tab1}, zonal day-night averages (T$_{am}$ + T$_{pm}$)/2 are plotted in Figures \ref{Tmean_60S-80S_season}-\ref{Tmean_20-20_season} as function of solar longitude L$_s$ for a selection of latitudinal bands: 20$^\circ$S-20$^\circ$N, 40$^\circ$N-60$^\circ$N, 60$^\circ$N-80$^\circ$N and 60$^\circ$S-80$^\circ$S.
Overall, those figures indicate that there is a small reduction in the magnitude of the biases at most seasons and latitudes, with some notable exceptions: (i) the mid-altitude negative bias at high southern latitudes in the first half of the year (northern spring and summer) is greatly reduced up to 20 K (Figure \ref{Tmean_60S-80S_season}), (ii) the mid-altitude positive bias at high northern latitudes in early winter (L$_s$= 240-270)  also decreases by 15-20 K (Figure \ref{Tmean_60N-80N_season}). Note that this period corresponds to the bulk of the dust storm season, yet the GW scheme improves the predicted thermal structure at those latitudes (iii) Positive biases in northern mid-latitudes (40$^\circ$-60$^\circ$N) are generally reduced up to 10 K at all altitudes and times of the year, except at very high altitudes around local winter solstice (Figure \ref{Tmean_40-60_season}) (iv) Large negative biases at low altitudes at high latitudes (in northern fall) and northern mid-latitudes (in northern fall and winter) are not improved by the non-orographic GW parameterization and may be due to incorrect dust distributions and/or interactions between dust and water ice clouds during the dustiest time of the year.
Remaining biases in the dust storm season (L$_s$= 180$^\circ$-330$^\circ$) between GCM simulations with the GW scheme activated and MCS observations can be attributed to aerosol effects. As circulation and vertical transport of dust are intertwined \citep{Kahre2015}, and presence of water ice clouds critically depends on temperature, complex feedbacks between modeled temperature and modeled fields of aerosols prevail, that need to be further explored in future model developments.

\begin{figure}[!htbp] 
	\centering
	\includegraphics[width=1.1\linewidth]{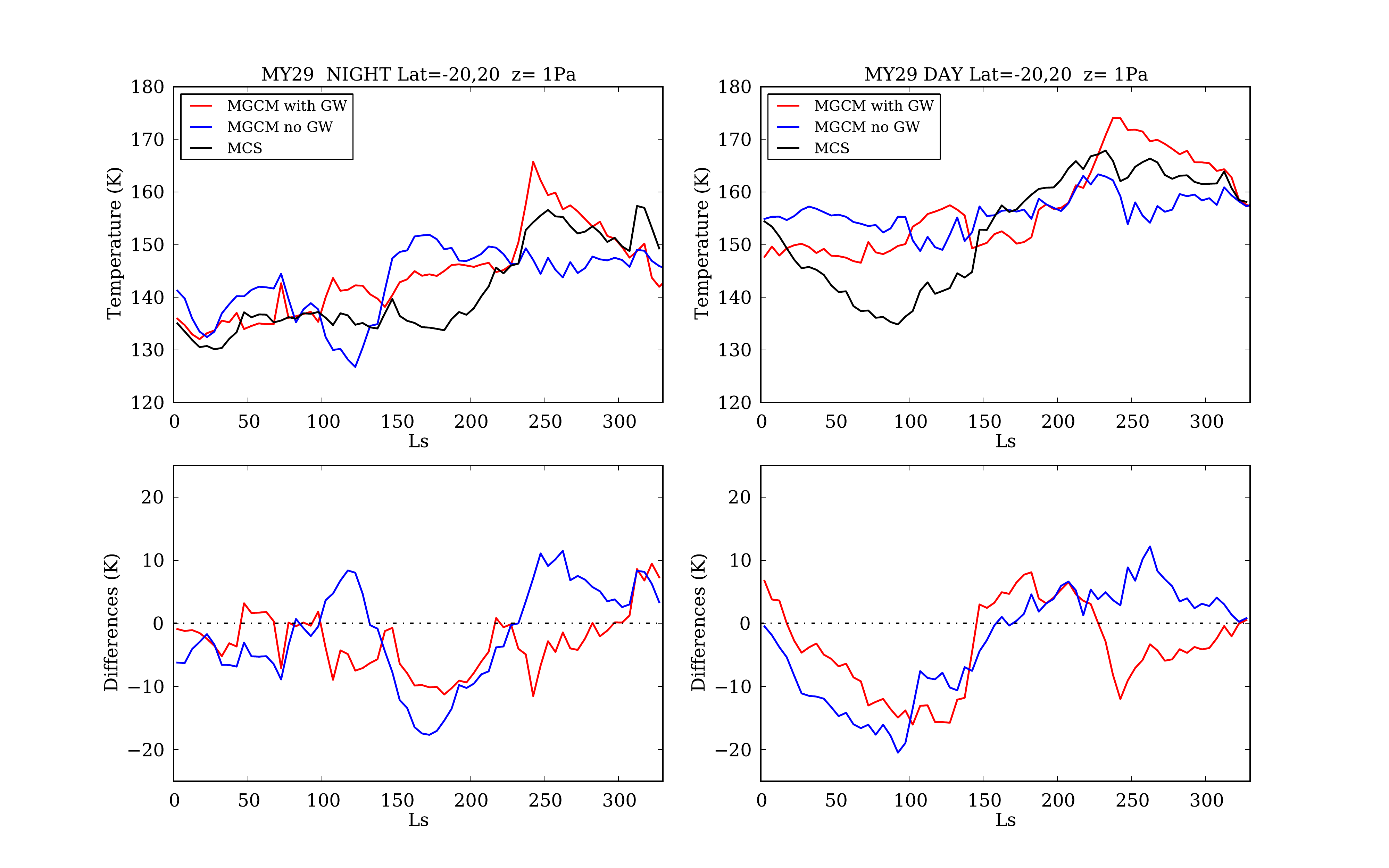}
	\caption{Seasonal variation of zonal temperature at pressure 1 Pa at nighttime LT=03:00 (top left) and daytime LT=15:00 (upper right), averaged at equatorial latitudes 20S-20N. MGCM simulations without (blue solid line) and with the GW scheme included (red solid line) using the subset of GW parameters as in Table \ref{tab1} are shown together with MCS observations (black solid lines). Temperature differences (MCS-MGCM) are plotted in K in the lower panels, for night (bottom left) and day time (bottom right)}	
	\label{Temp_Lsvar_20S-20N}	
\end{figure}

\begin{figure}[!htbp] 
	\centering
	\includegraphics[width=1.1\linewidth]{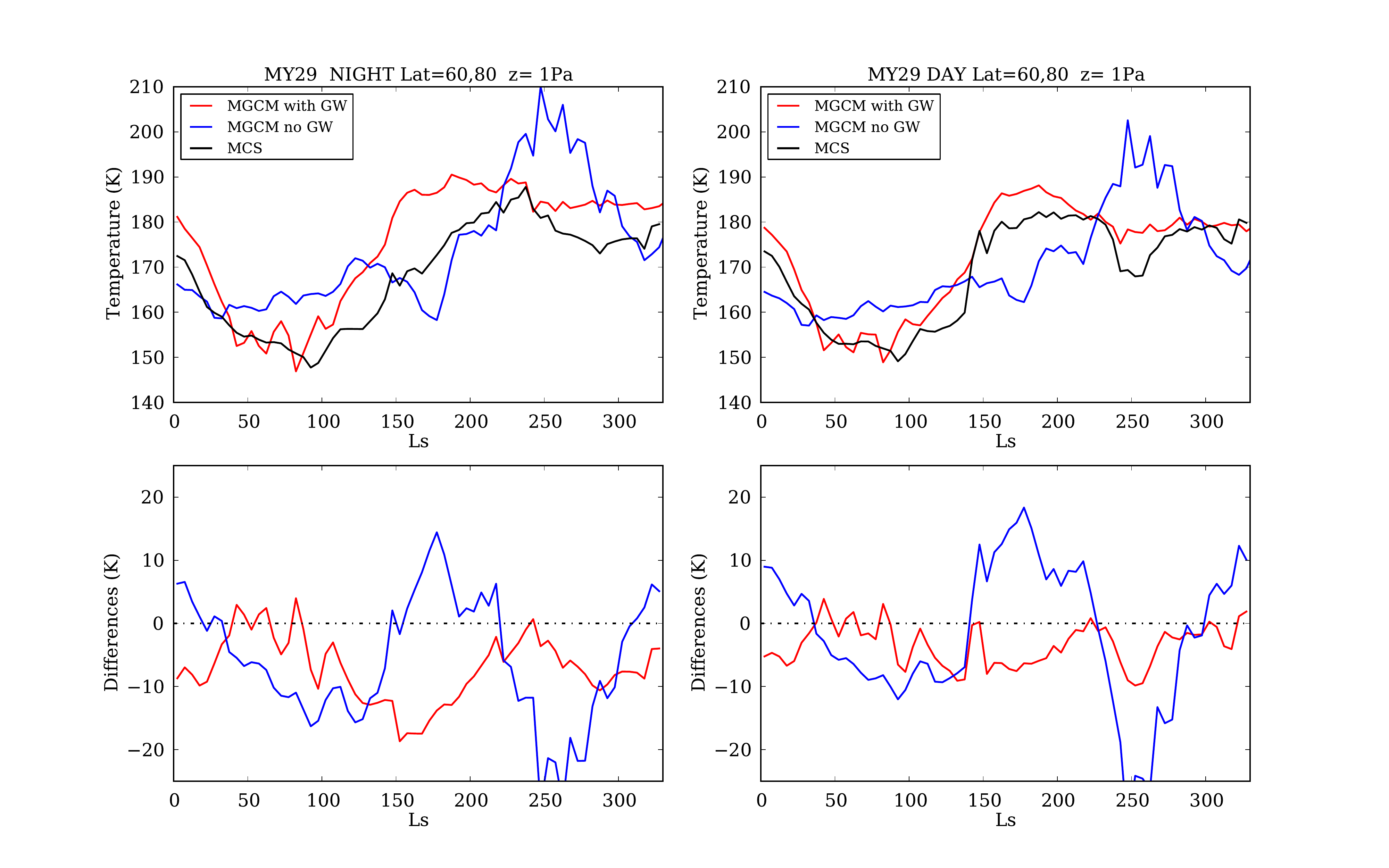}
	\caption{Same as Figure \ref{Temp_Lsvar_20S-20N} but for middle-high northern latitudes (60N-80N).}
	
	\label{Temp_Lsvar_60N-80N}	
\end{figure}

The improvement of mean temperature comes from the overall better representation of night and day temperatures, for most seasons. Two examples are shown in Figures \ref{Temp_Lsvar_20S-20N} and \ref{Temp_Lsvar_60N-80N}, together with MCS data.
In the equatorial region (Figure \ref{Temp_Lsvar_20S-20N}) nighttime temperatures are reduced up to 10 K near the perihelion and about 5-6 K at NH autumn (L$_{s}$=180$^\circ$). However, daytime temperatures are warmer ($\sim$ 8 K) than MCS around L$_{s}$=250$^\circ$, and only a few K closer to MCS data for L$_{s}$= 0-100$^\circ$.
At high to middle northern latitudes (60$^\circ$N-80$^\circ$N) the LMD-MGCM reproduces MCS daytime zonal mean temperatures more accurately for all season, in particular in the NH autumn and winter, with data-model biases reduced up to 15 K from L$_{s}\sim$150$^\circ$ to L$_{s}\sim$300$^\circ$ (during dust season) while no significant differences are obtained in the equatorial region. Nighttime zonal mean temperatures at high to middle latitudes are closer to MCS observations for all seasons when the GW scheme is activated.

\begin{figure}[!htbp] 
	\centering
	\includegraphics[width=1\linewidth]{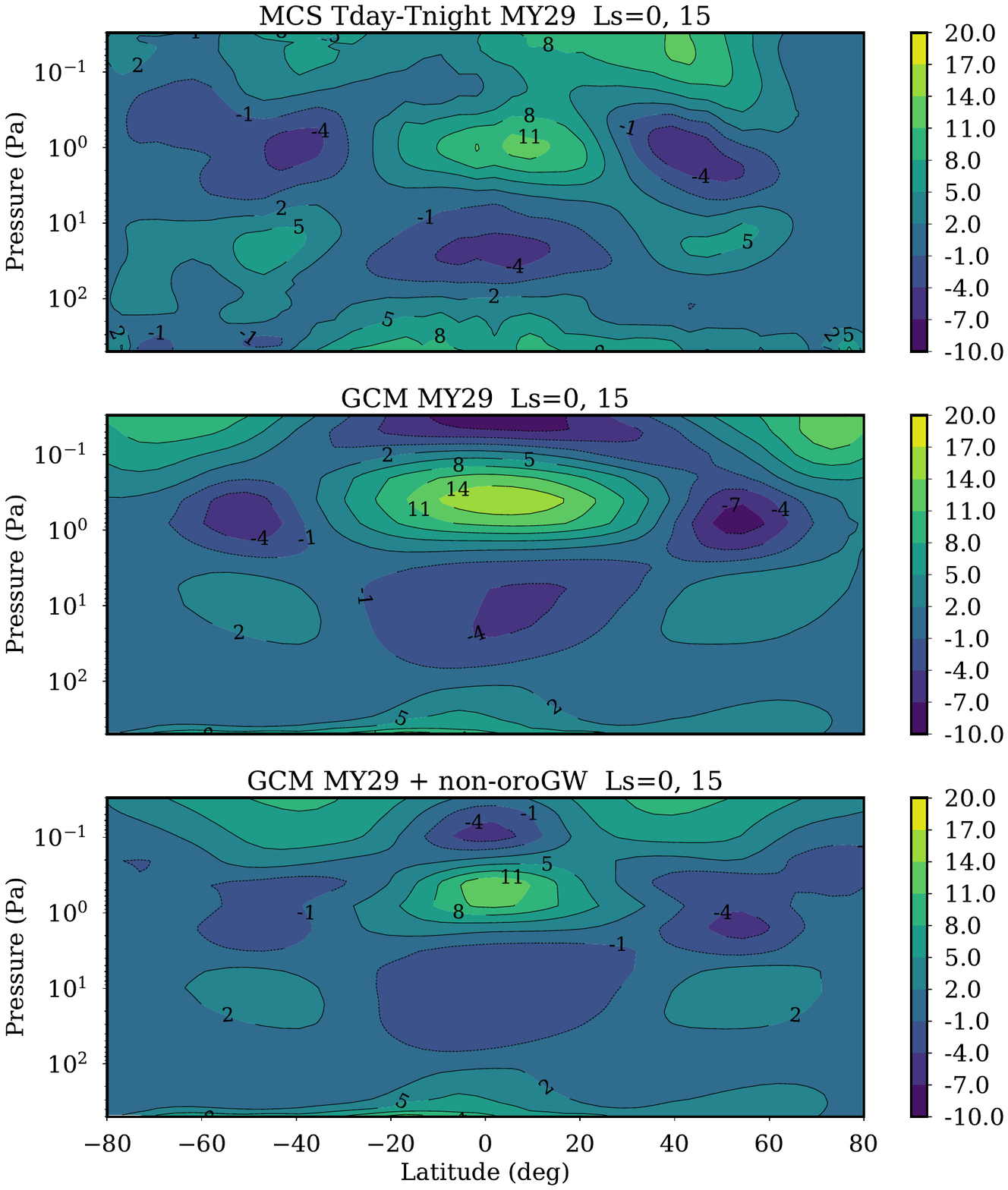}
	\caption{Zonal temperature differences (T$_{pm}$ - T$_{am}$)/2 between dayside ($\sim$ 15:00 h local time) and nightside ($\sim$ 03:00 h) observed by MCS (top panel); simulated by MGCM  without (middle panel) and with non-orographic GW (bottom panel). Data are binned in the range L$_{s}$=0$^\circ$-15$^\circ$, and baseline GW parameter values as in Table \ref{tab1} are used in our model simulations when the non-orographic GW scheme is on.}
	\label{Tides_Ls0}
\end{figure}

\begin{figure}[!htbp] 
	\centering
	\includegraphics[width=1\linewidth]{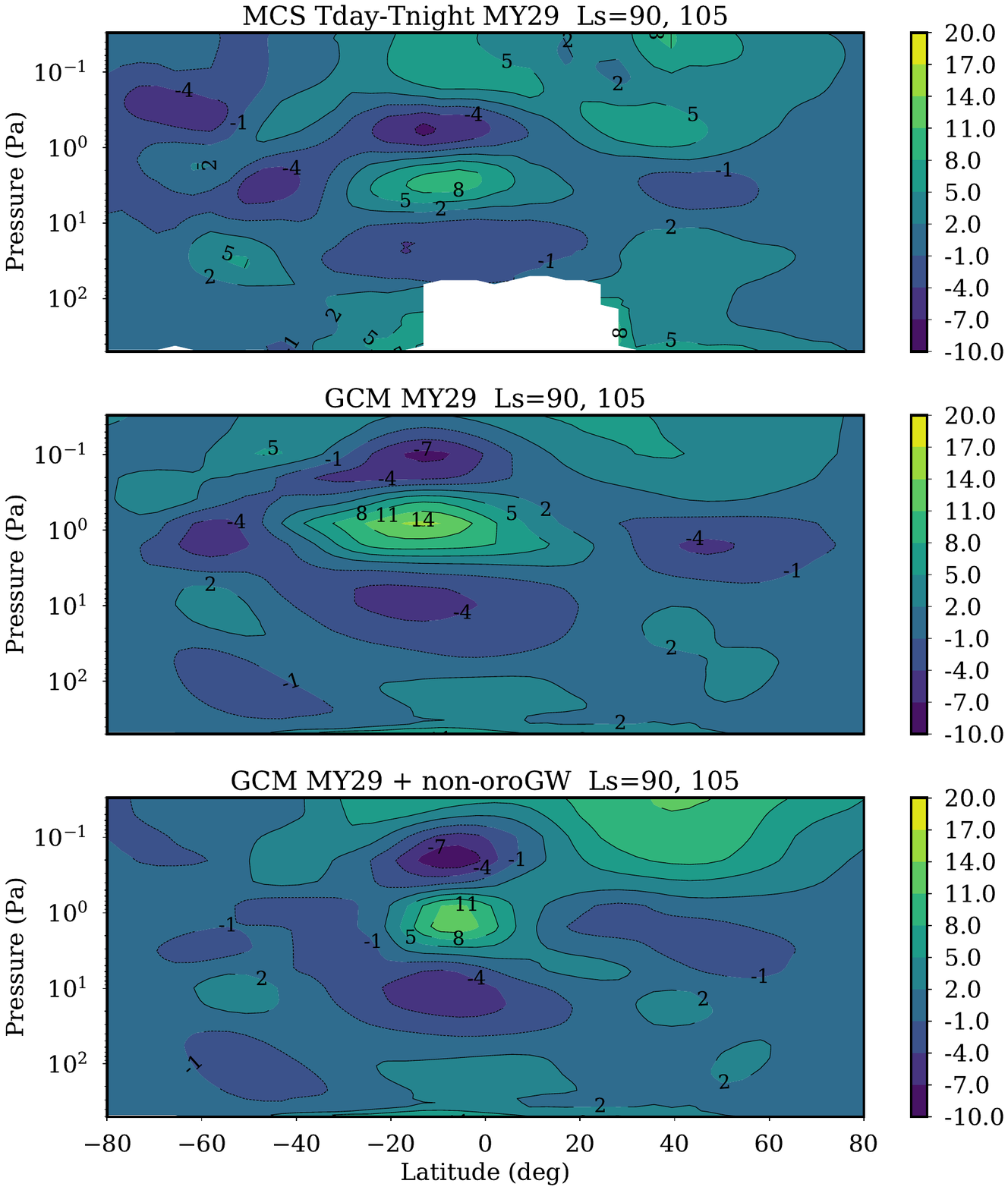}
	\caption{Example of diurnal tides as in Figure \ref{Tides_Ls0} but in the binned the range L$_{s}$=90$^\circ$-105$^\circ$. Note that the white spot at equatorial latitudes near the surface is due to the topography.
		
		}
	\label{Tides_Ls90}
\end{figure}

\begin{figure}[!htbp] 
	\centering
	\includegraphics[width=1\linewidth]{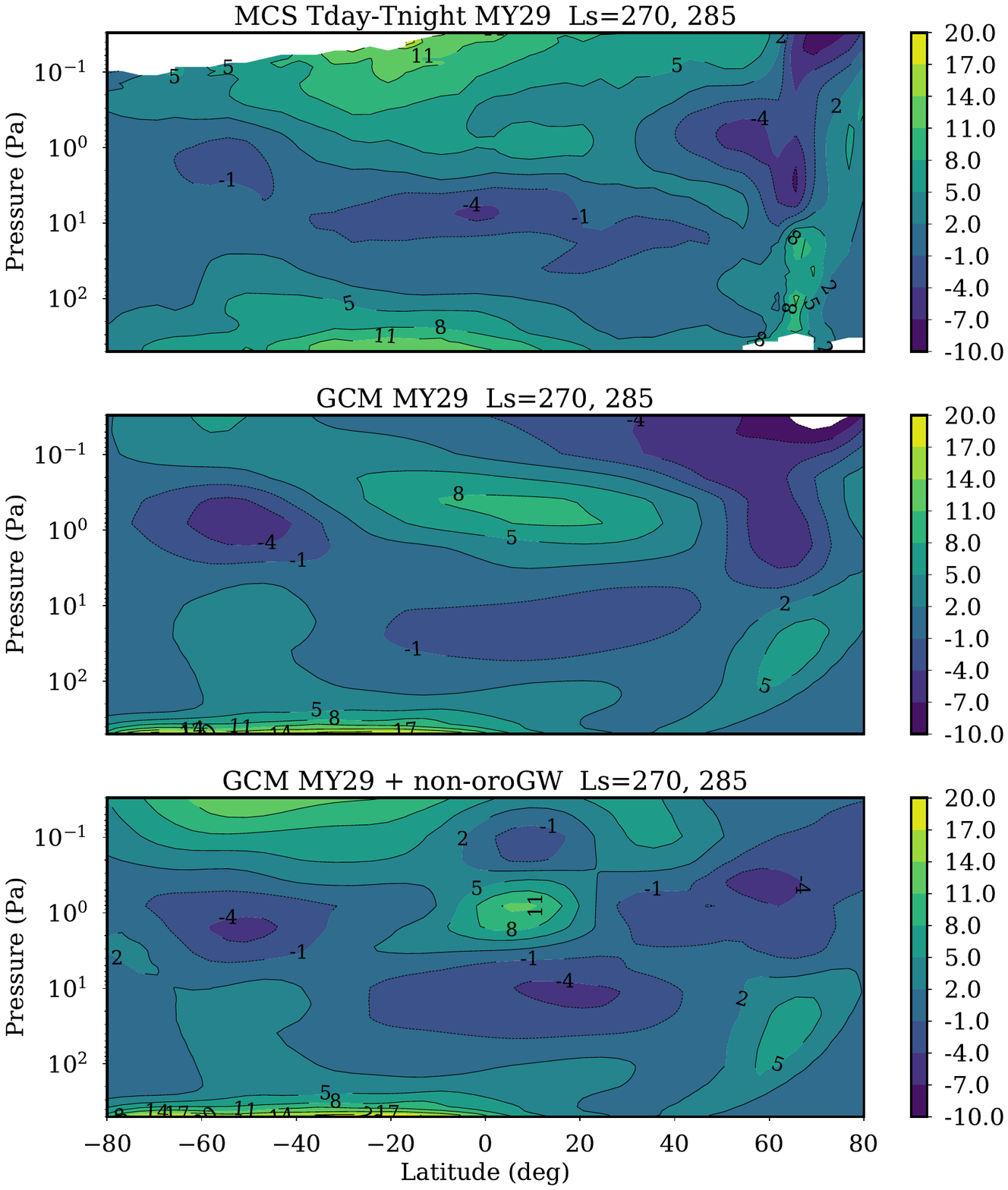}
	\caption{ As in Figure \ref{Tides_Ls0} but the data are binned in the range L$_{s}$=270$^\circ$-285$^\circ$.}
	\label{Tides_Ls270}
\end{figure}

\begin{figure}[!htbp] 
	\centering
	\includegraphics[width=1\linewidth]{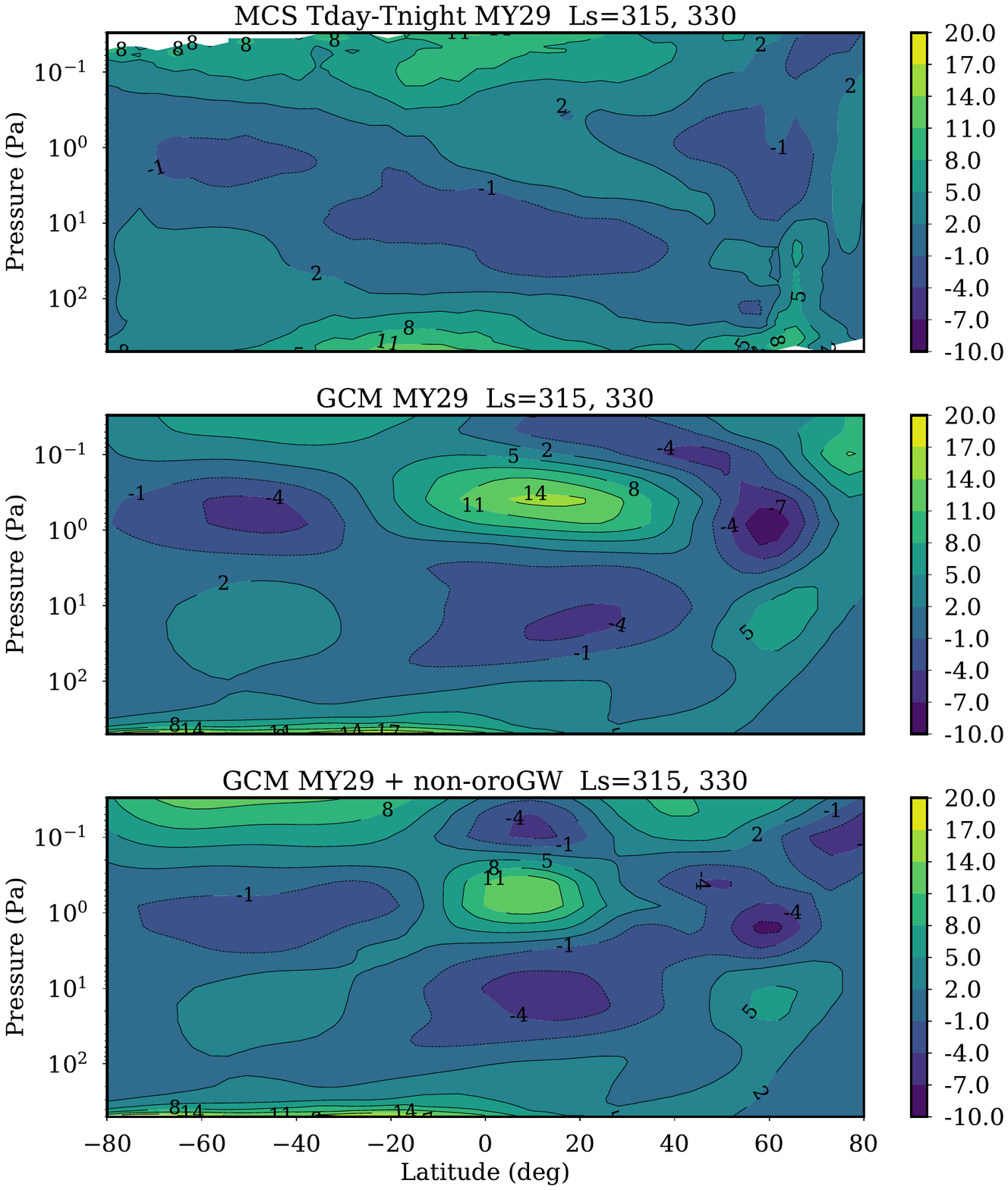}
	\caption{ As in Figure \ref{Tides_Ls0} but the data are binned in the range L$_{s}$=315$^\circ$-330$^\circ$.}
	\label{Tides_Ls315}
\end{figure}

\subsection{Improving the characterization of diurnal tides}
\label{diurnal_tides}

Thermal tides are atmospheric waves forced by the diurnal cycle of incoming sunlight and they have a major impact on temperature and wind variability in the Martian atmosphere \citep{WilsonHamilton1996}.  
The examples in Figures \ref{Tides_Ls0}, \ref{Tides_Ls90}, \ref{Tides_Ls270} and \ref{Tides_Ls315} show the quadrupole structure of the difference between dayside T$_{pm}$ and nightside T$_{am}$ temperature (T$_{diff}$ = (T$_{pm}$ - T$_{am}$)/2), as seen by  MCS and predicted by the MGCM,  for different solar longitudes, as indicated. The observed quadrupole of T$_{diff}$, centered roughly between 30$^\circ$S and 30$^\circ$N latitudes, is well known in the Martian atmosphere. It corresponds to the main response of the Hough mode of the diurnal tide, trapped between 22$^\circ$S and 22$^\circ$N latitudes and with a theoretical vertical wavelength of 30 km \citep{Lee2009, Zurek1976}. 
This T$_{diff}$ represents an important diagnostic quantity for the analysis of observations of the Martian atmosphere \citep{Lee2009, Guzewich2012}, which includes sun-synchronous tides and diurnal Kelvin waves.
The primary effect of GW is to damp the thermal tides by reducing the diurnal oscillation of the meridional and zonal wind (see Section \ref{GWmean_flow_forcing})
\\Overall, the thermal tides were correctly represented by the model, but with differences in the values of the amplitude and vertical phasing when compared with MCS \citep{Navarro2017}. The difference exceeded 15 K (the model being warmer than MCS between 1 and 10$^{-1}$ Pa), and the vertical phasing shifted slightly at higher altitudes.
Figure \ref{Tides_Ls0}, \ref{Tides_Ls90}, \ref{Tides_Ls270} and \ref{Tides_Ls315} clearly show that the there are at least two seasons (e.g. NH spring and winter) during which the improvement associated with the inclusion of non-orographic GW parameterization is significant, especially the thermal tides amplitude above 10 Pa at the equator. The non-northern winter seasons see improvement in the depiction of the tide, but only right over the equator. It also seems that the wave depiction beyond the tide's latitude singularities is worse with the GW parameterization. The poorer performance in northern winter (Figures \ref{Tides_Ls270} and \ref{Tides_Ls315}) suggests that the changes to the zonal wind driven by GW may be spurious as the tide is ducted into the northern hemisphere by the strong zonal wind in this season, and the baseline GCM simulations seem to show this better, at least in the overall structure.
With the non-orographic GW scheme activated, the maximum day-night difference value is around 10 K, comparable with observed values, and more interestingly the altitude of the peak of the tides is shifted down and its amplitude is also more realistic (between 22$^\circ$S and 22$^\circ$N latitudes, as observed). As discussed in the previous section, those improvements come from the overall better representation of  both the nighttime (LT= 3:00) and daytime (LT= 15:00) thermal structure of the Martian atmosphere above 10 Pa.

\subsubsection{Diurnal tide seasonal variation}
\label{diurnal seasonal_var}

\begin{figure}[!htbp] 
	\centering
	\includegraphics[width=\linewidth]{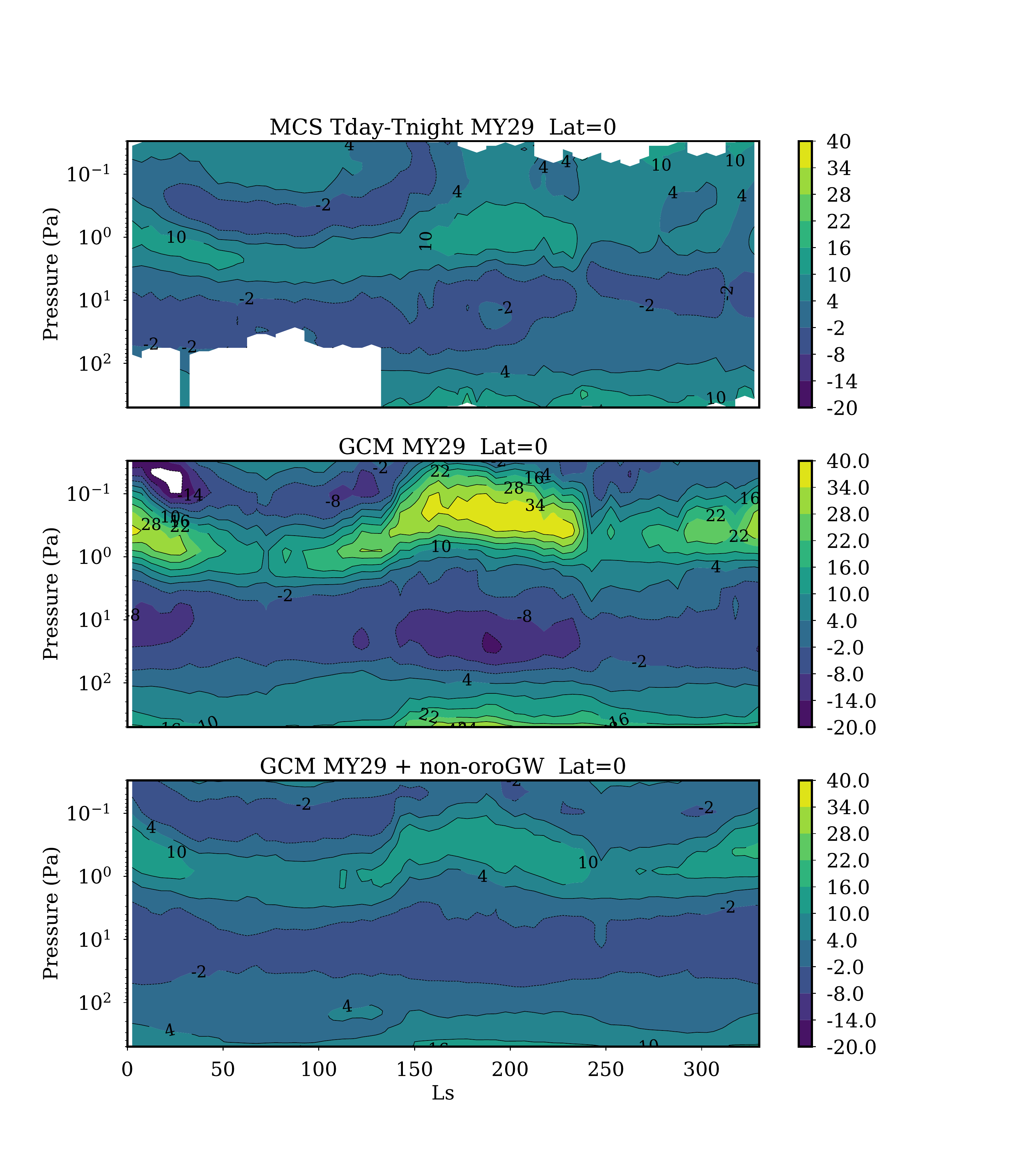}
	\caption{Zonal day-night temperature differences in the tropical band 20$^\circ$S-20$^\circ$N, as function of solar longitudes, seen by MCS (top panel) and predicted by MGCM without (middle panel) and with (bottom panel) GW using baseline GW parameter values in Table \ref{tab1}.}
	\label{Tides_tropic_Ls}
\end{figure}

The evolution of the main mode of the diurnal tide T$_{diff}$ in the tropical band 20$^\circ$S-20$^\circ$N over all Martian seasons (MY29) is shown in Figure \ref{Tides_tropic_Ls} for MCS data (top panel) and MGCM results (middle and bottom panels). The seasonal trend of T$_{diff}$ is better represented by the MGCM with the GW scheme turned on (bottom panel), confirming the overall improvement in characterizing thermal tides in both amplitude and intensity, as described in the previous section. 
However, MCS data show a clear feature (e.g. the slight increase of temperature around 20-30 Pa that shows up as a green blob) linked to a dust event at L$_s$= 240$^{\circ}$ which drastically changes the T$_{diff}$ structure, and that the simulations presented in this paper fail to reproduce. This is probably related to the fact that the Martian thermal structure during dust storms is not well represented by the GCM. The storm is clearly visible in the \citet{Montabone2015} map of column dust optical depth for this MY, so perhaps the vertical dust distribution, size distribution, and/or water ice interactions are not correctly represented in our model, and they are producing the wrong thermal structure at that L$_s$. Otherwise this suggests that during dust storms there are more GW propagating in the atmosphere, which could be taken into account in a future work by including a GW source varying with seasons in our scheme. It will be also crucial in the future to get the GW response to dust storms right when substantially larger storms are present (e.g. MYs 25, 28 and 34).

\begin{table}[!htbp]

	\centering
	\begin{tabular}{| c | c | c | c |}
		\hline 
		& |$c$|  &  F$^{0}$  &  S$_c$ \\

		& \small [m/s] &  \small [kg m$^{-1}$ s$^{-2}$ ] &   \\
		\hline 
		Case1  & [1-30]  &  [0-7$ \cdot 10^{-8}$]   &  1  \\
		Case2  & [1-30] &   [0-7$ \cdot 10^{-7}$]   &  1  \\
		Case3  & [1-30]  &  [0-7$ \cdot 10^{-6}$]   &  1   \\
		Case4  & [1-30]  &  [0-7$ \cdot 10^{-7}$]   &  0.1  \\
		Case5  & [1-30]  &  [0-7$ \cdot 10^{-7}$]   & 10  \\
		Case6  & [1-60]  &  [0-7$ \cdot 10^{-7}$]   &  1  \\
		Case7  & [1-10]  &  [0-7$ \cdot 10^{-7}$]   & 1   \\
		\hline
	\end{tabular}
	\caption{Selection of sensitivity tests described in this work, varying the upper limit of the probability distribution for three of the main tunable GW parameters (phase speed $c$, moment flux F$^{0}$ at the source, Saturation S$_c$)}
\label{tab2}
\end{table}

\section{Sensitivity tests: impact on thermal tides}

\begin{figure}[!htbp] 
	\centering
	\includegraphics[width=1.1\linewidth]{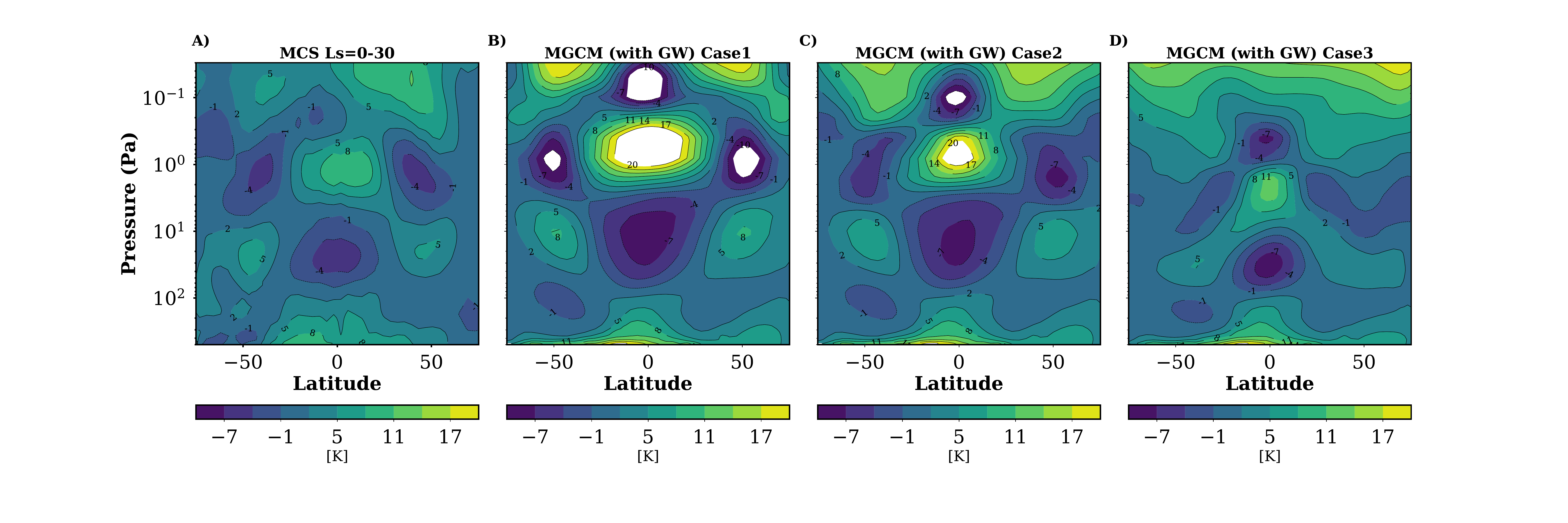}
	\includegraphics[width=1.1\linewidth]{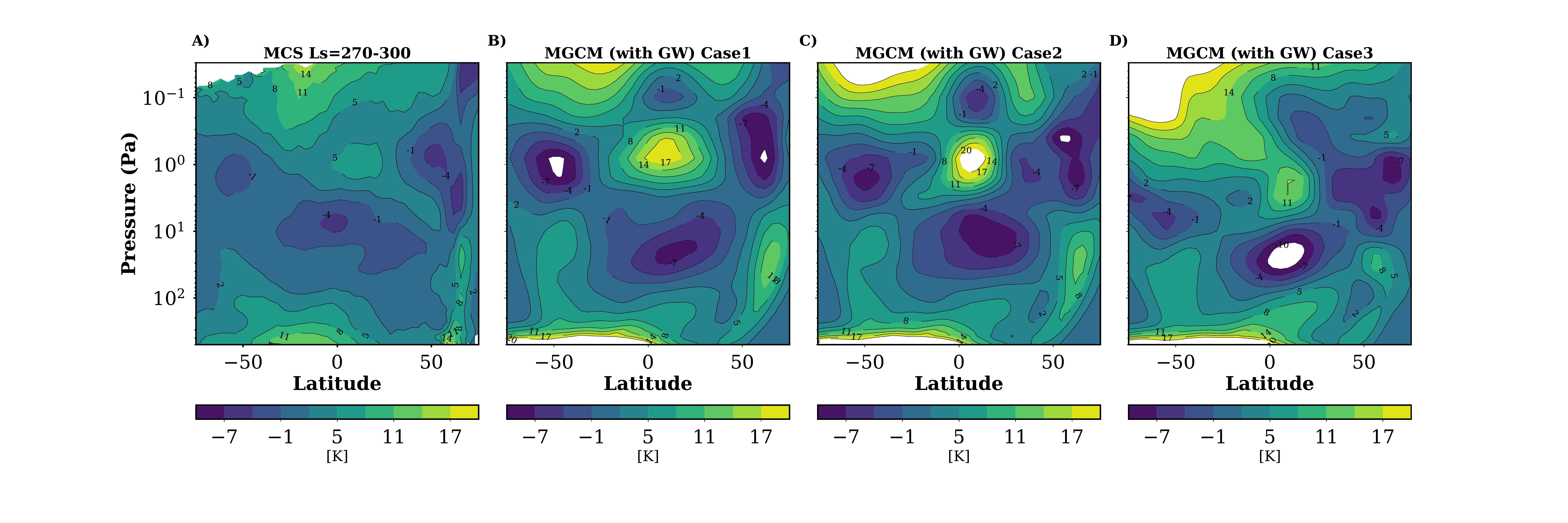}
	\caption{Selection of sensitivity tests performed in this work, varying only the upper limit of the probability distribution of the vertical momentum flux at the sources (F$^{0}_{max}$), as in Case 1 (Panel B), Case 2 (Panel C) and Case 3 (Panel D) in Table \ref{tab2} (see text for details). The contours are simulated day-night temperature differences T$_{diff}$ between dayside (15 $h$) and nightside (3 $h$), shown for two seasons: Ls= 0$^\circ$-30$^\circ$N (upper panels) and Ls= 270$^\circ$-300$^\circ$ (lower panels). The "best-case" simulation (Case 2 in Table \ref{tab2}), used as reference throughout this paper is represented in Panels C. MCS results for the same seasons (panels A) are  also shown for comparison.}
	\label{FigsensTest1}
\end{figure}

\begin{figure}[!htbp] 
	\centering
	\includegraphics[width=1.1\linewidth]{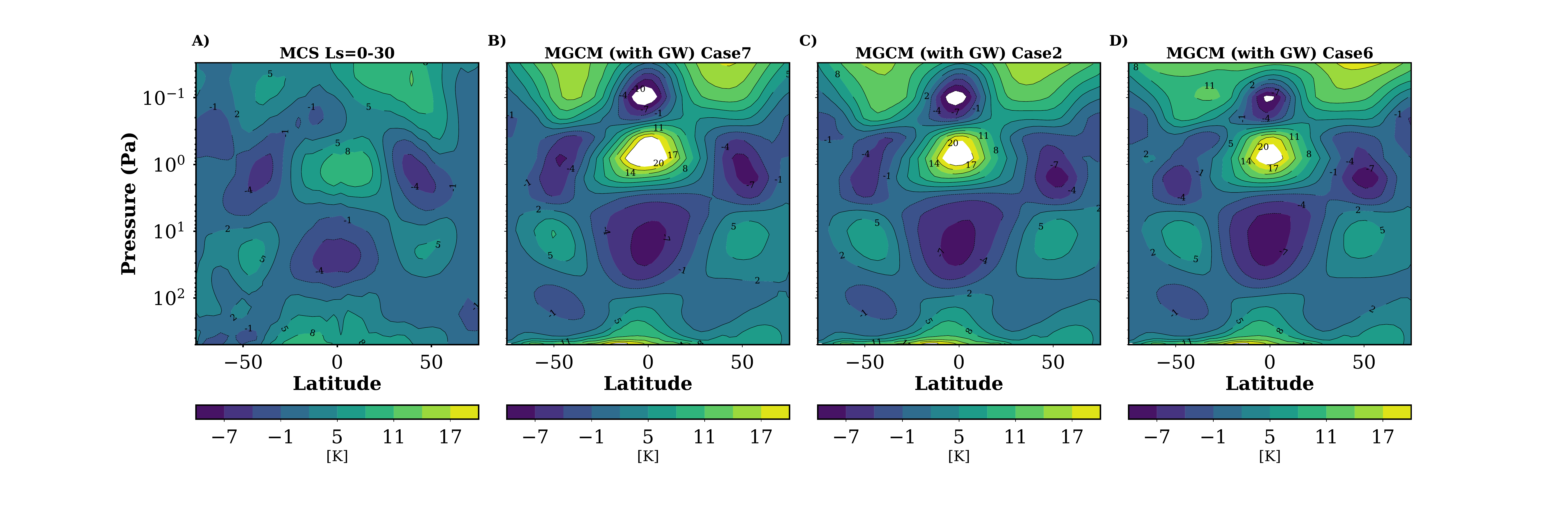}
	\includegraphics[width=1.1\linewidth]{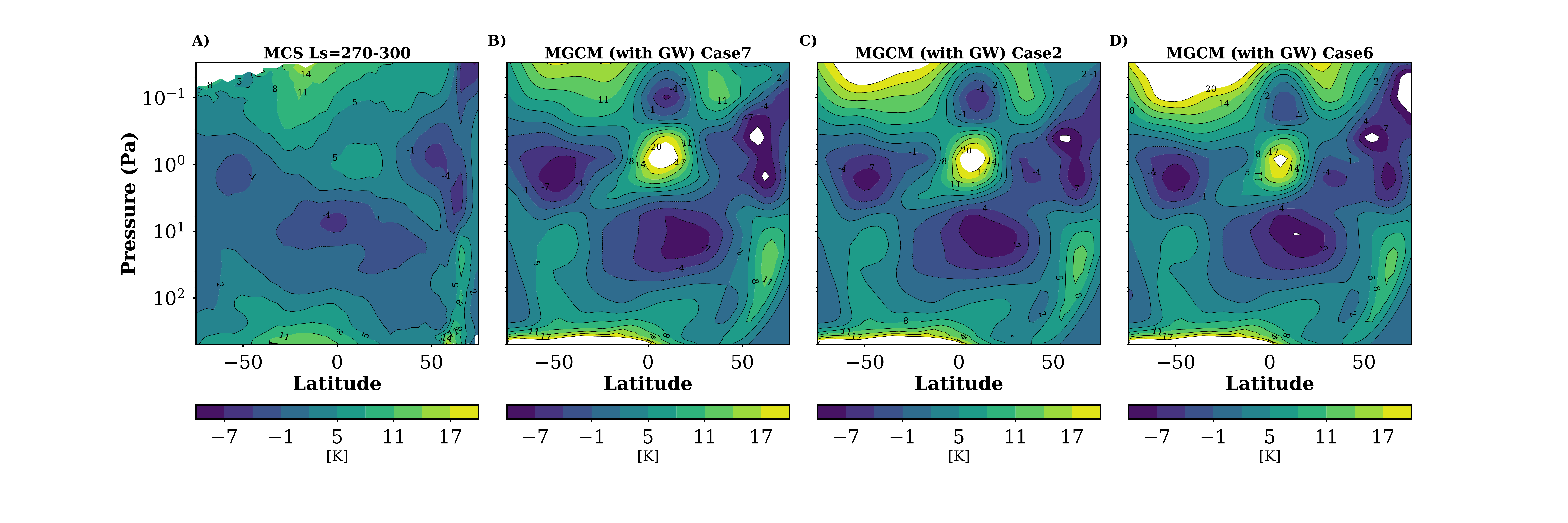}
	\caption{Sensitivity tests as in Figure \ref{FigsensTest1} but  varying only the upper limit of the probability distribution of phase speed $c_{max}$. The reference run (Case 2) is plotted in the panels C), while panels B) and D) represent Case 7 and Case 6, respectively.}
	\label{FigsensTest2}
\end{figure}

\begin{figure}[!htb] 
	\centering
	\includegraphics[width=1.1\linewidth]{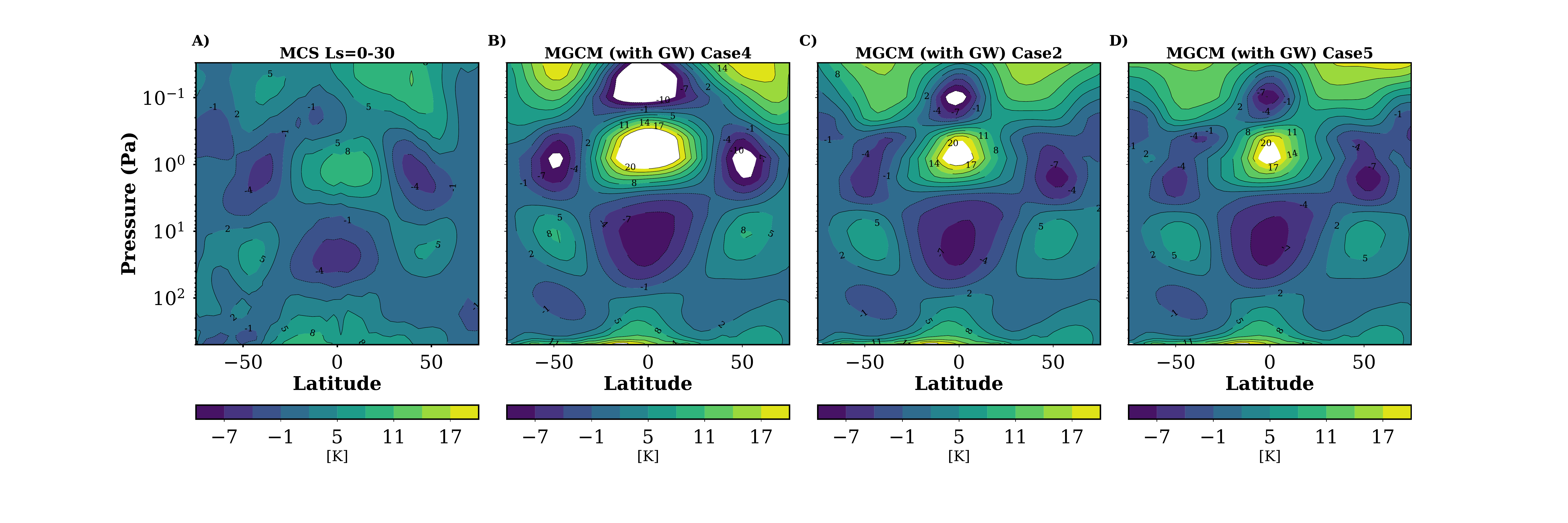}
	\includegraphics[width=1.1\linewidth]{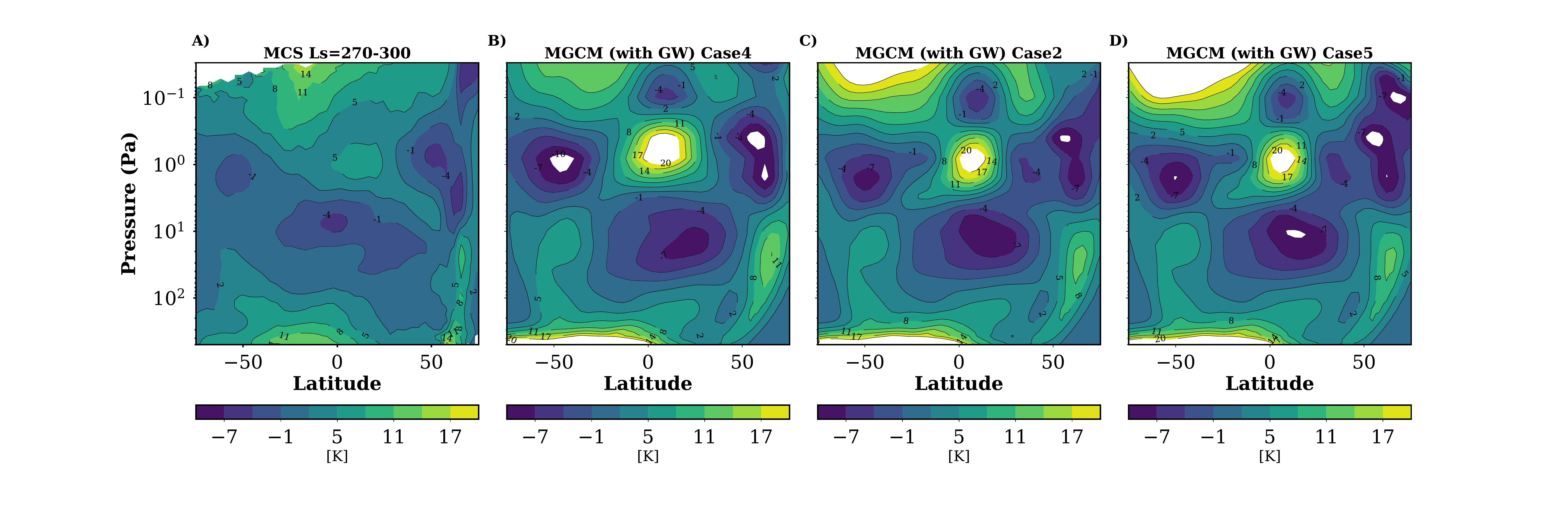}
	\caption{Sensitivity tests as in Figure \ref{FigsensTest1} but  varying the saturation parameter S$_c$. The reference run (Case 2)  is plotted in the panels C), while panels B) and D) represent Case 4 and Case 5, respectively.}
	\label{FigsensTest3}
\end{figure}

\label{sensTest}
A number of sensitivity tests were performed to further evaluate the effect of GW induced drag on the GCM zonal wind and temperature and we use thermal tides (e.g. T$_{diff}$) as a diagnostic. We found that our simulations, notably the zonal wind fields, were sensitive to all the set of parameters used in the scheme. Therefore the choice of the parameter ranges was done with caution, according to observational and theoretical constraints as described in Section \ref{Sec_inputs}. We select here 7 tests for a range of plausible values of gravity wave parameters and for different representations of the mean flow thermal forcing, listed in Table \ref{tab2}.
For this set of runs the horizontal wavenumber k$_h$ range was fixed as in Table \ref{tab1}, corresponding to horizontal wavelength $\lambda_h$ between 10 km and 300 km (consistent with available observations); the other parameters vary with respect to the "best-fit" case in Table \ref{tab1}, which corresponds to Case 2 in Table \ref{tab2}). Specifically, the F$^{0}_{max}$ was increased/decreased by one order of magnitude (Case 1 and Case 3 in Table \ref{tab2}, respectively) and the phase speed |$c_{max}$| multiplied/divided by a factor 2 and 3 (Case 6 and Case 7, respectively). We found that the phase speed ($c_{max}$) is not a useful knob to tune compared with the momentum flux, which is instead one of the most important parameters in our scheme. For instance, values of F$^{0}_{max}$ below 10$^{-8}$ kg m$^{-1}$ s$^{-2}$ produced a very negligible impact on the simulated field, while values larger than 10$^{-2}$ kg m$^{-1}$ s$^{-2}$ (not shown here) gave very unrealistic looking results.
For details on the upper/lower limits of these parameters' probability distributions, see  also Section \ref{Sec_inputs}.
\\Examples of sensitivity tests are shown in Figures \ref{FigsensTest1}, \ref{FigsensTest2} and \ref{FigsensTest3} for two seasons, NH Spring Ls= 0$^\circ$-30$^\circ$ and NH winter Ls = 270$^\circ$-300$^\circ$. 
MCS data are also plotted in all the figures (panels A), for the same periods. 
These tests confirm that the deposition of GW momentum flux into the Martian mesosphere may drastically change the thermal tides. This is particularly discernible in Figure \ref{FigsensTest1}  where we show the impact of the decreasing (Panels B: Case 1) and increasing (Panels D: Case 3) of the probability range of the momentum flux amplitude at the source (F$^{0}_{max}$) by one order of magnitude with respect the reference value (Panel C: Case 2). The larger the GW momentum deposition, the smaller the amplitude and the intensity of the thermal tide, and the more the main mode of diurnal tide is shifted at lower altitudes.
\\It is interesting to note that during the dust storm season Ls = 270$^\circ$-300$^\circ$ (Figure \ref{FigsensTest1} lower panels), stronger non-orographic GW momentum deposition (Panel D: Case 3) is required to better represent observations (e.g. an EP flux value between Case 2 and 3). This may indicate that winter jets add a further source of non-orographic GW and they could enhance the effect of waves propagation in the middle-upper atmosphere, suggesting that a source variable with season is required to fully reproduce the MCS fields (see also Section \ref{diurnal seasonal_var})
\\A similar test was done varying the phase speed, more precisely varying the maximum value of the probability distribution of the propagating GW (c$_{max}$) from 10 m/s to 60 m/s, as shown in Figure \ref{FigsensTest2} (Case 2, Case 6 and Case 7 in Table \ref{tab2}). Those values are comparable with background zonal wind ranges at the source level, and with typical values used in the literature \citep{Medvedev2011}. As shown in Section \ref{impact_jets}, GW propagate vertically and break in the presence of strong zonal jets (the higher the contrast between the mean zonal flow wind and the phase speed, the higher the drag).
In absence of wind jets, as in the southern hemisphere summer, higher phase speed (a factor 2 larger than the nominal values) reproduces slightly better the MCS results, since the waves can propagate vertically at higher altitudes before encountering critical levels and breaking. 
The best values identified for c$_{max}$ during these tests are between 30-40 m/s, which is within the realistic range.
The last set of tests (Case 2, 4 and 5 as in Table \ref{tab2}) justifies the selection of the saturation parameter $S_c = 1$ in our reference simulations (see Section \ref{gw_param}). The saturation controls the altitude at which the GW break. Values between 1 and 5 (not shown here) produce similar results. Smaller values (e.g. $S_c = 0.1$ as in Panel B) in our parameterization implies that only a small portion of wave energy is permitted to traverse the atmosphere above the source, and consequently it has a negligible impact on the GCM simulated fields (panels B can be compared with model results in Figures \ref{Tides_Ls0} and \ref{Tides_Ls270} for similar seasons.

\section{Summary and conclusions}
\label{conclusions}
A stochastic parameterization of non-orographic GW  was implemented into the Laboratoire de M\'et\'eorologie Dynamique (LMD) Mars General Circulation Model (LMD-MGCM), following an innovative scheme, as described in \citet{Lott2012,Lott2013}. 
According to those authors, this approach is one of the best choices to represent the unpredictable aspects of the sub-grid dynamics, considering that convection and fronts can generate waves throughout the full range of phase speeds, wave frequency and vertical and horizontal scales.
In addition, GW on Mars are still poorly constrained in terms of their basic parameters, including their sources of excitation, geographical distribution and vertical levels of dissipation. 
Given this uncertainty it is difficult to quantify the impact of non-orographic GW drag in the middle atmosphere of Mars using a unique set of parameters. 
\\The LMD-MGCM already included a parametrization of orographic GW \citep{Forget1999, Angelatsicoll2005} based on the GW drag scheme of \cite{LottMiller1997} and \cite{Miller1989}. At the time the scheme parameters were chosen conservatively and the effect of the orographic wave drag probably underestimated. Here, we added a more general scheme which includes non-orographic GW, but it is possible that this also accounts for the previously underestimated effect of orographic GW. At this stage it is difficult to separate the contribution of orographic sources to the wind drag from non-orographic ones, but this will be certainly the next step in a future study.
In our scheme the source is assumed to be located above typical convective cells ($\sim$ 250 Pa) and a we have included a multiwave parameterization based on a stochastic approach.
\\The main goal here is to understand  the role of non-orographic GW on the global circulation and  the thermal structure of the Martian middle atmosphere.
Our strategy was to identify a subset of "best-fit" GW characteristics by comparing LMD-MGCM simulations of the full MY29 and MCS/MRO temperature data, mainly between 1 and 0.01 Pa ($\sim$ 50 and 80 km), where data-model biases were systematically larger before implementing the non-orographic GW scheme and locally reaching 10 to 20 K. 
\\The results show that the inclusion of a key physical mechanism such as the propagation of GW from the convective region to the upper atmosphere may partially explain those discrepancies at mesospheric layers, and it improves the accuracy of the LMD-MGCM, in comparison with MCS observations.
Using the baseline parameters as in Table \ref{tab1},the GW drag occurs mostly above 1 Pa ($\sim$ 50 km altitude) and the main effect of GW is to slow down the winds everywhere. It predominantly reduces the westerly jets at high latitudes (negative drag) and weakens the equatorial easterly jet by accelerating the zonal wind (positive drag).
Despite the GW decelerating the the westerly jets at mid-high latitudes (from tens to hundreds of m/s) in the NH winter solstice, with GW drag reaching maximum magnitudes of the order of 1 m/s/sol around 10$^{-2}$ Pa, the direct GW drag on the retrograde (easterly) wind at low latitudes is about two order of magnitudes weaker. This is due to the damping of the diurnal tides by the GW, and thus of the mean zonal wind, since the thermal tides - mean flow interaction is the main forcing of this retrograde jet.
\\Interestingly those results show that, in spite of the strong impact of GW on zonal winds, there are only minor changes on the predicted temperature field and a small reduction in the magnitude of the biases between MCS and LMD-MGCM at most seasons and latitudes. The physical explanation is that the changes in the temperature fields are not directly linked to the zonal mean drag but with the smaller GW-induced acceleration/deceleration of the meridional flow. There are however some notable exceptions of major changes in the temperature: (i) The mid-altitude negative bias at high southern latitudes in the first half of the year (northern spring and summer) is greatly reduced up to 20 K (ii) The mid-altitude positive bias at high northern latitudes in early winter (L$_s$= 240-270)  also decreases by 15-20 K (iii) Positive biases in northern mid-latitudes (40-60$^\circ$N) are generally reduced up to 10 K at all altitudes and times of the year, except at very high altitudes around local winter solstice (iv) Large negative biases at low altitudes at high latitudes (in northern fall) and northern mid-latitudes (in northern fall and winter) are not improved by the non-orographic GW parameterization and may be due to incorrect dust distributions and/or interactions between dust and water ice clouds during the dustiest time of the year.
In addition, remaining biases with GW parameters may be attributed to aerosol effects, that need to be explored in further studies.
Regarding the diurnal tides, they are better represented by the LMD-GCM when the GW scheme is on, especially their amplitude above 10 Pa at the equator. As previously discussed, the primary effect of GW is to damp the thermal tides by reducing the diurnal oscillation of the meridional and zonal wind. With the non-orographic GW scheme activated in the model the maximum day-night difference value is around 20 K, comparable with observed values, and also the peak of the tides is shifted at lower altitudes, in better agreement with MCS data. 
However, the simulations presented in this paper, (both with or without the non-orographic GW scheme included), fail to reproduce  the slight increase of temperature  linked to the dust event at L$_{s}\sim$ 240$^\circ$ observed by MCS around 20-30 Pa, which correspond to a global increase of day and night temperature. This may be related to the fact that the vertical dust distribution, size distribution and/or water ice interactions are not correctly represented in our model, and they are producing the wrong thermal structure at that L$_s$. Otherwise this suggests  that during dust storms there are more GW propagating in the atmosphere, and a non-orographic GW source varying with seasons may be required to further improve the results.

%
%
%
%
%
%
%

\acknowledgments
This work was supported by the European Space Agency contract 4000122721/17/NL/LF/as and the Funda\c{c}\~{a}o para a Ci\^{e}ncia e a Tecnologia (FCT, Portugal) through national funds and from FEDER through COMPETE2020 by grants UID/FIS/04434/2013, POCI-01-0145-FEDER-007672, PTDC/FIS-AST/1526/2014, POCI-01-0145-FEDER-016886, and P-TUGA PTDC/FIS-AST/29942/2017. GG also received funding from the European Union$'$s Horizon 2020 research and innovation programme under the Marie Sklodowska-Curie grant agreement No 796923.
This work was granted access to the HPC resources of CINES (Grant A0040110391). 
MGCM and MCS data used in the paper are available at  https://doi.org/10.14768/20181126001.1


\begin{thebibliography}{71}
	\providecommand{\natexlab}[1]{#1}
	\expandafter\ifx\csname urlstyle\endcsname\relax
	\providecommand{\doi}[1]{doi:\discretionary{}{}{}#1}\else
	\providecommand{\doi}{doi:\discretionary{}{}{}\begingroup
		\urlstyle{rm}\Url}\fi
	
	\bibitem[{\textit{{Alexander} et~al.}(2010)\textit{{Alexander}, {Geller},
			{McLandress}, {Polavarapu}, {Preusse}, {Sassi}, {Sato}, {Eckermann}, {Ern},
			{Hertzog}, {Kawatani}, {Pulido}, {Shaw}, {Sigmond}, {Vincent}, and
			{Watanabe}}}]{Alexander2010}
	{Alexander}, M.~J., M.~{Geller}, C.~{McLandress}, S.~{Polavarapu},
	P.~{Preusse}, F.~{Sassi}, K.~{Sato}, S.~{Eckermann}, M.~{Ern}, A.~{Hertzog},
	Y.~{Kawatani}, M.~{Pulido}, T.~A. {Shaw}, M.~{Sigmond}, R.~{Vincent}, and
	S.~{Watanabe} (2010), Recent developments in gravity wave effects in climate
	models and the global distribution of gravity wave momentum flux from
	observations and models, \textit{Quarterly Journal of the Royal
		Meteorological Society}, \textit{136}(650), 1103--1124, \doi{10.1002/qj.637}.
	
	\bibitem[{\textit{{Angelats i Coll} et~al.}(2005)\textit{{Angelats i Coll},
			{Forget}, {L{\'o}pez-Valverde}, and
			{Gonz{\'a}lez-Galindo}}}]{Angelatsicoll2005}
	{Angelats i Coll}, M., F.~{Forget}, M.~A. {L{\'o}pez-Valverde}, and
	F.~{Gonz{\'a}lez-Galindo} (2005), {The first Mars thermospheric general
		circulation model: The Martian atmosphere from the ground to 240 km},
	\textit{Geophysical Research Letters}, \textit{32}, 4201--+.
	
	\bibitem[{\textit{{Barnes}}(1990)}]{Barnes1990}
	{Barnes}, J.~R. (1990), {Possible effects of breaking gravity waves on the
		circulation of the middle atmosphere of Mars}, \textit{Journal of Geophysical
		Research}, \textit{95}, 1401--1421, \doi{10.1029/JB095iB02p01401}.
	
	\bibitem[{\textit{{Cola{\"i}tis} et~al.}(2013)\textit{{Cola{\"i}tis}, {Spiga},
			{Hourdin}, {Rio}, {Forget}, and {Millour}}}]{Colaitis2013}
	{Cola{\"i}tis}, A., A.~{Spiga}, F.~{Hourdin}, C.~{Rio}, F.~{Forget}, and
	E.~{Millour} (2013), {A thermal plume model for the Martian convective
		boundary layer}, \textit{Journal of Geophysical Research (Planets)},
	\textit{118}, 1468--1487, \doi{10.1002/jgre.20104}.
	
	\bibitem[{\textit{{Collins} et~al.}(1997)\textit{{Collins}, {Lewis}, and
			{Read}}}]{Collins1997}
	{Collins}, M., S.~R. {Lewis}, and P.~L. {Read} (1997), {Gravity wave drag in a
		global circulation model of the martian atmosphere: parameterisation and
		validation}, \textit{Advances in Space Research}, \textit{19}, 1245--1254,
	\doi{10.1016/S0273-1177(97)00277-9}.
	
	\bibitem[{\textit{{Creasey} et~al.}(2006{\natexlab{a}})\textit{{Creasey},
			{Forbes}, and {Keating}}}]{Creasey2006b}
	{Creasey}, J.~E., J.~M. {Forbes}, and G.~M. {Keating} (2006{\natexlab{a}}),
	{Density variability at scales typical of gravity waves observed in Mars'
		thermosphere by the MGS accelerometer}, \textit{Geophysical Research
		Letters}, \textit{33}, L22814, \doi{10.1029/2006GL027583}.
	
	\bibitem[{\textit{{Creasey} et~al.}(2006{\natexlab{b}})\textit{{Creasey},
			{Forbes}, and {Hinson}}}]{Creasey2006a}
	{Creasey}, J.~E., J.~M. {Forbes}, and D.~P. {Hinson} (2006{\natexlab{b}}),
	{Global and seasonal distribution of gravity wave activity in Mars' lower
		atmosphere derived from MGS radio occultation data}, \textit{Geophysical
		Research Letters}, \textit{33}, L01803, \doi{10.1029/2005GL024037}.
	
	\bibitem[{\textit{{Forbes} et~al.}(2002)\textit{{Forbes}, {Bridger}, {Bougher},
			{Hagan}, {Hollingsworth}, {Keating}, and {Murphy}}}]{Forbes2002}
	{Forbes}, J.~M., A.~F.~C. {Bridger}, S.~W. {Bougher}, M.~E. {Hagan}, J.~L.
	{Hollingsworth}, G.~M. {Keating}, and J.~{Murphy} (2002), {Nonmigrating tides
		in the thermosphere of Mars}, \textit{Journal of Geophysical Research
		(Planets)}, \textit{107}(E11), 5113, \doi{10.1029/2001JE001582}.
	
	\bibitem[{\textit{{Forget} et~al.}(1998)\textit{{Forget}, {Hourdin}, and
			{Talagrand}}}]{Forget1998}
	{Forget}, F., F.~{Hourdin}, and O.~{Talagrand} (1998), {CO $_{2}$Snowfall on
		Mars: Simulation with a General Circulation Model}, \textit{Icarus},
	\textit{131}, 302--316, \doi{10.1006/icar.1997.5874}.
	
	\bibitem[{\textit{Forget et~al.}(1999)\textit{Forget, Hourdin, Fournier,
			Hourdin, Talagrand, Collins, Lewis, Read, and Huot.}}]{Forget1999}
	Forget, F., F.~Hourdin, R.~Fournier, C.~Hourdin, O.~Talagrand, M.~Collins,
	S.~R. Lewis, P.~L. Read, and J.-P. Huot. (1999), Improved general circulation
	models of the {Martian} atmosphere from the surface to above 80~km,
	\textit{J. Geophys. Res.}, \textit{104}, 24,155--24,176.
	
	\bibitem[{\textit{{Forget} et~al.}(2009)\textit{{Forget}, {Montmessin},
			{Bertaux}, {Gonz{\'a}lez-Galindo}, {Lebonnois}, {Qu{\'e}merais}, {Reberac},
			{Dimarellis}, and {L{\'o}pez-Valverde}}}]{Forget2009}
	{Forget}, F., F.~{Montmessin}, J.-L. {Bertaux}, F.~{Gonz{\'a}lez-Galindo},
	S.~{Lebonnois}, E.~{Qu{\'e}merais}, A.~{Reberac}, E.~{Dimarellis}, and M.~A.
	{L{\'o}pez-Valverde} (2009), {Density and temperatures of the upper Martian
		atmosphere measured by stellar occultations with Mars Express SPICAM},
	\textit{Journal of Geophysical Research (Planets)}, \textit{114}, E01004,
	\doi{10.1029/2008JE003086}.
	
	\bibitem[{\textit{{Forget} et~al.}(2017)\textit{{Forget}, {Wang}, {Pottier},
			{Millour}, {Gilli}, {Vals}, {Zakharov}, {Spiga}, {Navarro}, and
			{Montabone}}}]{Forget2017mamo}
	{Forget}, F., C.~{Wang}, A.~{Pottier}, E.~{Millour}, G.~{Gilli}, M.~{Vals},
	V.~{Zakharov}, A.~{Spiga}, T.~{Navarro}, and L.~{Montabone} (2017), {Major
		Challenges in Mars Climate Modeling: Dust, Clouds and Waves}, in \textit{The
		Mars Atmosphere: Modelling and observation}, edited by F.~{Forget} and
	M.~{Millour}, p. 1113.
	
	\bibitem[{\textit{{Fritts} and {Alexander}}(2003)}]{FrittsAlexander2003}
	{Fritts}, D.~C., and M.~J. {Alexander} (2003), {Gravity wave dynamics and
		effects in the middle atmosphere}, \textit{Reviews of Geophysics},
	\textit{41}, 1003, \doi{10.1029/2001RG000106}.
	
	\bibitem[{\textit{{Fritts} et~al.}(2006)\textit{{Fritts}, {Wang}, and
			{Tolson}}}]{Fritts2006}
	{Fritts}, D.~C., L.~{Wang}, and R.~H. {Tolson} (2006), {Mean and gravity wave
		structures and variability in the Mars upper atmosphere inferred from Mars
		Global Surveyor and Mars Odyssey aerobraking densities}, \textit{Journal of
		Geophysical Research (Space Physics)}, \textit{111}, A12304,
	\doi{10.1029/2006JA011897}.
	
	\bibitem[{\textit{{Gilli} et~al.}(2017)\textit{{Gilli}, {Lebonnois},
			{Gonz{\'a}lez-Galindo}, {L{\'o}pez-Valverde}, {Stolzenbach}, {Lef{\`e}vre},
			{Chaufray}, and {Lott}}}]{Gilli2017}
	{Gilli}, G., S.~{Lebonnois}, F.~{Gonz{\'a}lez-Galindo}, M.~A.
	{L{\'o}pez-Valverde}, A.~{Stolzenbach}, F.~{Lef{\`e}vre}, J.~Y. {Chaufray},
	and F.~{Lott} (2017), {Thermal structure of the upper atmosphere of Venus
		simulated by a ground-to-thermosphere GCM}, \textit{Icarus}, \textit{281},
	55--72, \doi{10.1016/j.icarus.2016.09.016}.
	
	\bibitem[{\textit{{Guzewich} et~al.}(2012)\textit{{Guzewich}, {Talaat}, and
			{Waugh}}}]{Guzewich2012}
	{Guzewich}, S.~D., E.~R. {Talaat}, and D.~W. {Waugh} (2012), {Observations of
		planetary waves and nonmigrating tides by the Mars Climate Sounder},
	\textit{Journal of Geophysical Research (Planets)}, \textit{117}, E03010,
	\doi{10.1029/2011JE003924}.
	
	\bibitem[{\textit{{Haberle} et~al.}(1999)\textit{{Haberle}, {Joshi}, {Murphy},
			{Barnes}, {Schofield}, {Wilson}, {Lopez-Valverde}, {Hollingsworth},
			{Bridger}, and {Schaeffer}}}]{Haberle1999}
	{Haberle}, R.~M., M.~M. {Joshi}, J.~R. {Murphy}, J.~R. {Barnes}, J.~T.
	{Schofield}, G.~{Wilson}, M.~{Lopez-Valverde}, J.~L. {Hollingsworth},
	A.~F.~C. {Bridger}, and J.~{Schaeffer} (1999), {General circulation model
		simulations of the Mars Pathfinder atmospheric structure
		investigation/meteorology data}, \textit{Journal of Geophysical Research},
	\textit{104}, 8957--8974, \doi{10.1029/1998JE900040}.
	
	\bibitem[{\textit{{Hartogh} et~al.}(2005)\textit{{Hartogh}, {Medvedev},
			{Kuroda}, {Saito}, {Villanueva}, {Feofilov}, {Kutepov}, and
			{Berger}}}]{Hartogh2005}
	{Hartogh}, P., A.~S. {Medvedev}, T.~{Kuroda}, R.~{Saito}, G.~{Villanueva},
	A.~G. {Feofilov}, A.~A. {Kutepov}, and U.~{Berger} (2005), {Description and
		climatology of a new general circulation model of the Martian atmosphere},
	\textit{Journal of Geophysical Research (Planets)}, \textit{110}, E11008,
	\doi{10.1029/2005JE002498}.
	
	\bibitem[{\textit{{Hauchecorne} et~al.}(1987)\textit{{Hauchecorne}, {Chanin},
			and {Wilson}}}]{Hauchecorne1987}
	{Hauchecorne}, A., M.~L. {Chanin}, and R.~{Wilson} (1987), {Mesospheric
		temperature inversion and gravity wave breaking}, \textit{Geophys Res.
		Lett.}, \textit{14}, 933--936, \doi{10.1029/GL014i009p00933}.
	
	\bibitem[{\textit{{Heavens} et~al.}(2011{\natexlab{a}})\textit{{Heavens},
			{Richardson}, {Kleinb{\"o}hl}, {Kass}, {McCleese}, {Abdou}, {Benson},
			{Schofield}, {Shirley}, and {Wolkenberg}}}]{Heavens2011a}
	{Heavens}, N.~G., M.~I. {Richardson}, A.~{Kleinb{\"o}hl}, D.~M. {Kass}, D.~J.
	{McCleese}, W.~{Abdou}, J.~L. {Benson}, J.~T. {Schofield}, J.~H. {Shirley},
	and P.~M. {Wolkenberg} (2011{\natexlab{a}}), {The vertical distribution of
		dust in the Martian atmosphere during northern spring and summer:
		Observations by the Mars Climate Sounder and analysis of zonal average
		vertical dust profiles}, \textit{Journal of Geophysical Research (Planets)},
	\textit{116}, E04003, \doi{10.1029/2010JE003691}.
	
	\bibitem[{\textit{{Heavens} et~al.}(2011{\natexlab{b}})\textit{{Heavens},
			{Richardson}, {Kleinb{\"o}hl}, {Kass}, {McCleese}, {Abdou}, {Benson},
			{Schofield}, {Shirley}, and {Wolkenberg}}}]{Heavens2011b}
	{Heavens}, N.~G., M.~I. {Richardson}, A.~{Kleinb{\"o}hl}, D.~M. {Kass}, D.~J.
	{McCleese}, W.~{Abdou}, J.~L. {Benson}, J.~T. {Schofield}, J.~H. {Shirley},
	and P.~M. {Wolkenberg} (2011{\natexlab{b}}), {Vertical distribution of dust
		in the Martian atmosphere during northern spring and summer: High-altitude
		tropical dust maximum at northern summer solstice}, \textit{Journal of
		Geophysical Research (Planets)}, \textit{116}, E01007,
	\doi{10.1029/2010JE003692}.
	
	\bibitem[{\textit{{Heavens} et~al.}(2011{\natexlab{c}})\textit{{Heavens},
			{McCleese}, {Richardson}, {Kass}, {Kleinb{\"o}hl}, and
			{Schofield}}}]{Heavens2011c}
	{Heavens}, N.~G., D.~J. {McCleese}, M.~I. {Richardson}, D.~M. {Kass},
	A.~{Kleinb{\"o}hl}, and J.~T. {Schofield} (2011{\natexlab{c}}), {Structure
		and dynamics of the Martian lower and middle atmosphere as observed by the
		Mars Climate Sounder: 2. Implications of the thermal structure and aerosol
		distributions for the mean meridional circulation}, \textit{Journal of
		Geophysical Research (Planets)}, \textit{116}, E01010,
	\doi{10.1029/2010JE003713}.
	
	\bibitem[{\textit{{Hines}}(1997)}]{Hines1997}
	{Hines}, C.~O. (1997), {Doppler-spread parameterization of gravity-wave
		momentum deposition in the middle atmosphere. Part 1: Basic formulation},
	\textit{Journal of Atmospheric and Solar-Terrestrial Physics}, \textit{59},
	371--386, \doi{10.1016/S1364-6826(96)00079-X}.
	
	\bibitem[{\textit{{Hinson} et~al.}(1999)\textit{{Hinson}, {Simpson}, {Twicken},
			{Tyler}, and {Flasar}}}]{Hinson1999}
	{Hinson}, D.~P., R.~A. {Simpson}, J.~D. {Twicken}, G.~L. {Tyler}, and F.~M.
	{Flasar} (1999), {Initial results from radio occultation measurements with
		Mars Global Surveyor}, \textit{Journal of Geophysical Research},
	\textit{104}, 26,997--27,012, \doi{10.1029/1999JE001069}.
	
	\bibitem[{\textit{{Hinson} et~al.}(2008)\textit{{Hinson}, {P{\"a}tzold},
			{Tellmann}, {H{\"a}usler}, and {Tyler}}}]{Hinson2008}
	{Hinson}, D.~P., M.~{P{\"a}tzold}, S.~{Tellmann}, B.~{H{\"a}usler}, and G.~L.
	{Tyler} (2008), {The depth of the convective boundary layer on Mars},
	\textit{Icarus}, \textit{198}, 57--66, \doi{10.1016/j.icarus.2008.07.003}.
	
	\bibitem[{\textit{{Holstein-Rathlou} et~al.}(2016)\textit{{Holstein-Rathlou},
			{Maue}, and {Withers}}}]{Holstein-Rathlou2016}
	{Holstein-Rathlou}, C., A.~{Maue}, and P.~{Withers} (2016), {Atmospheric
		studies from the Mars Science Laboratory Entry, Descent and Landing
		atmospheric structure reconstruction}, \textit{Planet Space Sci},
	\textit{120}, 15--23, \doi{10.1016/j.pss.2015.10.015}.
	
	\bibitem[{\textit{{Holton}}(1982)}]{Holton1982}
	{Holton}, J.~R. (1982), {The role of gravity wave induced drag and diffusion in
		the momentum budget of the mesosphere}, \textit{Journal of Atmospheric
		Sciences}, \textit{39}, 791--799,
	\doi{10.1175/1520-0469(1982)039<0791:TROGWI>2.0.CO;2}.
	
	\bibitem[{\textit{{Jakosky} and {Martin}}(1987)}]{JakoskyMartin1987}
	{Jakosky}, B.~M., and T.~Z. {Martin} (1987), {Mars - North-Polar atmospheric
		warming during dust storms}, \textit{Icarus}, \textit{72}, 528--534,
	\doi{10.1016/0019-1035(87)90050-9}.
	
	\bibitem[{\textit{{Joshi} et~al.}(1995)\textit{{Joshi}, {Lawrence}, and
			{Lewis}}}]{Joshi1995}
	{Joshi}, M.~M., B.~N. {Lawrence}, and S.~R. {Lewis} (1995), {Gravity wave drag
		in three-dimensional atmospheric models of Mars}, \textit{Journal of
		Geophysical Research}, \textit{100}, 21,235--21,246, \doi{10.1029/95JE02486}.
	
	\bibitem[{\textit{{Kahre} et~al.}(2015)\textit{{Kahre}, {Hollingsworth},
			{Haberle}, and {Wilson}}}]{Kahre2015}
	{Kahre}, M.~A., J.~L. {Hollingsworth}, R.~M. {Haberle}, and R.~J. {Wilson}
	(2015), {Coupling the Mars dust and water cycles: The importance of
		radiative-dynamic feedbacks during northern hemisphere summer},
	\textit{Icarus}, \textit{260}, 477--480, \doi{10.1016/j.icarus.2014.07.017}.
	
	\bibitem[{\textit{{Keating} et~al.}(1998)\textit{{Keating}, {Bougher}, {Zurek},
			{Tolson}, {Cancro}, {Noll}, {Parker}, {Schellenberg}, {Shane}, {Wilkerson},
			{Murphy}, {Hollingsworth}, {Haberle}, {Joshi}, {Pearl}, {Conrath}, {Smith},
			{Clancy}, {Blanchard}, {Wilmoth}, {Rault}, {Martin}, {Lyons}, {Esposito},
			{Johnston}, {Whetzel}, {Justus}, and {Babicke}}}]{Keating1998}
	{Keating}, G.~M., S.~W. {Bougher}, R.~W. {Zurek}, R.~H. {Tolson}, G.~J.
	{Cancro}, S.~N. {Noll}, J.~S. {Parker}, T.~J. {Schellenberg}, R.~W. {Shane},
	B.~L. {Wilkerson}, J.~R. {Murphy}, J.~L. {Hollingsworth}, R.~M. {Haberle},
	M.~{Joshi}, J.~C. {Pearl}, B.~J. {Conrath}, M.~D. {Smith}, R.~T. {Clancy},
	R.~C. {Blanchard}, R.~G. {Wilmoth}, D.~F. {Rault}, T.~Z. {Martin}, D.~T.
	{Lyons}, P.~B. {Esposito}, M.~D. {Johnston}, C.~W. {Whetzel}, C.~G. {Justus},
	and J.~M. {Babicke} (1998), {The Structure of the Upper Atmosphere of Mars:
		In Situ Accelerometer Measurements from Mars Global Surveyor},
	\textit{Science}, \textit{279}, 1672, \doi{10.1126/science.279.5357.1672}.
	
	\bibitem[{\textit{{Keating} et~al.}(2007)\textit{{Keating}, {Bougher},
			{Theriot}, {Zurek}, {Blanchard}, {Tolson}, and {Murphy}}}]{Keating2007AGU}
	{Keating}, G.~M., S.~W. {Bougher}, M.~E. {Theriot}, R.~W. {Zurek}, R.~C.
	{Blanchard}, R.~H. {Tolson}, and J.~R. {Murphy} (2007), {Mars Reconnaissance
		Orbiter Accelerometer Experiment Results}, \textit{AGU Spring Meeting
		Abstracts}, P23A-03.
	
	\bibitem[{\textit{{Kleinb{\"o}hl} et~al.}(2009)\textit{{Kleinb{\"o}hl},
			{Schofield}, {Kass}, {Abdou}, {Backus}, {Sen}, {Shirley}, {Lawson},
			{Richardson}, {Taylor}, {Teanby}, and {McCleese}}}]{Kleinboehl2009}
	{Kleinb{\"o}hl}, A., J.~T. {Schofield}, D.~M. {Kass}, W.~A. {Abdou}, C.~R.
	{Backus}, B.~{Sen}, J.~H. {Shirley}, W.~G. {Lawson}, M.~I. {Richardson},
	F.~W. {Taylor}, N.~A. {Teanby}, and D.~J. {McCleese} (2009), {Mars Climate
		Sounder limb profile retrieval of atmospheric temperature, pressure, and dust
		and water ice opacity}, \textit{Journal of Geophysical Research (Planets)},
	\textit{114}, E10006, \doi{10.1029/2009JE003358}.
	
	\bibitem[{\textit{{Kleinb{\"o}hl} et~al.}(2013)\textit{{Kleinb{\"o}hl}, {John
				Wilson}, {Kass}, {Schofield}, and {McCleese}}}]{Kleinbohl2013}
	{Kleinb{\"o}hl}, A., R.~{John Wilson}, D.~{Kass}, J.~T. {Schofield}, and D.~J.
	{McCleese} (2013), {The semidiurnal tide in the middle atmosphere of Mars},
	\textit{Geophys Res. Lett.}, \textit{40}, 1952--1959,
	\doi{10.1002/grl.50497}.
	
	\bibitem[{\textit{{Kleinb{\"o}hl} et~al.}(2017)\textit{{Kleinb{\"o}hl},
			{Friedson}, and {Schofield}}}]{Kleinboehl2017}
	{Kleinb{\"o}hl}, A., A.~J. {Friedson}, and J.~T. {Schofield} (2017),
	{Two-dimensional radiative transfer for the retrieval of limb emission
		measurements in the martian atmosphere}, \textit{\textit{Journal of
			Quantitative Spectroscopy and Radiative transfer}}, \textit{187}, 511--522,
	\doi{10.1016/j.jqsrt.2016.07.009}.

\bibitem[{\textit{{Kuroda} et~al.}(2015)\textit{{Kuroda}, {Medvedev},
		{Yi{\v{g}}it}, and {Hartogh}}}]{Kuroda2015}
{Kuroda}, T., A.~S. {Medvedev}, E.~{Yi{\v{g}}it}, and P.~{Hartogh} (2015), {A
	global view of gravity waves in the Martian atmosphere inferred from a
	high-resolution general circulation model}, \textit{Geophys Res.
	Lett.}, \textit{42}(21),
9213--9222, \doi{10.1002/2015GL066332}.

\bibitem[{\textit{{Kuroda} et~al.}(2016)\textit{{Kuroda}, {Medvedev},
		{Yi{\u{g}}it}, and {Hartogh}}}]{Kuroda2016}
{Kuroda}, T., A.~S. {Medvedev}, E.~{Yi{\u{g}}it}, and P.~{Hartogh} (2016),
{Global Distribution of Gravity Wave Sources and Fields in the Martian
	Atmosphere during Equinox and Solstice Inferred from a High-Resolution
	General Circulation Model}, \textit{Journal of Atmospheric Sciences},
\textit{73}(12), 4895--4909, \doi{10.1175/JAS-D-16-0142.1}.

\bibitem[{\textit{{Kuroda} et~al.}(2019)\textit{{Kuroda}, {Yi{\v{g}}it}, and
		{Medvedev}}}]{Kuroda2019}
{Kuroda}, T., E.~{Yi{\v{g}}it}, and A.~S. {Medvedev} (2019), {Annual Cycle of
	Gravity Wave Activity Derived From a High-Resolution Martian General
	Circulation Model}, \textit{Journal of Geophysical Research (Planets)},
\textit{124}(6), 1618--1632, \doi{10.1029/2018JE005847}.
	
	\bibitem[{\textit{{Lee} et~al.}(2009)\textit{{Lee}, {Lawson}, {Richardson},
			{Heavens}, {Kleinb{\"o}hl}, {Banfield}, {McCleese}, {Zurek}, {Kass},
			{Schofield}, {Leovy}, {Taylor}, and {Toigo}}}]{Lee2009}
	{Lee}, C., W.~G. {Lawson}, M.~I. {Richardson}, N.~G. {Heavens},
	A.~{Kleinb{\"o}hl}, D.~{Banfield}, D.~J. {McCleese}, R.~{Zurek}, D.~{Kass},
	J.~T. {Schofield}, C.~B. {Leovy}, F.~W. {Taylor}, and A.~D. {Toigo} (2009),
	{Thermal tides in the Martian middle atmosphere as seen by the Mars Climate
		Sounder}, \textit{Journal of Geophysical Research (Planets)}, \textit{114},
	E03005, \doi{10.1029/2008JE003285}.
	
	\bibitem[{\textit{{Lindzen}}(1981)}]{Lindzen1981}
	{Lindzen}, R.~S. (1981), {Turbulence and stress owing to gravity wave and tidal
		breakdown}, \textit{Journal of Geophysical Research}, \textit{86},
	9707--9714, \doi{10.1029/JC086iC10p09707}.
	
	\bibitem[{\textit{{Lindzen} and {Holton}}(1968)}]{LindzenHolton1968}
	{Lindzen}, R.~S., and J.~R. {Holton} (1968), {A Theory of the Quasi-Biennial
		Oscillation.}, \textit{Journal of Atmospheric Sciences}, \textit{25},
	1095--1107, \doi{10.1175/1520-0469(1968)025<1095:ATOTQB>2.0.CO;2}.
	
	\bibitem[{\textit{{Lopez-Valverde} et~al.}(2016)\textit{{Lopez-Valverde},
			{Montabone}, {Sornig}, and {Sonnabend}}}]{Valverde2016}
	{Lopez-Valverde}, M.~A., L.~{Montabone}, M.~{Sornig}, and G.~{Sonnabend}
	(2016), {On the Retrieval of Mesospheric Winds on Mars and Venus from
		Ground-based Observations at 10 um}, \textit{The Astrophysical Journal},
	\textit{816}(2), 103, \doi{10.3847/0004-637X/816/2/103}.
	
	\bibitem[{\textit{{Lott} and {Guez}}(2013)}]{Lott2013}
	{Lott}, F., and L.~{Guez} (2013), {A stochastic parameterization of the gravity
		waves due to convection and its impact on the equatorial stratosphere},
	\textit{Journal of Geophysical Research (Atmospheres)}, \textit{118},
	8897--8909, \doi{10.1002/jgrd.50705}.
	
	\bibitem[{\textit{{Lott} and {Miller}}(1997)}]{LottMiller1997}
	{Lott}, F., and {Miller} (1997), {A new subgrid-scale orographic drag
		parametrization: Its formulation and testing}, \textit{Quarterly Journal of
		the Royal Meteorological Society}, \textit{123}, 101--127,
	\doi{10.1002/qj.49712353704}.
	
	\bibitem[{\textit{{Lott} et~al.}(2012)\textit{{Lott}, {Guez}, and
			{Maury}}}]{Lott2012}
	{Lott}, F., L.~{Guez}, and P.~{Maury} (2012), {A stochastic parameterization of
		non-orographic gravity waves: Formalism and impact on the equatorial
		stratosphere}, \textit{Geophysical Research Letters}, \textit{39}, L06807,
	\doi{10.1029/2012GL051001}.
	
	\bibitem[{\textit{{Madeleine} et~al.}(2011)\textit{{Madeleine}, {Forget},
			{Millour}, {Montabone}, and {Wolff}}}]{Madeleine2011}
	{Madeleine}, J.-B., F.~{Forget}, E.~{Millour}, L.~{Montabone}, and M.~J.
	{Wolff} (2011), {Revisiting the radiative impact of dust on Mars using the
		LMD Global Climate Model}, \textit{Journal of Geophysical Research
		(Planets)}, \textit{116}, E11010, \doi{10.1029/2011JE003855}.
	
	\bibitem[{\textit{{Madeleine} et~al.}(2012)\textit{{Madeleine}, {Forget},
			{Millour}, {Navarro}, and {Spiga}}}]{Madeleine2012}
	{Madeleine}, J.-B., F.~{Forget}, E.~{Millour}, T.~{Navarro}, and A.~{Spiga}
	(2012), {The influence of radiatively active water ice clouds on the Martian
		climate}, \textit{Geophys Res. Lett.}, \textit{39}, L23202,
	\doi{10.1029/2012GL053564}.
	
	\bibitem[{\textit{{Madeleine} et~al.}(2014)\textit{{Madeleine}, {Head},
			{Forget}, {Navarro}, {Millour}, {Spiga}, {Cola{\"i}tis},
			{M{\"a}{\"a}tt{\"a}nen}, {Montmessin}, and {Dickson}}}]{Madeleine2014}
	{Madeleine}, J.-B., J.~W. {Head}, F.~{Forget}, T.~{Navarro}, E.~{Millour},
	A.~{Spiga}, A.~{Cola{\"i}tis}, A.~{M{\"a}{\"a}tt{\"a}nen}, F.~{Montmessin},
	and J.~L. {Dickson} (2014), {Recent Ice Ages on Mars: The role of radiatively
		active clouds and cloud microphysics}, \textit{Geophys Res. Lett.},
	\textit{41}, 4873--4879, \doi{10.1002/2014GL059861}.
	
	\bibitem[{\textit{{Magalh{\~a}es} et~al.}(1999)\textit{{Magalh{\~a}es},
			{Schofield}, and {Seiff}}}]{Magalhaes1999}
	{Magalh{\~a}es}, J.~A., J.~T. {Schofield}, and A.~{Seiff} (1999), {Results of
		the Mars Pathfinder atmospheric structure investigation}, \textit{Journal of
		Geophysical Research}, \textit{104}, 8943--8956, \doi{10.1029/1998JE900041}.
	
	\bibitem[{\textit{{McCleese} et~al.}(2007)\textit{{McCleese}, {Schofield},
			{Taylor}, {Calcutt}, {Foote}, {Kass}, {Leovy}, {Paige}, {Read}, and
			{Zurek}}}]{McCleese2007}
	{McCleese}, D.~J., J.~T. {Schofield}, F.~W. {Taylor}, S.~B. {Calcutt}, M.~C.
	{Foote}, D.~M. {Kass}, C.~B. {Leovy}, D.~A. {Paige}, P.~L. {Read}, and R.~W.
	{Zurek} (2007), {Mars Climate Sounder: An investigation of thermal and water
		vapor structure, dust and condensate distributions in the atmosphere, and
		energy balance of the polar regions}, \textit{Journal of Geophysical Research
		(Planets)}, \textit{112}, E05S06, \doi{10.1029/2006JE002790}.
	
	\bibitem[{\textit{{McCleese} et~al.}(2010)\textit{{McCleese}, {Heavens},
			{Schofield}, {Abdou}, {Bandfield}, {Calcutt}, {Irwin}, {Kass},
			{Kleinb{\"o}hl}, {Lewis}, {Paige}, {Read}, {Richardson}, {Shirley}, {Taylor},
			{Teanby}, and {Zurek}}}]{McCleese2010}
	{McCleese}, D.~J., N.~G. {Heavens}, J.~T. {Schofield}, W.~A. {Abdou}, J.~L.
	{Bandfield}, S.~B. {Calcutt}, P.~G.~J. {Irwin}, D.~M. {Kass},
	A.~{Kleinb{\"o}hl}, S.~R. {Lewis}, D.~A. {Paige}, P.~L. {Read}, M.~I.
	{Richardson}, J.~H. {Shirley}, F.~W. {Taylor}, N.~{Teanby}, and R.~W. {Zurek}
	(2010), {Structure and dynamics of the Martian lower and middle atmosphere as
		observed by the Mars Climate Sounder: Seasonal variations in zonal mean
		temperature, dust, and water ice aerosols}, \textit{Journal of Geophysical
		Research (Planets)}, \textit{115}(E14), E12016, \doi{10.1029/2010JE003677}.
	
	\bibitem[{\textit{{Medvedev} and {Yi{\v g}it}}(2012)}]{MedvedevYigit2012}
	{Medvedev}, A.~S., and E.~{Yi{\v g}it} (2012), {Thermal effects of internal
		gravity waves in the Martian upper atmosphere}, \textit{Geophys Res. Lett.},
	\textit{39}, L05201, \doi{10.1029/2012GL050852}.
	
	\bibitem[{\textit{{Medvedev} et~al.}(2011)\textit{{Medvedev}, {Yi{\v g}it},
			{Hartogh}, and {Becker}}}]{Medvedev2011}
	{Medvedev}, A.~S., E.~{Yi{\v g}it}, P.~{Hartogh}, and E.~{Becker} (2011),
	{Influence of gravity waves on the Martian atmosphere: General circulation
		modeling}, \textit{Journal of Geophysical Research (Planets)},
	\textit{116}(E15), E10004, \doi{10.1029/2011JE003848}.
	
	\bibitem[{\textit{{Medvedev} et~al.}(2015)\textit{{Medvedev},
			{Gonz{\'a}lez-Galindo}, {Yi{\v g}it}, {Feofilov}, {Forget}, and
			{Hartogh}}}]{Medvedev2015}
	{Medvedev}, A.~S., F.~{Gonz{\'a}lez-Galindo}, E.~{Yi{\v g}it}, A.~G.
	{Feofilov}, F.~{Forget}, and P.~{Hartogh} (2015), {Cooling of the Martian
		thermosphere by CO$_{2}$ radiation and gravity waves: An intercomparison
		study with two general circulation models}, \textit{Journal of Geophysical
		Research (Planets)}, \textit{120}, 913--927, \doi{10.1002/2015JE004802}.
	
	\bibitem[{\textit{{Miller} et~al.}(1989)\textit{{Miller}, {Palmer}, and
			{Swinbank}}}]{Miller1989}
	{Miller}, M.~J., T.~N. {Palmer}, and R.~{Swinbank} (1989), {Parametrization and
		influence of subgridscale orography in general circulation and numerical
		weather prediction models}, \textit{Meteorology and Atmospheric Physics},
	\textit{40}, 84--109, \doi{10.1007/BF01027469}.
	
	\bibitem[{\textit{{Montabone} et~al.}(2015)\textit{{Montabone}, {Forget},
			{Millour}, {Wilson}, {Lewis}, {Cantor}, {Kass}, {Kleinb{\"o}hl}, {Lemmon},
			{Smith}, and {Wolff}}}]{Montabone2015}
	{Montabone}, L., F.~{Forget}, E.~{Millour}, R.~J. {Wilson}, S.~R. {Lewis},
	B.~{Cantor}, D.~{Kass}, A.~{Kleinb{\"o}hl}, M.~T. {Lemmon}, M.~D. {Smith},
	and M.~J. {Wolff} (2015), {Eight-year climatology of dust optical depth on
		Mars}, \textit{\textit{Icarus}}, \textit{251}, 65--95,
	\doi{10.1016/j.icarus.2014.12.034}.
	
	\bibitem[{\textit{{Montmessin} et~al.}(2007)\textit{{Montmessin}, {Gondet},
			{Bibring}, {Langevin}, {Drossart}, {Forget}, and {Fouchet}}}]{Montmessin2007}
	{Montmessin}, F., B.~{Gondet}, J.-P. {Bibring}, Y.~{Langevin}, P.~{Drossart},
	F.~{Forget}, and T.~{Fouchet} (2007), {Hyperspectral imaging of convective
		CO$_{2}$ ice clouds in the equatorial mesosphere of Mars}, \textit{Journal of
		Geophysical Research (Planets)}, \textit{112}, E11S90,
	\doi{10.1029/2007JE002944}.
	
	\bibitem[{\textit{{Navarro} et~al.}(2014)\textit{{Navarro}, {Madeleine},
			{Forget}, {Spiga}, {Millour}, {Montmessin}, and
			{M{\"a}{\"a}tt{\"a}nen}}}]{Navarro2014}
	{Navarro}, T., J.-B. {Madeleine}, F.~{Forget}, A.~{Spiga}, E.~{Millour},
	F.~{Montmessin}, and A.~{M{\"a}{\"a}tt{\"a}nen} (2014), {Global climate
		modeling of the Martian water cycle with improved microphysics and
		radiatively active water ice clouds}, \textit{Journal of Geophysical Research
		(Planets)}, \textit{119}, 1479--1495, \doi{10.1002/2013JE004550}.
	
	\bibitem[{\textit{Navarro et~al.}(2017)\textit{Navarro, Forget, Millour,
			Greybush, Kalnay, and Miyoshi}}]{Navarro2017}
	Navarro, T., F.~Forget, E.~Millour, S.~J. Greybush, E.~Kalnay, and T.~Miyoshi
	(2017), The challenge of atmospheric data assimilation on mars, \textit{Earth
		and Space Science}, \textit{4}(12), 690--722, \doi{10.1002/2017EA000274}.
	
	\bibitem[{\textit{{Parish} et~al.}(2009)\textit{{Parish}, {Schubert}, {Hickey},
			and {Walterscheid}}}]{Parish2009}
	{Parish}, H.~F., G.~{Schubert}, M.~P. {Hickey}, and R.~L. {Walterscheid}
	(2009), {Propagation of tropospheric gravity waves into the upper atmosphere
		of Mars}, \textit{Icarus}, \textit{203}, 28--37,
	\doi{10.1016/j.icarus.2009.04.031}.
	
	\bibitem[{\textit{{Rafkin} et~al.}(2001)\textit{{Rafkin}, {Haberle}, and
			{Michaels}}}]{Rafkin2001}
	{Rafkin}, S.~C.~R., R.~M. {Haberle}, and T.~I. {Michaels} (2001), {The Mars
		Regional Atmospheric Modeling System: Model Description and Selected
		Simulations}, \textit{\textit{Icarus}}, \textit{151}, 228--256,
	\doi{10.1006/icar.2001.6605}.
	
	\bibitem[{\textit{{Sonnabend} et~al.}(2012)\textit{{Sonnabend}, {Sornig},
			{Kroetz}, and {Stupar}}}]{Sonnabend2012}
	{Sonnabend}, G., M.~{Sornig}, P.~{Kroetz}, and D.~{Stupar} (2012), {Mars
		mesospheric zonal wind around northern spring equinox from infrared
		heterodyne observations of CO $_{2}$}, \textit{Icarus}, \textit{217}(1),
	315--321, \doi{10.1016/j.icarus.2011.11.009}.
	
	\bibitem[{\textit{{Spiga} et~al.}(2010)\textit{{Spiga}, {Forget}, {Lewis}, and
			{Hinson}}}]{Spiga2010}
	{Spiga}, A., F.~{Forget}, S.~R. {Lewis}, and D.~P. {Hinson} (2010), {Structure
		and dynamics of the convective boundary layer on Mars as inferred from
		large-eddy simulations and remote-sensing measurements}, \textit{Quarterly
		Journal of the Royal Meteorological Society}, \textit{136}, 414--428.
	
	\bibitem[{\textit{{Spiga} et~al.}(2012)\textit{{Spiga}, {Gonzalez-Galindo},
			{Lopez-Valverde}, and {Forget}}}]{Spiga2012}
	{Spiga}, A., F.~{Gonzalez-Galindo}, M.~{Lopez-Valverde}, and F.~{Forget}
	(2012), {Gravity waves, cold pockets and CO$_{2}$ clouds in the Martian
		mesosphere}, \textit{Geophysical Research Letters}, \textit{39}, L02201,
	\doi{10.1029/2011GL050343}.
	
	\bibitem[{\textit{{Spiga} et~al.}(2013)\textit{{Spiga}, {Faure}, {Madeleine},
			{M{\"a}{\"a}tt{\"a}nen}, and {Forget}}}]{Spiga2013}
	{Spiga}, A., J.~{Faure}, J.-B. {Madeleine}, A.~{M{\"a}{\"a}tt{\"a}nen}, and
	F.~{Forget} (2013), {Rocket dust storms and detached dust layers in the
		Martian atmosphere}, \textit{Journal of Geophysical Research (Planets)},
	\textit{118}, 746--767, \doi{10.1002/jgre.20046}.
	
	\bibitem[{\textit{{Terada} et~al.}(2017)\textit{{Terada}, {Leblanc},
			{Nakagawa}, {Medvedev}, {Yi{\v g}it}, {Kuroda}, {Hara}, {England},
			{Fujiwara}, {Terada}, {Seki}, {Mahaffy}, {Elrod}, {Benna}, {Grebowsky}, and
			{Jakosky}}}]{Terada2017}
	{Terada}, N., F.~{Leblanc}, H.~{Nakagawa}, A.~S. {Medvedev}, E.~{Yi{\v g}it},
	T.~{Kuroda}, T.~{Hara}, S.~L. {England}, H.~{Fujiwara}, K.~{Terada},
	K.~{Seki}, P.~R. {Mahaffy}, M.~{Elrod}, M.~{Benna}, J.~{Grebowsky}, and B.~M.
	{Jakosky} (2017), {Global distribution and parameter dependences of gravity
		wave activity in the Martian upper thermosphere derived from MAVEN/NGIMS
		observations}, \textit{Journal of Geophysical Research (Space Physics)},
	\textit{122}, 2374--2397, \doi{10.1002/2016JA023476}.
	
	\bibitem[{\textit{{Theodore} et~al.}(1993)\textit{{Theodore}, {Lellouch},
			{Chassefiere}, and {Hauchecorne}}}]{Theodore1993}
	{Theodore}, B., E.~{Lellouch}, E.~{Chassefiere}, and A.~{Hauchecorne} (1993),
	{Solstitial temperature inversions in the Martian middle atmosphere -
		Observational clues and 2-D modeling}, \textit{Icarus}, \textit{105}, 512,
	\doi{10.1006/icar.1993.1145}.
	
	\bibitem[{\textit{{Wilson} and {Hamilton}}(1996)}]{WilsonHamilton1996}
	{Wilson}, R.~J., and K.~{Hamilton} (1996), {Comprehensive model simulation of
		thermal tides in the Martian atmosphere.}, \textit{Journal of Atmospheric
		Sciences}, \textit{53}, 1290--1326,
	\doi{10.1175/1520-0469(1996)053<1290:CMSOTT>2.0.CO;2}.
	
	\bibitem[{\textit{{Withers}}(2006)}]{Withers2006}
	{Withers}, P. (2006), {Mars Global Surveyor and Mars Odyssey Accelerometer
		observations of the Martian upper atmosphere during aerobraking},
	\textit{Geophys Res. Lett.}, \textit{33}, L02201, \doi{10.1029/2005GL024447}.
	
	\bibitem[{\textit{{Withers} and {Smith}}(2006)}]{WithersSmith2006}
	{Withers}, P., and M.~D. {Smith} (2006), {Atmospheric entry profiles from the
		Mars Exploration Rovers Spirit and Opportunity}, \textit{Icarus},
	\textit{185}, 133--142, \doi{10.1016/j.icarus.2006.06.013}.
	
	\bibitem[{\textit{{Yi{\v g}it} and {Medvedev}}(2010)}]{Yigit2010}
	{Yi{\v g}it}, E., and A.~S. {Medvedev} (2010), {Internal gravity waves in the
		thermosphere during low and high solar activity: Simulation study},
	\textit{Journal of Geophysical Research (Space Physics)}, \textit{115},
	A00G02, \doi{10.1029/2009JA015106}.
	
	\bibitem[{\textit{{Yi{\v g}it} et~al.}(2015)\textit{{Yi{\v g}it}, {England},
			{Liu}, {Medvedev}, {Mahaffy}, {Kuroda}, and {Jakosky}}}]{Yigit2015}
	{Yi{\v g}it}, E., S.~L. {England}, G.~{Liu}, A.~S. {Medvedev}, P.~R. {Mahaffy},
	T.~{Kuroda}, and B.~M. {Jakosky} (2015), {High-altitude gravity waves in the
		Martian thermosphere observed by MAVEN/NGIMS and modeled by a gravity wave
		scheme}, \textit{Geophysical Research Letters}, \textit{42}, 8993--9000,
	\doi{10.1002/2015GL065307}.
	
	\bibitem[{\textit{{Zurek}}(1976)}]{Zurek1976}
	{Zurek}, R.~W. (1976), {Diurnal tide in the Martian atmosphere},
	\textit{Journal of Atmospheric Sciences}, \textit{33}, 321--337,
	\doi{10.1175/1520-0469(1976)033<0321:DTITMA>2.0.CO;2}.
	
	\bibitem[{\textit{{Zurek} and {Smrekar}}(2007)}]{ZurekSmekar2007}
	{Zurek}, R.~W., and S.~E. {Smrekar} (2007), {An overview of the Mars
		Reconnaissance Orbiter (MRO) science mission}, \textit{Journal of Geophysical
		Research (Planets)}, \textit{112}, E05S01, \doi{10.1029/2006JE002701}.
	
\end{thebibliography}
\end{document}